\documentclass{article}
\usepackage{graphicx} 
\usepackage[utf8]{inputenc}
\usepackage{amsmath,amsbsy,amssymb,array,bm}
\usepackage{nomencl}
\usepackage{subfig}
\usepackage{pdflscape}
\usepackage{caption}
\usepackage{float}
\usepackage{multirow,multicol,tabularx,booktabs}
\usepackage{graphicx,placeins,url} 
\usepackage[dvipsnames,table]{xcolor}
\usepackage{amsthm}
\usepackage[a4paper]{geometry}
\usepackage{authblk}
\usepackage{algorithm} 
\usepackage{algpseudocode} 
\usepackage{adjustbox}
\usepackage{graphicx}
\usepackage{amsmath}
\usepackage{cite}
\usepackage[colorinlistoftodos]{todonotes}
\usepackage{tikz}
\usetikzlibrary{arrows.meta, positioning}
\usepackage{ circuitikz }
\usepackage{acro}
\usepackage[acronym, nonumberlist, nogroupskip]{glossaries}
\usepackage{pgfplots}
\usepackage{rotating}
\pgfplotsset{compat=1.18}
\usepackage{lscape}

\newtheorem{thm}{Theorem}
\newtheorem{cor}[thm]{Corollary}
\newtheorem{prop}[thm]{Proposition}

\newtheorem{defi}{Definition}

\newtheorem{assump}{Assumption}

\def\Pb{{\mathbb P}}

\def\Esp{\mathbb{E}}

\def\Span{\mbox{Span}}

\title{Synthetic Data for Portfolios:\\A Throw of the Dice Will Never Abolish Chance}
\author{Adil Rengim CETINGOZ\thanks{Université Paris 1 Panthéon-Sorbonne, Centre d’Economie de la Sorbonne, 106 Boulevard de l’Hôpital, 75642 Paris Cedex 13, France, rengimcetingoz@gmail.com} \quad Charles-Albert LEHALLE \thanks{Ecole Polytechnique, Centre de mathématiques appliquées, Route de Saclay, 91128, Palaiseau Cedex, France, charles-albert.lehalle@polytechnique.edu}}      
\date{}

\makenomenclature

\newacronym{ACF}{ACF}{Autocorrelation Function}
\newacronym{CI}{CI}{Confidence Interval}
\newacronym{Sig-W1}{Sig-W1}{Signature Wasserstein-1 metric}
\newacronym{KS}{KS}{Kolmogorov–Smirnov test}
\newacronym{ES}{ES}{Expected Shortfall}
\newacronym{VaR}{VaR}{Value-at-Risk}
\newacronym{HH}{HH}{Herfindahl–Hirschman index}
\newacronym{TC}{TC}{Transaction Costs}
\newacronym{WD}{WD}{Wasserstein Distance}
\newacronym{DY}{DY}{Dragulescu-Yakovenko metric}
\newacronym{JS}{JS}{Jensen-Shanon divergence}
\newacronym{DS}{DS}{Discriminative Score}
\newacronym{SR}{SR}{Sharpe Ratio}
\newacronym{HM}{HM}{Higher Moments}
\newacronym{MAE}{MAE}{Mean Absolute Error}
\newacronym{VC}{VC}{Volatility Clustering}
\newacronym{LE}{LE}{Leverage Effect}
\newacronym{SBT}{SBT}{Score-Based Test}
\newacronym{CT}{CT}{Coverage Test}
\newacronym{MMD}{MMD}{Maximum Mean Discrepancy}
\newacronym{VRT}{VRT}{Variance Ratio Test}
\newacronym{G/LA}{G/LA}{Gain/Loss Asymmetry}
\newacronym{CFVC}{CFVC}{Coarse-Fine Volatility Correlation}

\begin{document}

\maketitle

\begin{abstract}
  Simulation methods have always been instrumental in finance, and data-driven methods with minimal model specification---commonly referred to as \textit{generative models}---have attracted increasing attention, especially after the success of deep learning in a broad range of fields.
  However, the adoption of these models in financial applications has not matched the growing interest, probably due to the unique complexities and challenges of financial markets.
  This paper contributes to a deeper understanding of the limitations of generative models, particularly in portfolio and risk management.
  To this end, we begin by presenting theoretical results on the importance of initial sample size, and point out the potential pitfalls of generating far more data than originally available.
  We then highlight the inseparable nature of model development and the desired uses by touching on a paradox: usual generative models inherently \textit{care less} about what is important for constructing portfolios (in particular the long-short ones). 

\medskip
  
Based on these findings, we propose a pipeline for the generation of multivariate returns that meets conventional evaluation standards on a large universe of US equities while being compliant with stylized facts observed in asset returns and turning around the pitfalls we previously identified.
  Moreover, we insist on the need for more accurate evaluation methods, and suggest, through an example of mean-reversion strategies, a method designed to identify \textit{poor} models for a given application based on \textit{regurgitative} training, i.e. retraining the model using the data it has itself generated, which is commonly referred to in statistics as \emph{identifiability}.

\end{abstract}
\vfill
\begin{flushleft}
\end{flushleft}
\newpage

\tableofcontents

\newpage

\section{Introduction}

In 1887, the French poet Stéphane Mallarmé composed \emph{``Un coup de dés jamais n'abolira le~hasard''} (in English \emph{A Throw of the Dice will Never Abolish Chance}), which was a first step in the direction of concrete poetry. 
In a context of the flourishing scientific developments in Europe, especially in probability and statistics (think about Adolphe Quetelet and Bernoulli's and Poisson's work on the \emph{``loi des grands nombres''}, the law of large numbers), the message of the poem was as simple as its title: in the context of randomness, drawing one realization does not cancel the reality of the underlying stochastic phenomenon.
This is the topic of our paper: in the non-stationary and high-dimensional realm of returns of financial instruments, drawing synthetic data made of the informational content of a small sample cannot eliminate the depth of this initial randomness.
\medskip

The practice of drawing samples from a model to address specific financial tasks dates back to at least the 1960s, when \cite{hertz1964risk} proposed using simulations for investment risk analysis. The potential of this approach was quickly used for the valuation of options, with one of the earliest applications appearing in \cite{RePEc:eee:jfinec:v:4:y:1977:i:3:p:323-338}. The adoption of such techniques, along with the effort to develop more \textit{realistic} models, has grown significantly with the increasing complexity of derivatives pricing (e.g., \cite{broadie1996estimating} and \cite{carriere1996valuation}), the need for stress-testing and accurate risk estimation (e.g., Value-at-Risk) for large portfolios of non-linear instruments (e.g., \cite{berkowitz1999coherent} and \cite{jamshidian1996scenario}), and the optimization of portfolios with sensitivity to extreme losses \cite{rockafellar2000optimization}, among many other applications across various areas in finance. 

\medskip
The conventional approach, known as Monte Carlo methods (see \cite{glasserman2004monte} and \cite{jackel2002monte}), involves random sampling from a parametric model specified by the user to ensure that it accurately represents the process, distribution, or environment being modeled. The parameters of the specified model are typically calibrated using the available historical data. Despite the interpretability provided by the parametric nature of the model, a major limitation of this approach is its inherent dependence on the initial assumptions about the model design, which might not accurately reflect the underlying reality. Another drawback is its questionable scalability as the complexity of the system increases, requiring the user to have an extremely deep understanding of the environment---a task that can be practically impossible in some cases. 

\medskip
Therefore, machine learning methods that make minimal assumptions about the underlying distribution and allow the data to \textit{speak for itself} have been gaining increasing attention within the financial community \cite{capponi2023machine}. This growing interest is likely driven not only by the limitations of traditional methods but also by the remarkable achievements in deep generative modeling, particularly in areas like text and image generation (e.g., \cite{brown2020language} and \cite{ramesh2022hierarchical}) although it is worth mentioning that recent academic papers identify drawbacks and degenerate behaviors that stem from synthetic data  \cite{shumailov2024ai,angelopoulos2023prediction}.

\medskip
At the heart of generative modeling lies the idea of training a model that generates samples from the same distribution as the training data. Aside from early examples like energy-based models~\cite{hinton1983optimal}, recent approaches such as variational autoencoders (VAE)~\cite{kingma2013auto}, generative adversarial networks (GAN)~\cite{goodfellow_generative_2014}, and diffusion models~\cite{ho2020denoising} achieve this by learning a function, typically a neural network, that can transform (often) lower-dimensional random inputs (noise) into realistic higher-dimensional outputs. These methods differ primarily in how they learn the parameters of the generative network that handles this transformation.

\newpage

Academics and practitioners have not hesitated to test these approaches in various financial applications. GANs have emerged as the most common architecture, being used for simulating financial return time series~\cite{wiese_quant_2020}, implied volatility surfaces~\cite{vuletic2023volgan}, equity correlation matrices~\cite{marti2020corrgan}, and limit order books~\cite{cont2023limit}, not to mention other applications. Similarly, VAEs have been applied to generate paths~\cite{buehler2020data}, for credit portfolio risk modeling~\cite{caprioli_quantifying_2023}, and for multivariate time series generation~\cite{desai_timevae_2021}. Diffusion models have also found applications, such as in financial tabular data generation~\cite{sattarov2023findiff}. 

\medskip
However, these applications have not yet proven to be as successful as their counterparts in text and image generation. This is likely due to the unique complexities and characteristics inherent in financial markets. When focusing on multivariate asset returns, which is central to this paper, certain differences become apparent compared to text and image data. Firstly, asset prices are influenced not only by the activity of informed trades but also by irrelevant elements perceived as information~\cite{black1986noise}. This introduces noise into prices, and consequently, returns, which can obscure any statistically relevant patterns that the model should detect. Secondly, asset returns exhibit specific statistical properties, both at the marginal and joint levels, referred to as the \textit{stylized facts} of asset returns \cite{cont_empirical_nodate}, which are not straightforward to capture. For example, asset return distributions are often heavy-tailed and asymmetric \cite{mandelbrot1997variation}. Additionally, there is an intertemporal dependence structure between different time points, as evidenced by volatility clustering and leverage effects (see \cite{ding1993long}, \cite{bouchaud2001leverage} and \cite{zumbach_time_2007}). On a joint level, assets can exhibit a complex and dynamic co-movement structure, where the number of cross-asset relationships increases polynomially with the number of assets studied \cite{plerou1999universal}. Ideally, a generative model should capture these properties effectively at both the marginal and joint levels, as well as any other relevant characteristics not explicitly known to the user.

\medskip
The limited availability of data (30 years of daily returns amount to only about 7,500 observations, to be confronted with the thousands of companies to jointly model) adds another layer of complexity compared to applications in text and images, where vast datasets are readily available. Moreover, using the entire historical dataset may not always be ideal, as older data can be less relevant due to the non-stationarity of financial markets and the influence of macroeconomic regime shifts \cite{issler1996common} and business cycles \cite{romer1999changes} on asset returns. These constraints make brute-force approaches, like scaling up model parameters and dataset sizes, less feasible in finance compared to their effectiveness in other domains \cite{kaplan2020scaling}. 

\medskip
Some papers recognize the challenges of simply training a state-of-the-art generative model on financial data. As a result, they \textit{engineer} these models to make them more suitable for financial applications, giving them a better chance to perform well despite the difficulties mentioned earlier. For instance, \cite{liao2024sig} demonstrates that using a mathematical property of signatures can reduce the challenging min-max problem required for training GANs to a simpler supervised problem, significantly easing the training process. \cite{cont_tail-gan_2022} replaces the classical loss functions used in GANs with a financially relevant one based on the joint elicitability of a risk measure pair, thereby \textit{forcing} the model to better learn the tails of return distributions. Another approach involves modifications during the data processing phase. For example, \cite{wiese_quant_2020} employs a Lambert-W transformation \cite{goerg2015lambert} to normalize the training data and eliminate heavy-tails, while \cite{pena_modified_2023} transforms multivariate data to obtain orthogonal variables, thereby simplifying the model's task of handling cross-dependencies.
These approaches question the potential generality of such models: if one must structure a generative model differently for each usage, does this person has enough knowledge of her or his task at hand to shortcut the machinery of GANs and build a dedicated Monte Carlo approach?

\medskip
Another very crucial aspect of generative modeling is the evaluation of the \textit{quality} of generate data, although there is no consensus on how to evaluate and validate what is produced by the trained model (see \cite{borji_pros_2018} for a review of evaluation measures).
It is natural to expect different evaluation measures in different domains, but the lack of a common set of \textit{intra-domain} measures makes it difficult to compare and horse-race models that have been instrumental in the theoretical progress of deep learning over the last decade, as we have seen with ImageNet \cite{deng2009imagenet}.
The considerable efforts being made to benchmark and evaluate LLMs also underline the importance of the issue \cite{hendrycks2020measuring}. In finance, given the difficulty of evaluating returns for a high-dimensional universe of assets, the evaluation process is often poor.
It typically involves checking if the generated data qualitatively reproduces a small subset of \textit{stylized facts} and is distributionally close to the training set, with often no out-of-sample check and a minimal analysis with respect to financial applications.

\medskip
In this paper, we explore to what extent and how generative models can be efficiently utilized in finance, particularly for portfolio construction.
We aim to identify the requirements for more reliable and effective development and evaluation of these models in an environment where there is growing demand for their use in highly specialized tasks, such as optimizing the hyperparameters of long-short quantitative strategies to balance the trade-off between expected return and risk \cite{lopez2019risk}.

\medskip
To this end, we begin by theoretically underlining the crucial relationship that must be considered between the initial sample size of the training set and the amount of generated data. While the endless generation of synthetic data is often taken for granted in other domains, in finance, the initial sample size must remain a constant point of concern both before and after training.

\medskip
Next, we demonstrate why a \textit{plug-and-play} approach with classical generative models is not suitable for finance, emphasizing the need for model design to be tailored to the specific use case.
We illustrate this through an example that can serve as a basis for future approaches. Specifically, we show why the following does not work: using a generative model trained under generic loss functions to construct mean-variance portfolios.
 
\medskip
In light of these results, we propose a generative pipeline designed to address the limitations discussed earlier.
This pipeline begins with appropriate data transformations and decompositions, followed by the use of GANs and parametric models to generate high-dimensional financial multivariate time series. We evaluate our generative pipeline using a real dataset of US stocks, subjecting it to rigorous tests with a demanding and detail-oriented approach, though initially without direct consideration of the final application. 

\medskip
Finally, we propose a methodology for evaluating generative models for financial time series, in two steps.
First, we focus the evaluation on a stylized fact that crosses the time scales (equity mean-reversion that has a positive predictive power at a shorter frequency than a week \cite{avellaneda2010statistical,yeo2017risk}, then a negative one at a frequency around one year, to finally reach the positive domain again at time scales around three to five years \cite{asness2013value}).
Second, we propose a way to measure the capability of a model to identify a data-specified task. This involves \emph{regurgitative training} ---a term coined by \cite{zhang2024regurgitative} to describe the process of training large language models (LLMs) using synthetic data. This approach assesses whether the model can effectively learn about the underlying application from its own generated data and might be of use to detect poor models in the sense that they are not able to identify the underlying model that generated the data, even when it belongs to their own class. In statistics, it is commonly referred to as the \emph{identifiability} of the approach (cf. \cite{picci1977some} and references therein).

\medskip
In Section~\ref{sec:chap32}, we present theoretical results on the influence of the initial sample size and on potential pitfalls of constructing portfolios using synthetic data generated by a generic generative model. Section~\ref{sec:chap33} begins with a review of existing generative models in the literature and then details the generative pipeline we propose. In Section~\ref{sec:a_fit_on_spx}, we report on the evaluation of the data generated by our pipeline, applied to a specific dataset of US stocks, and provide insight into the stylized facts of asset returns that support our design choices. Finally, Section~\ref{sec:evaluating_generative_models} proposes an in-depth evaluation of generative models of multivariate times series of financial returns. We propose to use a list of benchmark use-cases and provide one: the Sharpe ratio of the Equity Mean-Reversion factor across time scales.

\section{The essentials of generative modeling in finance}
\label{sec:chap32}
\subsection{The influence of the initial sample size}
\label{sec:sample_size}

In this section, we demonstrate that, when using a generative model to estimate a statistic, \emph{the initial sample size cannot be ignored}. Specifically, we will show that generating an excessive number of samples using the generative model does not necessarily improve the accuracy of the estimated statistic. Instead, if the number of generated samples $\Tilde{n}$ becomes too large, it can introduce a bias into the results.
\medskip

It is not new in statistics that one cannot generate too much data without perturbing the results of any estimation task performed using this generated data.
The most flourishing period for studying the limitations of data generation was focused on data imputation: one has a dataset of triplets of observations $(X,Z,R)_n$ and aims to build a model for $X=f(Z)$. The variable $R\in\{0,1\}$ indicates whether $(X,Z)$ has been fully observed or not \cite{rubin1976inference}.
Different approaches have been proposed to nevertheless use as many data points as possible, even when only part of $X$ or $Z$ is observable. Qualitatively, the result has been that data imputation ignoring the future use of the replaced data is not efficient (for instance, compared to careful bootstrapping \cite{efron1994missing}), and new, more non-parametric approaches have been proposed (see \cite{seaman2018introduction} and references therein).
The conclusion, nevertheless, has been that it is very difficult to obtain better confidence intervals than those from fully known data, and that the process of generating new information for incomplete data suffers from a bias, except in asymptotic conditions in which you have a lot of fully observed data.

\medskip

This approach has been recently revived by \emph{prediction-powered inference} and similar approaches in machine learning to counter the exact same bias appearing in GAN-generated data for medical, environmental or social applications \cite{angelopoulos2023prediction}.
The key element of reasoning to link missing data to generative model is to observe that a GAN maps observations $Z$ from a latent space (typically distributed as a Gaussian), to observations $X$ in the space of interest. This mapping is learned on a small subset (where $R=1$) and then more data are generated using new samples (i.e. observations drawn from the same Gaussian) $Z$ that produces synthetic versions of $X$. This process is identical to data imputation, which has been proven to be, in general, biased.

\medskip
For this paper, let us consider the class of \textit{U-statistics} to stay as generic as possible (see \cite{serfling2009approximation} for a general introduction). A $U$-statistic is a type of estimator defined as the average value of a symmetric function (kernel) applied to all possible increasing tuples of a fixed size.\footnote{In a more general case, the kernel can also be chosen to be non-symmetric.} Precisely, for a sample of $n$ independent and identically-distributed random variables $X_1, \dots, X_n \sim \mathcal{L}$,

$$ U_{n}  = f_n(X_1, \dots, X_n) = \frac{1}{\binom nr} \sum_{(i_1, \ldots, i_r) \in \mathcal{I}_{r, n}} f(X_{i_1}, \dots, X_{i_r}) $$

where $\mathcal{I}_{r, n}$ is a set of $r$-tuples (distinct and increasing) of indices from $\{1, \dots, n\}$ with $\mathbb E[U_{n}| \mathcal{L}] = \theta$ and $\text{Var}(U_{n}| \mathcal{L}) = \sigma^2 < \infty$ and the law 
$\mathcal{L}$ is assumed to ensure the existence of finite moments up to the required order. 

\medskip
By selecting different kernels $f(x_1, \dots, x_r)$, a wide range of useful estimators can be derived, such as the sample mean ($r=1$, $f(x_1)=x_1$), sample variance ($r=2$, $f(x_1, x_2)=(x_1- x_2)^2/2$), estimators for the $k^{th}$ moments and the variance-covariance matrix for multivariate data, among many others. Moreover, the $U$-statistic $ U_{n} $ satisfies the asymptotic normality property \cite{10.1214/aoms/1177730196}:
\begin{equation*}
\label{property:asym_normal}
    \sqrt{n} (U_{n} - \theta) \rightarrow \mathcal{N}(0, r^2\sigma_1^2)
\end{equation*}
    where the variance $ \sigma_1^2 \leq \sigma^2$ is a strictly positive value that can be computed using the Hoeffding decomposition, precisely it is the variance of $g(X_1)$ where $g(x) = \mathbb E[f(x, X_2, \dots, X_r)| \mathcal{L}]$.
    
\medskip

Berry-Esseen-type error bounds for $U$-statistics have been studied extensively (see \cite{bentkus2009normal} and \cite{chen2007normal}). If $\mathbb E[\lvert g(X_1) \rvert ^\beta| \mathcal{L}] < \infty$ and $\mathbb E[\lvert f(X_1, \dots, X_r)\rvert^\beta| \mathcal{L}]<\infty$ for $\beta \in (2,3]$, 

\begin{equation}
\label{eq:berry_esseen}
    \left| \mathbb{P}\left( \frac{\sqrt{n}(U_{n} - \theta)}{r\sigma_1} \leq x \right) - \Phi(x) \right| \leq \frac{c}{(1+|x|)^\beta\sqrt{n-r+1}}
\end{equation}

where $ \Phi(x) $ is the cumulative density function of the standard normal distribution
and $c>0$ represents a value that does not depend on $x$ or $n$.\footnote{Precisely, $$c = \frac{c_0(\sqrt{r}\mathbb E[\lvert f(X_1, \dots, X_r)\rvert^\beta| \mathcal{L}]+\mathbb E[\lvert g(X_1) \rvert ^\beta| \mathcal{L}])}{\sigma_1^\beta}$$ where $c_0>0$ represent a constant that does not depend on the specific variables or parameters of the problem. While tighter bounds are possible, the bound provided by Inequality~(\ref{eq:berry_esseen}) is sufficient for our analysis.}


\medskip

Suppose we aim to estimate a statistic $ \theta $, a true characteristic of an unknown distribution $ \mathcal{L} $, using a generative model. To do so, we first train the generative model on a dataset of size $ n $ sampled from $ \mathcal{L} $. After training, the generative model produces samples from a distribution $ \Tilde{\mathcal{L}} $, and the corresponding statistic under this distribution is $ \Tilde{\theta}_n := \mathbb{E}[U_{\Tilde{n}}| \Tilde{\mathcal{L}}] $, which serves as an estimate of the true statistic $ \theta $.

\medskip

However, due to the finite size $n$ of the initial training sample and associated biases it introduces in the learning process, an inherent discrepancy $ a_n := \Tilde{\theta}_n - \theta $, referred to as the \emph{learning accuracy}, arises between $ \theta $ and $ \Tilde{\theta}_n $. Given this context, when the goal is to estimate $ \theta $ within a desired tolerance $ b $, a natural question arises: can increasing the size $ \Tilde{n} $ of the synthetic dataset generated by the model improve the estimate of $ \theta $, or are there fundamental limitations imposed by the initial training sample size $ n $ that cannot be mitigated by simply enlarging $ \Tilde{n} $?

\medskip

\begin{figure}[!ht]
\centering
\begin{tikzpicture}
    \begin{axis}[
        no markers, 
        domain=-5:5, 
        samples=100, 
        axis y line=none, 
        axis x line=none, 
        xlabel style={at={(axis description cs:1,0)},anchor=north, font=\bfseries},
        xlabel=$ $, 
        ylabel={},
        height=4cm, 
        width=16cm,
        xtick=\empty, 
        ytick=\empty,
        enlargelimits=false, 
        clip=false, 
        axis on top,
        grid = none,
    ]
        \node[below] at (axis cs:-1,0) {\large$\theta$};
        
        \addplot[thick,black] {exp(-(x-1)^2/2/2.25)/sqrt(2*pi)/1.5};
        \node[below] at (axis cs:1,0) {\large$\Tilde{\theta}_n$};

        \def\xLabel{2} 
        \def\yLabel{0.25} 
        \node[above] at (axis cs:\xLabel,\yLabel) {\large$U_{\Tilde{n}}$};

            \def\xLabeltwo{-1.5} 
        \def\yLabeltwo{.4} 
        
        \draw[thick, black] (axis cs:-1,-.01) -- (axis cs:-1,.01);

        \draw[thick, black] (axis cs:1,-.01) -- (axis cs:1,.01);
        \addplot [
            draw=none,
            fill=gray, 
            fill opacity=0.2
        ][domain=-1.5:-.5] 
        {exp(-(x-1)^2/2/2.25)/sqrt(2*pi)/1.5}
        \closedcycle ;
        
        \draw[-, thin, dashed] (axis cs:-.8,-0.05) -- (axis cs:.7,-0.05) node[midway,below] {\small$a_n$};

        \draw[-, thin, dotted] (axis cs:-1.5,-0.05) -- (axis cs:-1.1,-0.05) node[midway,below] {\small$b$};
        
        \draw[ultra thick, -] (-5,0) -- (5,0) node[right] {\bfseries $ $};
    \end{axis}
\end{tikzpicture}
\caption{Illustration of the mentioned variables in the context of estimating a statistic using a generative model. The figure demonstrates how the generative model can only produce estimates clustered around the mean of $U_{\Tilde{n}}$, when $X_1, \dots, X_{\Tilde{n}}$ follows the estimated law $\Tilde{\mathcal{L}}$, represented by $\Tilde{\theta}_n$. The goal is to maximize the shaded area, which represents the probability of $U_{\Tilde{n}}$
being within a distance $b$ from the true statistic $\theta$. As the number of generated samples $\Tilde{n}$ increases, the variance of $U_{\Tilde{n}}$ computed on the synthetic sample decreases, causing its distribution to peak around $\Tilde{\theta}_n$ and reducing the probability of being within a distance $b$ of $\theta$ to nearly zero.}
\label{fig:diag:distances}
\end{figure}
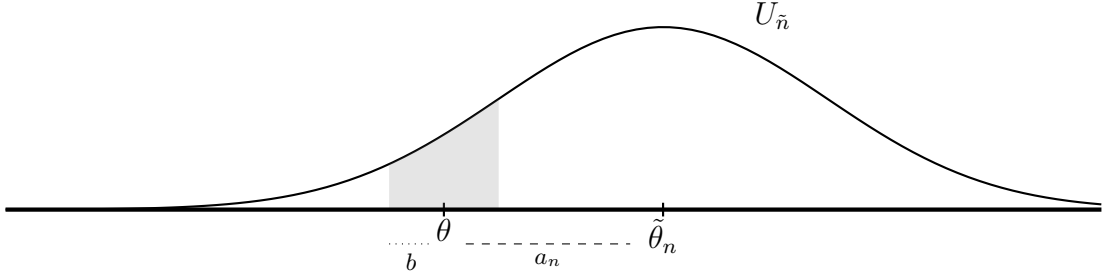

\begin{assump}
\label{assumption:distance}
The learning accuracy $a_n$ is inversely related to a power of the initial sample size $n$.
\end{assump}

Assumption~\ref{assumption:distance} essentially states that with more data, one will obtain a model that \textit{better} represents reality. 
Indeed, when the number of parameters of the model increases, the sample size needed to guarantee a given accuracy increases too.
The paradox of universal approximators like neural networks is that to guarantee their convergence via results like \cite{hornik1989multilayer}, an infinite number of parameters is needed, and hence an infinite number of data is required.
Moreover, recent results like \cite{arous2019landscape} show that it is not enough to have an infinite number of data points if the ratio of the number of neurons over the sample size is maintained over a certain threshold.


\medskip

Using Inequality~(\ref{eq:berry_esseen}), we can write the following error bound for a $U$-statistic computed using synthetic dataset of size $\Tilde{n}$ drawn from $\Tilde{\mathcal{L}}$:

\begin{equation*}
\label{eq:berry_esseen_syn}
    \left| \mathbb{P}\left( \frac{\sqrt{\Tilde{n}}(U_{\Tilde{n}} - \Tilde{\theta}_n)}{r\hat{\sigma}_1} \leq x \right) - \Phi(x) \right| \leq \frac{\hat{c}}{(1+|x|)^\beta\sqrt{\Tilde{n}-r+1}}
\end{equation*}
where $\hat{c}>0$ can be considered the counterpart to the constant $c$ that appears in Inequality~(\ref{eq:berry_esseen}) under the generative model and $\hat{\sigma}^2_1 $ is an estimate of the variance $ \sigma^2_1$. 

\medskip

While increasing the number of synthetic data points $\Tilde{n}$ will cause $U_{\Tilde{n}}$ to converge toward $ \Tilde{\theta}_n $, the ultimate goal is to improve the estimate of $ \theta $. To formalize, we are interested in the probability that $U_{\Tilde{n}}$ is within a distance $ b >0 $ of $ \theta $, given that $\Tilde{\theta}_n$ is $a_n$ away from $ \theta $, where $a_n$ is strongly related to the initial sample size under Assumption~\ref{assumption:distance}. This probability is represented by the shaded area in Figure~\ref{fig:diag:distances}. The bounds for this probability can be given by the following proposition.

\begin{prop}[Probability bounds for synthetic data estimators approximating true statistics]
Let $\Tilde{n}$ denote the size of synthetic dataset generated by a generative model trained on an initial dataset of size $ n $, and let $ \Tilde{\theta}_n $ represent the estimate of the true statistic $\theta$ under the generative model. Assume that $ a_n= \Tilde{\theta}_n - \theta $ denotes the learning accuracy. The probability that $U_{\Tilde{n}}$, the $U$-statistic computed using the synthetic dataset, is within a distance $b > 0$ of $\theta$ satisfies the following bounds:
\begin{equation}
\label{eq:target_distance_bounded}
\left| \mathbb{P}\left(|U_{\Tilde{n}} - \theta| \leq b \right) - \left[ \Phi\left(\frac{(a_n+b)\sqrt{\Tilde{n}}}{r\hat{\sigma}_1}\right) - \Phi\left(\frac{(a_n-b)\sqrt{\Tilde{n}}}{r\hat{\sigma}_1}\right)\right] \right| \leq \Tilde{c}(a_n,b,\Tilde{n}) + \Tilde{c}(a_n,-b,\Tilde{n})
\end{equation}
where $\Tilde{c}(x,y,z) = \frac{\hat{c}}{\left(1+\left|\frac{(y-x)\sqrt{z}}{r\hat{\sigma}_1}\right|\right)^\beta\sqrt{z-r+1}}$ and $r$, $\hat{\sigma}_1$, $\hat{c}$ and $\beta$ are constants related to the statistics of interest, as defined above.
\end{prop}

The proof of this result can be found in Appendix~\ref{appendix:error_bounds}.  This expression illustrates how the accuracy of $U_{\Tilde{n}}$ as an estimator for $ \theta $ depends on several factors: the sample size $ \Tilde{n} $ that we generate, the learning accuracy $a_n$ of $\Tilde{\theta}_n$ from $ \theta $ that is often unknown, the target distance $ b $ within which we want $ U_{\Tilde{n}} $ to fall, and other parameters which are related to the underlying statistics of interest and the distribution of the random variable on which this statistic is computed. However, a more interesting result emerges in the limiting case.

\begin{cor}[Excessive synthetic data generation leads to biased conclusions]\label{cor:moreisless}
If $|a_n| \geq b$ and $U_{\Tilde{n}}$ is computed using samples from the learned generative model, 
$$\lim_{\Tilde{n}\rightarrow\infty} \mathbb{P}\left(|U_{\Tilde{n}} - \theta| \leq b \right) = 0. $$
\end{cor}

Therefore, generating more data points with the generative model does not ensure that the estimated statistics will be closer to the true values. Instead, it introduces a bias (towards the small sample size available dataset) and creates a discrepancy from the true statistics.
The only way to avoid this effect is to have an arbitrarily large dataset to learn the generative process on (since it may decrease $|a_n| $).

\paragraph{Beyond U-statistics.} 
It is possible to extend such results to $L$-statistics (cf. \cite{aaronson1996strong} for asymptotics). 
In finance, this is not a mere detail, as widely used risk measures like Value-at-Risk and Expected Shortfall belong to this class of statistics.


\paragraph{Short discussion on Assumption \ref{assumption:distance}.}
This assumption states that a statistic computed on synthetic data produced by a (generative) model learned on a finite dataset lies between its true underlying value and its empirical value on the finite dataset.
In more formal terms: whatever $\Tilde{n}$ the number of synthetic data, $U_{\Tilde{n}}$ computed on the generative model learned from $X_1,\ldots,X_n$ is close to $U_{n}$ that is known only thanks to the initial sample of $n$ observations.
\medskip

Of course some methods are known to reduce a part of this bias (for instance the jackknife, for some specific dependence of bias on the sample size),
or regularisation methods can be used to drive estimators away from their biased versions.
But the role of regularisation is primarily to prevent excessive non-linearity to appear in the model, not to give it a specific shape.
When the true underlying model is unknown, it is not reasonable to think specific regularisation choices dedicated to the nature of this true model will be made.
This discussion is well known in the machine learning community and is usually referred to as the \emph{no free lunch theorem argument}, see \cite{wolpert1996lack}.\footnote{For readers familiar with statistics, let us cite \cite{wang1998large}: ``\emph{[For a consistent asymptotically linear] estimator $\theta$ of $\theta_0$ [with influence function $D$] $\sqrt{n} (\theta - \theta_0) = (1/\sqrt{n})\sum_i D_i + o_p(1)$, where $D$ has a zero mean and a finite covariance matrix and $o_p(1)$ denotes a random variable converging to zero in probability [when $n$ goes to infinity].}''
  We are commenting the existence of the random variable $o_p(1)$ that goes to zero only when the known sample size goes to infinity.}${}^,$\footnote{For readers familiar with the recent machine learning literature, we refer to how papers like \cite{angelopoulos2023prediction} reintroduce explicitly this bias in estimators to produce larger, but not misleading, confidence intervals via a ``rectifier'' term.}

\medskip

To conclude on the aspect of drawing $\Tilde{n}$ examples from a model based on an initial sample size of $n$: without any good reason to have chosen a specific generative model that is close to the way economics shapes the returns of financial instruments, Corollary \ref{cor:moreisless} stands and tells \emph{generating too many synthetic data gets the estimated statistics on these data away from their true values}.
In statistics, and especially with the context of the bootstrap, practitioners seem to have a rule of thumb that is to generate the same order of magnitude of points as the original sample size. In light of Inequality (\ref{eq:target_distance_bounded}), we can say that one should preserve the distance between the empirical value of the $U$-statistic $U_n$ and its theoretical counterpart $\Esp[U_n|{\cal L}]$ in the hope that $\Esp[U_{\Tilde{n}}|\Tilde{\cal L}]$ will not be too close to this empirical mean.
\paragraph{Small sample size also makes it difficult to evaluate the model.} Beyond the difficulty of training generative models on small datasets, an equally important challenge is evaluating the resulting models and their ability to capture the underlying data generation process. This evaluation is often performed by comparing the synthetic data to a test (out-of-sample) set that is assumed to be representative of the true distribution. When the out-of-sample dataset is large enough, it can reasonably be considered a good representation of the underlying distribution, and the comparison can be interpreted as meaningful. However, when the out-of-sample set itself contains only a limited number of observations, its representativeness becomes questionable. In this case, what is the informative value of poor out-of-sample performance if the test set is not big enough to accurately reflect the properties of the actual data generation process? Should a model be rejected solely on the basis of such an evaluation? We address these questions in Section~\ref{sec:test_for_iden} and propose an evaluation procedure that can be considered a necessary, but not sufficient, condition for generative models. We preferred to tackle the important question of model evaluation in the numerical section, where the discussion can be made more concrete with illustrations from real-world applications.

\subsection{Intrinsic complexity of portfolio construction for generative models}
\label{sec:learning_and_application}

This subsection focuses on how expected returns interact with modern portfolio construction in three steps:
\begin{itemize}
\item We first remind that Markowitz-like portfolio optimizations essentially multiply the vector of expected returns by the inverse of the covariance matrix of the instruments' returns.
\item We then show that, as a consequence: if a risk model is rewritten in the space of its PCs (principal components), the contribution of expected returns to portfolio construction predominantly resides in the subspace spanned by the lower-variance PCs.
\item Lastly, we demonstrate that generative models using standard loss functions (based on comparisons of distributions) learn high-variance PCs better than lower-variance ones, that is the opposite of what is needed for modern portfolio construction.
\end{itemize}


\paragraph{Basics of portfolio construction.}
First recall the basics of portfolio construction (see \cite{boyd2024markowitz} for details).
Given the vector of expected return $\bm{\mu}$ and the matrix of expected covariances between $d$ stocks being $\bm{\Sigma}$, most portfolio constructions boil down to
\begin{equation}
    \label{eq:mark:portf}
    \begin{array}{ll}
    \mbox{maximize}   & \bm{\mu}^\top \mathbf{w} \\
    \mbox{subject to} & \mathbf{w}^\top\bm{\Sigma} \mathbf{w} \leq s^2,
    \end{array}
\end{equation}
where $s$ represents the maximum portfolio volatility that can be tolerated by the user.

\medskip
Assume that $\bm{\Sigma}$ is derived from a statistical factor model constructed as follows. Let $\bm{V}$ be an initial covariance matrix estimated from a sample or obtained through other means. We perform principal component analysis (PCA) and select the first $m$ eigenvectors (stacked in the matrix $\mathbf{P}$ of dimension $d \times m$) as the \textit{significant} components or factors. These eigenvectors are associated with the $m$ largest eigenvalues $\lambda_1, \dots, \lambda_m$, which are stored in the diagonal matrix $\bm{\Delta}$ of dimension $m \times m$. The remaining $d-m$ eigenvectors are stacked in the $d \times (d-m)$ matrix $\mathbf{Q}$, and the remaining eigenvalues are replaced with a fixed value $\lambda_c$, ensuring that the trace of the original matrix is preserved. This process is equivalent to applying eigenvalue clipping to $\bm{V}$.

\medskip
It corresponds to having the following risk model: 

\begin{equation}\label{eq:risk:model}
\bm{\Sigma} = \bm{\Omega} + \lambda_c \mathbf{U}
\end{equation}
where $\bm{\Omega} = \mathbf{P}\bm{\Delta} \mathbf{P}^\top$ and $\mathbf{U}  = \mathbf{Q} \mathbf{I}_{d-m}\mathbf{Q}^\top$.

\medskip

It is important to note that $\lambda_c$ is smaller than the smallest eigenvalue in $\bm{\Delta}$:
\begin{equation*}\label{eq:sigma:bound}
    \lambda_c < \lambda_m.
\end{equation*}

\medskip

The following definitions will be useful for simplifying the expression of Problem~(\ref{eq:mark:portf}):

\begin{defi}[Natural space]
We define the \emph{natural space} of a d-dimensional random vector $X$ as the space of its coordinates, where the squared distance between $X$ and $\tilde{X}$ (from the same set) is given by $\|X - \tilde{X}\|^2$.
\end{defi}

\begin{defi}[Principal space]
We define the \emph{principal space} of the same vector, with a covariance matrix $\bm{\Omega} =\mathbf{P}\bm{\Delta} \mathbf{P}^\top$, as the space resulting from a change of coordinates to the principal components and rescaling by the inverse of the square roots of the eigenvalues. In this space, the squared distance between $X$ and $\tilde{X}$ is given by $\|\bm{\Delta}^{-1/2} (\mathbf{P}^\top X - \mathbf{P}^\top\tilde{X})\|^2$.
\end{defi}

In the context of Model (\ref{eq:risk:model}), the principal space is formed by the concatenation (i.e., horizontal stacking $[\mathbf{P},\mathbf{Q}]$) of $\mathbf{P}$ and $\mathbf{Q}$, as
\begin{equation}
    \bm{\Sigma} = [\mathbf{P},\mathbf{Q}] \begin{bmatrix}
                        \bm{\Delta} & 0\\
                        0 & \lambda_c \mathbf{I}_{d-m}
                    \end{bmatrix} [\mathbf{P},\mathbf{Q}]^\top.
\end{equation}

As a result, the associated distance between two vectors $X$ and $\tilde{X}$ in the principal space is given by
\[
\|\bm{\Delta}^{-1/2} (\mathbf{P}^\top X - \mathbf{P}^\top \tilde{X})\|^2 + \frac{1}{
\lambda_c}\|\mathbf{Q}^\top X - \mathbf{Q}^\top \tilde{X}\|^2.
\]

\medskip

Once we express Problem (\ref{eq:mark:portf}) in the principal space, setting $\bm{\mu}=[\mathbf{P},\mathbf{Q}][\mathbf{y_P};\mathbf{y_Q}]$ and $\mathbf{w}=[\mathbf{P},\mathbf{Q}] [\mathbf{v_P};\mathbf{v_Q}]$ with $[\cdot,\cdot]$ to note an horizontal stacking and $[\cdot;\cdot]$ to note a vertical stacking 
it reads
\begin{equation}\label{eq:mark:portf:P}
    \begin{array}{llll}
    \mbox{maximize}   & \mathbf{y_P}^\top \mathbf{v_P} &+ \phantom{\lambda_c}\, \mathbf{y_Q}^\top \mathbf{v_Q} \\[.3em]
    \mbox{subject to} & \mathbf{v_P}^\top \Delta \mathbf{v_P} &+ \lambda_c\, \mathbf{v_Q}^\top \mathbf{v_Q} &\leq s^2,
    \end{array}
\end{equation}

Using Lagrange multipliers, the solution comes immediately as
\begin{equation}\label{eq:mark:princ:sol}
    \mathbf{v_P} = \frac{1}{\gamma} \bm{\Delta}^{-1} \mathbf{y_P},\quad \mathbf{v_Q} = \frac{1}{\gamma\lambda_c} \mathbf{y_Q} \quad \text{where} \quad %
    \gamma = \frac{1}{s}\sqrt{\mathbf{y_P}^\top \bm{\Delta}^{-1} \mathbf{y_P} + \frac{1}{\lambda_c} \mathbf{y_Q}^\top\mathbf{y_Q}}.
\end{equation}

Keeping in mind that the portfolio weights $\mathbf{w}$ is the transformation by $[\mathbf{P},\mathbf{Q}]^\top$ of the vertical concatenation $[\mathbf{v_P};\mathbf{v_Q}]$ of the two upper vector, the following property becomes immediate:

\begin{prop}[Rescaling of the expected returns operated by the portfolio construction]
Expressed in the principal space, the portfolio weights $[\mathbf{v_P};\mathbf{v_Q}]$ are made of the first $d$ expected returns (or signal) $[\mathbf{y_P};\mathbf{y_Q}]$ multiplied by the inverse of the variances of the principal components: $\mathbf{v_P} \propto \bm{\Delta}^{-1} \mathbf{y_P}$, and for the second and last part by an exact copy of the expected returns $\mathbf{v_Q}\propto \mathbf{y_Q}$.
\end{prop}

This can be expressed coordinate by coordinate as follows.
\begin{prop}[Reducing the exposure to large eigenvectors]\label{prop:down:scale}
As long as a covariance matrix $ \bm{\Sigma}$ can be expressed by Model (\ref{eq:risk:model}), the multiplication of its inverse by an arbitrary vector $\mathbf{z}$ reads:
\begin{equation}
    \label{eq:Sigma:simple:inv}
    \bm{\Sigma}^{-1} \mathbf{z} = \frac{1}{\lambda_c}\left(\sum_{i\leq m} \frac{\lambda_c}{\lambda_i}\,\langle \mathbf{z},\mathbf{P}_{:,i} \rangle\, \mathbf{P}_{:,i} + \sum_{j\leq d-m} \langle \mathbf{z},\mathbf{Q}_{:,j} \rangle\,\mathbf{Q}_{:,j} \right).
\end{equation}
In particular, the optimal weights $\mathbf{w}$ of Problem (\ref{eq:mark:portf}), given expected returns $\bm{\mu}=[\mathbf{P},\mathbf{Q}][\mathbf{y_P};\mathbf{y_Q}]=[\mathbf{P},\mathbf{Q}][\mathbf{y}_k]_{1\leq k\leq d}$ reads
\begin{equation*}
    \label{eq:MPO:simple}
    \forall k: \mathbf{w}_k = \frac{1}{\lambda_c\gamma}\left(\sum_{i\leq m} \frac{\lambda_c}{\lambda_i}\,\mathbf{y}_i\, \mathbf{P}_{:,i} + \sum_{j\leq d-m} \mathbf{y}_j\,\mathbf{Q}_{:,j} \right),
\end{equation*}
where $\gamma$ is the renormalizing constant of (\ref{eq:mark:princ:sol}) to reach the target risk $s^2$.
\end{prop}
Qualitatively, it means the portfolio construction \emph{scales down the exposures to the first $m$ eigenvectors proportionally to their eigenvalues}, and it does not change the remaining components. Keep in mind that $\lambda_c<\inf_k \lambda_k$, that implies $\lambda_c/\lambda_k$ is smaller than one.

\medskip
These observations have implications for the generation of synthetic time series of returns that will be used for portfolio construction. Proposition~\ref{prop:down:scale} implies that the generated data should be more accurate along specific directions, particularly those corresponding to smaller $\lambda_i$, within the subspace spanned by $\mathbf{P}$. In other words, for portfolio construction, components associated with smaller variances are more critical than those with larger variances.

\paragraph{The way generic generative models learn.} On the other hand, generative models are typically trained to minimize the distance between the real data distribution and the synthetic one in the natural space. Such objective can be formalized as 

\begin{equation}\label{eq:gan_objective}
   \min_{g(.) \in \mathcal{G}} 
   W_p(\mathbb P_X, \mathbb P_{g(Z)}) 
\end{equation}

where $g:\mathbb R^m\rightarrow\mathbb R^d$ is a function from the set $\mathcal{G}$ and $W_p$ the $p$-Wasserstein distance \cite{kantorovich1960mathematical}.\footnote{Formally, it is defined as $$W_p(\mu, \nu) = \left(\inf_{\pi \in \Gamma(\mu, \nu)} \int_{\mathcal{X} \times \mathcal{X}} c(x, y)^p \, d\pi(x, y) \right)^{\frac{1}{p}}$$
where $\mu$ and $\nu$ are two probability distributions defined on a metric space $(\mathcal{X}, c)$, $c(x, y)$ is the distance between points $x$ and $y$ in $\mathcal{X}$ and $\Gamma(\mu, \nu)$ is the set of all couplings of $\mu$ and $\nu$, i.e., the set of joint distributions on $\mathcal{X} \times \mathcal{X}$ with marginals $\mu$ and $\nu$.} 

\medskip

However, generative models trained to minimize the Wasserstein distance between the real data distribution and the synthetic one tend to make larger errors on the eigenvectors associated with small eigenvalues compared to those with large eigenvalues. As a result, the synthetic data generated for portfolio construction are exposed to uncontrolled errors. This issue is related to the following theorem.

\begin{thm}[Theorem 1 in \cite{feizi2017understanding}]\label{gans_learning}
Let $X \sim \mathcal{N}(0, \bm{\Sigma})$, where $\bm{\Sigma}$ is a full-rank covariance matrix, and let $g$ be a generator function that maps an $k$-dimensional noise vector $Z \sim \mathcal{N}(0, \mathbf{I}_k)$ to a $d$-dimensional space with $k \leq d$. Assume $p = 2$ and that $\mathcal{G}$ is the family of linear functions. Then, in the population setting, the solution $g^*$ to Problem~(\ref{eq:gan_objective}) satisfies 
$ \Tilde{X} = g^*(Z) $, 
where 
$ \bm{\Tilde{\Sigma}} = \mathbb{E}[\Tilde{X} \Tilde{X}^\top] $
is a rank-$k$ matrix. Moreover, the eigenvectors of $\bm{\Tilde{\Sigma}}$ coincide with the top-$k$ eigenvectors of $\bm{\Sigma}$, and its eigenvalues correspond to the top-$k$ eigenvalues of $\bm{\Sigma}$.
\end{thm}

 Theorem~\ref{gans_learning} suggests that the generic loss functions of generative models drive to focus more on learning the components that explain the largest portion of total variance in the system. This can also be interpreted as these models giving less priority to, or paying less attention to, factors with smaller variance. 

\medskip

Although simplified, the task described in Theorem~\ref{gans_learning} is quite similar to what most generative models aim to achieve. In many applications, learning the latent factors with high variance is more crucial for generating results that appear realistic. This is often the case even when evaluating generative models that produce synthetic asset returns, where the synthetic returns are assessed based on their \textit{similarity} to historical data. However, in portfolio construction, the introduction of the inverse of the covariance matrix (or similar effects) makes learning the components with lower variance at least as important as those with higher variance. 

\paragraph{Portfolio sensitivity to eigenvector perturbations.}
To better understand the impact of a generative model's error on an eigenvector, consider the following perturbation in the column $k>1$ of $\mathbf{P}$:
\begin{equation*}
    \mathbf{\Tilde{P}}_{:,k} = \cos \epsilon \cdot \mathbf{P}_{:,k} + \sin \epsilon \cdot \mathbf{h},
\end{equation*}
where $\mathbf{h}$ is a unit vector belonging to the subspace spanned by $\mathbf{P}_{1:k-1}$ and $\epsilon$ is the error term controlling the magnitude of the perturbation.

\medskip
Without the loss of generality, let $\mathbf{h}=\mathbf{P}_{:,1}$. In this case, the first eigenvector should also be adjusted in a way that we conserve orthogonality, such as, $$ \mathbf{\Tilde{P}}_{:,1} = -\sin \epsilon \cdot \mathbf{P}_{:,k} + \cos \epsilon \cdot \mathbf{h}.$$  

We denote the resulting covariance matrix under this perturbation as 

\begin{equation}\label{new_cov}
\bm{\Tilde{\Sigma}}_{(k)} = \lambda_1\,\mathbf{\Tilde{P}}_{:,1}\mathbf{\Tilde{P}}_{:,1}^\top+ \sum_{\substack{1<i\leq m \\ i\neq k}} \lambda_i\,\mathbf{P}_{:,i}\mathbf{P}_{:,i}^\top+\lambda_k\,\mathbf{\Tilde{P}}_{:,k}\mathbf{\Tilde{P}}_{:,k}^\top+\lambda_c\sum_{i\leq d-m}\mathbf{Q}_{:,i}\mathbf{Q}_{:,i}^\top.
\end{equation} 

To measure how much we deviate, at the portfolio level, due to an $\epsilon$ perturbation in the $k$th eigenvector of $\bm{\Sigma}$ under a given vector $\mathbf{z}$, we can define
$$\delta_{(k)}(\epsilon, \mathbf{z}) = \bigl\| \bm{\Tilde{\Sigma}}_{(k)}^{-1} \mathbf{z} - \bm{\Sigma}^{-1} \mathbf{z} \bigl\|^2.$$

\begin{prop}[Overall effect of perturbing a specific eigenvector in a covariance matrix]
Consider perturbing, as described above, the $k$th eigenvector $(k>1)$ of a covariance matrix $ \bm{\Sigma}$, expressed as in Model (\ref{eq:risk:model}), by $\epsilon$ that specifies the magnitude of the perturbation. The resulting error, under an arbitrary vector $\mathbf{z}$, is given by:
$$\delta_{(k)}(\epsilon, \mathbf{z})=\left(\frac{1}{\lambda_1}-\frac{1}{\lambda_k}\right)^2\sin^2\epsilon\left(\langle \mathbf{z},\mathbf{P}_{:,k}\rangle^2 + \langle \mathbf{z},\mathbf{P}_{:,1}\rangle^2\right).$$
\end{prop}

The proof of this result is provided in Appendix~\ref{appendix:perturb}. 

\begin{cor}[Errors on low-variance factors are magnified]\label{corr:perturb}
If $\lambda_2 \gg \lambda_m$ (which is often the case, in finance, for covariance matrices of returns), for a given vector of expected returns $\bm{\mu}$ satisfying $|\mathbf{y}_2| \approx |\mathbf{y}_m|$ (recall that $\mathbf{y}_i=\langle \bm{\mu},\mathbf{P}_{:,i}\rangle$), meaning there is not a significant difference between expected returns in the principal space, the following inequality holds:
$$
\delta_{(m)}(\epsilon, \bm{\mu}) \gg \delta_{(2)}(\epsilon, \bm{\mu}),
$$
where $m$ corresponds to a low-variance factor and $2$ corresponds to a higher-variance factor.
\end{cor}

\medskip

Corollary~\ref{corr:perturb} highlights the fact that the same amount of error made in different eigenvectors has different impacts on the final portfolio. Specifically, errors in low-variance factors lead to significantly larger deviations between the calculated portfolio and the true optimal portfolio. Consequently, learning schemes that introduce larger errors in the subspace spanned by low-variance factors, such as generic generative models, are more likely to lead to significant errors at the portfolio level.

\begin{figure}[ht]
    \centering
    \includegraphics[width=15cm]{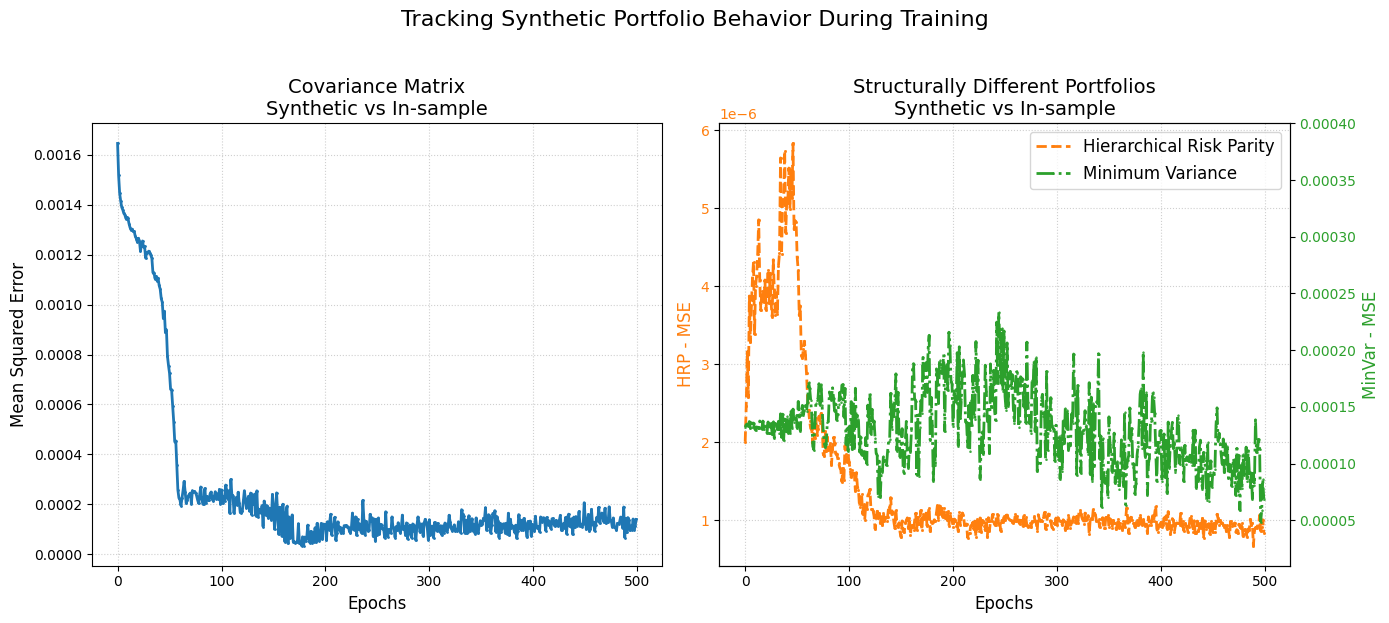}
    \caption{Left: Mean squared error between the synthetic and in-sample covariance matrices over training epochs, showing that the GAN learns the dominant correlation structure.
Right: Mean squared error of portfolio returns under synthetic data versus in-sample returns for HRP (orange, left axis) and minimum variance (green, right axis) portfolios. While covariance accuracy improves, the synthetic data fails to reproduce the structure needed for accurate minimum variance portfolios, highlighting a disconnect between statistical and portfolio-relevant learning.}
    \label{fig:learning_pcs}
\end{figure}

\paragraph{A short note on the perturbation methodology.}
When perturbing an eigenvector with another vector that resides in the subspace spanned by a set of remaining eigenvectors, one must also adjust these eigenvectors to maintain orthogonality with the perturbed eigenvector. This adjustment introduces additional impacts on the portfolio, complicating a marginal analysis focused on a specific eigenvector. For this reason, we use the first eigenvector, expected to have less influence on the final portfolio, to introduce errors into other eigenvectors belonging to $\mathbf{P}$. If one chooses $\mathbf{h} \in \Span(\mathbf{Q})$, the impact of the necessary adjustments to $\mathbf{Q}$ dominates the error introduced by a perturbation applied to $\mathbf{P}_{:,k}$. This motivates our choice of $\mathbf{h} \in \Span(\mathbf{P}_{1:k-1})$ when analyzing the impact of a perturbation in $\mathbf{P}_{:,k}$.
\paragraph{Empirical illustration of theoretical claims.} So far in this section, we have presented theoretical arguments explaining why a standard generative model may not produce multivariate asset return data suitable for Markowitz-type portfolio optimization. However, as in most theoretical analyses, our arguments relied on simplifying assumptions, such as model linearity and a specific risk model, in order to make the problem tractable. We now wish to conclude this section by empirically examining whether our theoretical claims correspond to what actually happens when training generative models with market data, particularly from a portfolio construction perspective. To this end, we train a GAN on the historical multivariate returns of S\&P 500 stocks and verify whether the model is able to capture the covariance structure between assets during training.\footnote{We use a very simple network architecture, whose details are omitted for brevity. The repository below provides the full implementation, including all code used for the empirical analysis as well as figure and table generation for the rest of the paper: \url{https://github.com/rengimcetingoz/synthetic_data_for_portfolios}.} As shown in the left-hand graph in Figure~\ref{fig:learning_pcs}, the model, particularly during the first half of training, rapidly improves its fit to the in-sample covariance matrix (measured by mean squared error) and appears to learn the dominant dependency structure in the data. However, this improvement is likely due to the model learning the high-variance factors of the return space (i.e., the leading principal components), which, although statistically dominant, are of limited use in constructing a minimum variance portfolio, as illustrated by the right-hand graph in Figure~\ref{fig:learning_pcs}. Although it captures the covariance structure, the model, during the first half of learning, fails to generate data for which the minimum variance portfolios resemble their in-sample counterparts. This limitation is less problematic for approaches such as hierarchical risk parity (HRP) \cite{lopez2016building}, which do not require the inversion of the covariance matrix and are instead designed to operate directly on the covariance matrix itself. It is noteworthy that the improvement observed in the second part of the training process for the minimum variance portfolio is roughly linear, unlike the exponential decay generally observed in neural network losses (and here for HRP portfolios). This behaviour is also consistent with our theoretical conclusions that, from a portfolio construction perspective, the principal components have a similar impact on the portfolio, regardless of the magnitude of their eigenvalues or associated variances. This brief experiment therefore supports the idea that generative models, when trained in a standard manner, may prioritise statistically dominant structures that are not aligned with the objectives of modern portfolio theory, and that evaluating training solely through the lens of the loss function may be misleading from a portfolio construction perspective.

\section{Generative models for financial time series}
\label{sec:chap33}

Now that previous section establishes that
it is counterproductive to generate too many synthetic data points, and that the accuracy of the model should not be focused on directions carrying more variance, we can address the practical aspect of existing generative models and propose ourselves one approach.

\medskip

That for, we will first review existing models, then we will propose an architecture turning around the propensity of standard generative models to focus on directions carrying the more variance.

\subsection{Review of existing models}

While several papers review generative models for financial time series (e.g., \cite{assefa_generating_2020}, \cite{eckerli_generative_2021},  \cite{ericson2024deep}, \cite{horvath2023harnessing}, \cite{potluru2023synthetic}), no mention is made of two particularly important aspects of portfolio construction:
\begin{itemize} 
\item The multivariate nature of the model: Are the assets generated independently, or is there a dependency structure? 
\item The type of financial instruments and the length of the historical data used for training. 
\end{itemize}
Moreover, in existing papers, the mode of evaluation of the synthetic approach can be qualitative (it is often the case) or quantitative, and there can be (or not) an out-of-sample assessment.


\begin{sidewaystable}
{\def\prcite#1{\parbox[t]{8em}{\footnotesize\raggedleft\cite{#1}}}
\def\pmodel#1{\parbox[t]{8em}{\footnotesize\raggedleft #1}}
\def\pinstr#1{\parbox[t]{8em}{\tiny\raggedleft #1}}
\def\phist#1{\parbox[t]{4em}{\tiny\raggedleft #1}}
\def\vmetrics#1{\parbox[t]{8em}{\footnotesize\raggedleft #1}}
\def\qmetrics#1{\parbox[t]{8em}{\footnotesize\raggedleft #1}}
\small 
\setlength{\tabcolsep}{1pt}
\centering\rowcolors{2}{gray!25}{white}
\scalebox{.95}{\begin{tabular}{ccccccrcrc}\rowcolor{gray!50}
Reference & Journal & Model & Multi-asset & Cond. & Data & Visual Metrics* & Quant. Metrics* & OoS?\\
\prcite{liao2024sig} & MathFin & \pmodel{Sig-CWGAN}  & Yes & Yes & \pinstr{S\&P500 and DJI prices and realized volatilities (daily, -), BTC-USD (hourly, 2020-2021)} & \vmetrics{corr., return dist.} & \qmetrics{\acrshort{Sig-W1}} & Yes\\
\prcite{pena_modified_2023} & QFin & \pmodel{Modified-CTGAN} & Yes & Yes & \pinstr{10 market indices (daily, 2008-2022)}  & \vmetrics{pair-plot, corr.} & \qmetrics{\acrshort{KS}, \acrshort{ES}, \acrshort{HH}, and \acrshort{TC}} & Yes\\
\prcite{sun2023decision} & ACM & \pmodel{DAT-CGAN} & Yes & Yes & \pinstr{4 U.S. ETFs (weekly, 1999-2016)}  & --- & \qmetrics{\acrshort{WD}} & Yes\\
\prcite{10.1145/3501305} & ACM TOMM & \pmodel{GMMN (+others)} & Yes & Yes & \pinstr{1,506 components of S\&P500 (2000-2020)} & \vmetrics{HM, ACF, VC, tails, trend ratio} & \qmetrics{KL, JS, WD} & No\\

\prcite{wiese_quant_2020} & QFin & \pmodel{Quant GAN} & No & No & \pinstr{S\&P500 (daily, 2009-2018)} & \vmetrics{return dist., \acrshort{ACF} (with \acrshort{CI})} & \qmetrics{\acrshort{WD}, \acrshort{DY}, \acrshort{ACF}, \acrshort{VC}, \acrshort{LE}} & No\\
\prcite{yoon_time-series_2019} & NeurIPS & \pmodel{TimeGAN} & Yes & No & \pinstr{Google stock open, close, high, low, adj. close, volume (daily, 2004-2019)}  & \vmetrics{t-SNE} & \qmetrics{\acrshort{MAE}, \acrshort{DS}} & No\\
\prcite{takahashi_modeling_2019} & Phys.A & \pmodel{FIN-GAN} & No & No & \pinstr{486 S\&P500 components (daily, 1962-2016)} & \vmetrics{return dist., \acrshort{ACF}, \acrshort{VC}, \acrshort{LE}, \acrshort{G/LA}, \acrshort{CFVC}} & --- & No \\
\hline \hline
\prcite{chen2025diffusion} & arXiv & \pmodel{Diffusion models} & Yes & No & \pinstr{U.S. stocks (2001-2024)} & \vmetrics{return dist.} & \qmetrics{PCA, portfolio stats.} & Yes\\
\prcite{fu2022simulating} & arXiv & \pmodel{TA-GAN, TT-GAN} & No & No & \pinstr{S\&P500 (2009-2020)} & \vmetrics{---} & \qmetrics{\acrshort{WD}, \acrshort{HM}, \acrshort{ACF}, \acrshort{VC}, \acrshort{LE} (no \acrshort{CI})} & No\\
\prcite{cont_tail-gan_2022} & arXiv & \pmodel{Tail-GAN} & Yes & No & \pinstr{AAPL, AMZN, GOOG, JPM, and QQQ (intraday, Nov 2019-Dec 2019)} & \vmetrics{corr., \acrshort{ACF}} & \qmetrics{\acrshort{VaR}, \acrshort{ES}, \acrshort{SBT}, \acrshort{CT}} & Yes\\
\prcite{lezmi_improving_2020} & arXiv & \pmodel{Cond. Bernoulli RBM/W-GAN} & Yes & Yes & \pinstr{6 futures contracts (daily, 2007-2019)} & \vmetrics{QQ-plot, return dist., \acrshort{ACF}} & \qmetrics{mean, vol., quantiles, \acrshort{SR}} & No\\
\prcite{buehler2020data} & arXiv & \pmodel{VAE-Signature} & No & Yes & \pinstr{S\&P500 (daily, weekly, monthly, -)} & \vmetrics{paths, return dist., signature projections} & \qmetrics{\acrshort{MMD}, \acrshort{KS}} & No\\
\prcite{da2019style} & arXiv & \pmodel{DAE+CNN} & No & No & \pinstr{AUD-USD (daily, 2009-2018)} & \vmetrics{\acrshort{ACF}, technical analysis} & \qmetrics{mean, vol., \acrshort{HM}, \acrshort{KS}, \acrshort{VRT}} & No \\
\prcite{de2019enriching} & --- & \pmodel{WGAN-GP/RaGAN} & 2-d & Yes & \pinstr{S\&P500 and VIX, 2000-2016; 2004-2019} & \vmetrics{corr., \acrshort{ACF}, \acrshort{HM}} & --- & No\\
\prcite{kondratyev_market_2019} & SSRN & \pmodel{Bernoulli RBM} & Yes & Yes & \pinstr{4 currency pairs: EUR-USD, GBP-USD, USD-JPY, USD-CAD (1999-2019)} & \vmetrics{QQ-plot, return dist., \acrshort{ACF}} & \qmetrics{corr., vol., quantiles} & No\\
\hline
\end{tabular}}
\caption{Comparison of papers introducing generative models for financial time series.*ACF (Autocorrelation Function), CI (Confidence Interval), Sig-W1 (Signature Wasserstein-1 metric), KS (Kolmogorov–Smirnov test), ES (Expected Shortfall), VaR (Value-at-Risk), HH (Herfindahl–Hirschman index), TC (Transaction Costs), WD (Wasserstein Distance), DY (Dragulescu-Yakovenko metric), JS (Jensen-Shanon divergence), DS (Discriminative Score), SR (Sharpe Ratio), HM (Higher Moments), MAE (Mean Absolute Error), VC (Volatility Clustering), LE (Leverage Effect), SBT (Score-Based Test), CT (Coverage Test), MMD (Maximum Mean Discrepancy), VRT (Variance Ratio Test), G/LA (Gain/Loss Asymmetry), CFVC (Coarse-Fine Volatility Correlation), KL (Kullback-Leibler divergence).}

\label{tab:compare:papers}
}
\end{sidewaystable}

\medskip

Table \ref{tab:compare:papers} summarizes the elements of papers proposing generative models for financial time series. We began with existing review papers and examined their bibliographies. We excluded papers that did not focus on time series of financial instrument returns, those centered on return prediction (e.g., \cite{kaastra1996designing}, \cite{kim2019gan}, \cite{koshiyama_generative_2019}, \cite{lezmi2023time}, \cite{mariani2019pagan}, \cite{saad1998comparative} or \cite{vuletic_fin-gan_2023}), and those focusing on implied volatility (e.g., \cite{henry2019generative}, \cite{limmer_robust_2023}, \cite{vuletic2023volgan} or \cite{wiese2019deep}). Although we find them very interesting, we exclude studies like~\cite{morel_path_2023} and \cite{parent2024factorial}, which do not rely on neural networks, to maintain the focus and feasibility of this review.

\medskip

Only seven of the 15 reviewed papers have been published in journals or conference proceedings.\footnote{\cite{pardo2019mitigating} is not reviewed, but \cite{de2019enriching} is; their models are very similar.} These are listed in the upper part of the table, sorted by publication year from the most recent to the oldest. The following elements are documented:

\begin{itemize} 
\item The multivariate aspect: Are the time series of $d$ financial instruments generated independently or not? Most models are univariate or bivariate, with the notable exception of \cite{10.1145/3501305}, which generates all 1,506 time series simultaneously. 
\item The presence of a conditioner that can synchronize the generated time series. Typically, this is a volatility regime indicator (e.g., instructing the model to generate \textit{high volatility} or \textit{low volatility} time series). Note that conditioning can introduce correlation among time series; for example, \cite{sun2023decision} uses the same random seed for four US ETF time series, creating a dependency, and \cite{de2019enriching} first generates one time series corresponding to the first PCA component of returns, then conditions the other two series on this first step. 
\item The number of time series and the historical period: It is important to note whether no historical dataset is used, the sample is unspecified, or fewer than 10 time series are utilized. \item While some papers only provide a visual check; we list the quantitative metrics used in other papers. \item Only five papers use out-of-sample data. Surprisingly, only three of the seven published papers provide out-of-sample metrics. 
\end{itemize}

In conclusion, only three papers, \cite{10.1145/3501305}, \cite{flaig_validation_2023} and \cite{chen2025diffusion}, generate more than 10 time series in a correlated way. Additionally, while many authors check in-sample distances between generated and reference data (often part of the loss function), there is a consensus on testing the stylized facts of financial returns. See \cite{cont_stylized_2010} for a description of these properties: autocorrelations, heavy tails, volatility clustering, very short memory of returns, long memory of squared returns, and skewness. These characteristics are expected in generated financial time series to validate the use of the generative model for further applications.

\medskip

A few papers extend this analysis to check stylized facts on combinations of returns, such as profits and losses of portfolios, which are linear combinations of returns. For example, \cite{lezmi_improving_2020} checks risk parity portfolios, and \cite{pena_modified_2023} checks long-only mean-variance portfolios. We believe this is a valuable metric, as stylized facts on mean-reversion and momentum are well-known, and papers like \cite{bryzgalova_forest_2019} may provide a method for constructing numerous portfolios to test synthetic data from multivariate market generators.

\subsection{A proposal of a generative model architecture}
\label{sec:training_pipeline}

In this section, we propose a generative pipeline for portfolios that addresses the two main issues outlined in the first part of the paper: keep the number of parameters of the model low despite the number of stock returns being generated (we use the stocks of the S\&P500 in the application in Section~\ref{sec:a_fit_on_spx}),
and prevent the model from focusing its accuracy on principal components with large variance.


\paragraph{High-level view of the model.}
Let $\{X(t)\}_{t \in T}$ be a $d$-dimensional stochastic process of asset returns.\footnote{The index set $T = \mathbb{Z}$ is used to maintain coherence in the analysis presented in the following sections.} The objective of generative modeling is to develop a model that captures the true characteristics of the return process using a limited sample of $n$ realizations / observations from the process, denoted by $\mathbf{X} \in \mathbb R^{n \times d}$, along with any available prior knowledge about its behavior. 

\medskip 
The basis of our framework is to model $X(t)$, as a function of two underlying processes: 
\begin{equation}
\label{eq:main_model}
X(t) = (\bm{\beta}F(t) + Z(t)) \odot \bm{\sigma} + \bm{\mu}
\end{equation}

where $\bm{\beta} \in \mathbb R^{d \times m}$ is a projection matrix, $F(t)$ and $Z(t)$ are random variables associated with $\{F(t)\}_{t \in T}$, an $m$-dimensional process of factor returns and $\{Z(t)\}_{t \in T}$, a $d$-dimensional process of residual returns, respectively. The vectors $\bm{\mu}$ and $\bm{\sigma}$ represent the means and volatilities of individual asset returns, respectively. The underlying processes are assumed to be independent and have zero mean, i.e. $\mathbb E[F(t)] =\mathbf{0}_m$ and $\mathbb E[Z(t)] = \mathbf{0}_d$. Furthermore, we assume that $\text{diag}^{-1}(\mathbf{C})=\mathbf{1}_d$ where $\mathbf{C} = \mathbb E[(\bm{\beta}F(t) + Z(t))(\bm{\beta}F(t) + Z(t))^\top]$. These assumptions ensure that the individual asset means and volatilities are preserved as specified by $\bm{\mu}$ and $\bm{\sigma}$.

\medskip
In the following subsections, we present a framework to transform $\mathbf{X}$ in a way that we can separately learn models for the underlying components $\{F(t)\}_{t \in T}$ and $\{Z(t)\}_{t \in T}$ to effectively capture the dynamics of $\{X(t)\}_{t \in T}$.

\subsubsection{Extracting factors from noisy returns}
\label{extract_factors}

Let us first standardize the sample $\mathbf{X}$ using the vectors of sample mean $\hat{\bm{\mu}}$ and volatility $\hat{\bm{\sigma}}$ to obtain a sample of standardized asset returns $\bar{\mathbf{X}}$ with columns of zero means and unit variances. The covariance matrix of standardized asset returns (or the correlation matrix of $\mathbf{X}$) is computed by $\hat{\mathbf{\Sigma}}=\frac{1}{n}\bar{\mathbf{X}}^\top\bar{\mathbf{X}}$, which can be expressed in terms of its eigen decomposition as $$\hat{\mathbf{\Sigma}} = \mathbf{P}\bm{\Delta}\mathbf{P}^\top$$ where $\bm{\Delta}$ is a diagonal matrix containing the eigenvalues $(\lambda_1, \dots, \lambda_d)$, arranged in decreasing order, and $\mathbf{P}$ is a matrix comprising orthogonal columns that holds the corresponding eigenvectors.\footnote{Note that $\mathbf{P}^\top=\mathbf{P}^{-1}$, due to orthogonality.}

\medskip
A sample of $m \leq d$ uncorrelated factor returns can therefore be obtained through the linear transformation $$\mathbf{F} = \bar{\mathbf{X}}\mathbf{P}_{1:m}$$ where $\mathbf{F} \in \mathbb{R}^{n \times m}$ is actually a sample of the first $m$ principal component (or factor) returns which are uncorrelated with variances of $\lambda_1, \dots, \lambda_m$, respectively. 

\medskip

By applying an inverse transformation, we can achieve a partial (or complete, when $m = d$) reconstruction of $\bar{\mathbf{X}}$ which allows for obtaining the remaining part in the form of residuals $$\mathbf{Z} = \bar{\mathbf{X}} - \mathbf{F}\mathbf{P}_{1:m}^\top$$ where $\mathbf{Z} \in \mathbb{R}^{n \times d}$ is a sample of residual returns.

\medskip
Consequently, the initial sample can be written as 
\begin{equation}
\label{eq:sample_return_decomp}
\mathbf{X} = (\mathbf{F}\mathbf{P}_{1:m}^\top + \mathbf{Z}) \odot \mathbf{1}_d \hat{\bm{\sigma}}^\top + \mathbf{1}_d \hat{\bm{\mu}}^\top.    
\end{equation}
This encourages us to learn and estimate the variables and parameters that appear in Equation~(\ref{eq:main_model}) using the above decomposition, given the obvious similarity between the two forms. Indeed, for instance, the projection matrix is estimated by $\hat{\bm{\beta}} = \mathbf{P}_{1:m}$.

\medskip

One critical point in this approach is the choice of $m$, which determines the part of asset returns that will be associated with the factors. The choice of $m$ is in fact free and can be obtained discretionally or using a statistical methodology of a specific kind. We use results from random matrix theory to determine the optimal number of factors $m$.

\medskip

In simplest terms, as $n, d \rightarrow \infty$ with the ratio $\frac{n}{d} = q \geq 1$ remaining constant, the distribution of eigenvalues of a covariance matrix computed from an $n \times d$ matrix, whose columns consist of i.i.d. random entries with variance $\sigma^2$, converges almost surely to the Marcenko-Pastur distribution \cite{marchenko1967distribution}
:
\begin{equation}
\label{eq:mp_density}
f_\lambda(x)=\frac{q}{2 \pi \sigma^2} \frac{\sqrt{\left(\lambda_+-x\right)\left(x-\lambda_-\right)}}{x}
\end{equation}
with $x\in [\lambda_-, \lambda_+]$ where $\lambda_- = \sigma^2\Big(1 - \sqrt{\frac{1}{q}}\Big)^2$ and $\lambda_+ = \sigma^2\Big(1 + \sqrt{\frac{1}{q}}\Big)^2$.

\medskip 

The value $\lambda_+$ sets an upper bound for the eigenvalues of covariance matrices computed on i.i.d. samples. Based on this fact, we treat, in our analysis, the eigenvalues exceeding this threshold as objects related to factors that break the \textit{i.i.d-ness}, leading to a deviation from the Marcenko-Pastur distribution. Therefore, we simply use $\lambda_+$ to choose $m$. Basically, $m$ is equal to the number of eigenvalues $\lambda_1, \dots, \lambda_d$ which are greater than $\lambda_+ = (1 + \sqrt{d/n})^2$.\footnote{Note that $\sigma^2=1$ in our case since $\hat{\mathbf{\Sigma}}$ is computed using standardized returns.} 

\medskip

 In this subsection, we introduced a modeling framework that views $X(t)$ as a function of $F(t)$ and $Z(t)$ and a means of decomposing the sample of asset returns $\mathbf{X}$ into two samples, $\mathbf{F}$ and $\mathbf{Z}$. In the next subsections, we will focus on the tasks of modeling these components using the obtained samples while treating them differently in terms of modeling approaches.

\subsubsection{Capturing memory using Generative Adversarial Networks}
\label{sec:GANs}

\paragraph{Reducing the exposure to high-variance components.}
Thanks to the orthogonal decomposition mentioned in the previous section, we can afford to model the components independently. Since the amplitude of their respective eigenvalue is factored out, this step prevents the model to be exposed to the weakness detailed in Section \ref{sec:learning_and_application}.


\paragraph{Preserving the memory of principal components.}
We use the Generative Adversarial Network (GAN) framework proposed by \cite{goodfellow_generative_2014} to model factors. A GAN is typically composed of two neural networks, namely the generator and the discriminator, trained simultaneously on a sample of input data in order to find the set of parameters for the generator network that is capable of producing simulated data (from noise) whose distribution matches that of the input data.

\medskip
Let us consider the first factor $F_1(t)$ which is associated with the univariate stochastic process of returns of the first factor with a sample of realizations $\mathbf{F}_{:,1}=\bm{\xi} = (\xi_1, \dots, \xi_n) \in \mathbb R^n$ in the form of a univariate time series. We cannot directly use this sample for training since we want to model the relationships across time. We therefore need to build a training set out of $\bm{\xi}$ by slicing out windows of the size we assume to be that of the length of the \textit{memory}, denoted by $s$, in the process. Consequently, we get the training set $\bm{\Xi} = (\bm{\xi}_{1:s}, \bm{\xi}_{2:s+1}, \dots, \bm{\xi}_{n-s+1:n})\in \mathbb R^{(n-s+1) \times s}$. The generator is therefore supposed to learn the joint distribution of an $s$-dimensional random variable $\Xi$ following a probability distribution $\mathbb P_{\Xi}$ from which $\bm{\Xi}$ is assumed to be sampled.

\medskip
If we denote the generator by the function $g: \mathbb R^k \times \mathcal{X} \rightarrow \mathbb R^s$ where $\mathcal{X}$ is the parameter space for the function parameters, the goal is to find $\bm{\Theta}^*_g \in \mathcal{X}$ such that $g(Z, \bm{\Theta}^*_g) \sim \mathbb P_{\Xi}$ where $Z \sim \mathbb P_Z$ is a $k$-dimensional random variable, so-called the noise, generally assumed to follow a uniform or normal distribution.\footnote{The noise does not necessarily have to be characterized as a vector of fixed size. It can also be characterized as a matrix or higher-dimensional objects of variable size, depending on the network architecture.} The optimal parameters $\bm{\Theta}^*_g$, under which $g$ can transform random inputs into trajectories with distributional properties close to the available sample,
are found through adversarial training of the generator against the discriminator denoted by the function $d: \mathbb R^s \times \mathcal{Y} \rightarrow \mathbb [0,1]$ whose output can be interpreted as the probability that the input vector is coming from $\mathbb P_{\Xi}$.\footnote{$\mathcal{Y}$ represents the parameter space for the discriminator parameters, which do not have to belong to the same space as the generator parameters, since the generator and discriminator can have different architectures.} During adversarial training, together with $\bm{\Theta}_g$, the discriminator parameters $\bm{\Theta}_d \in \mathcal{Y}$ are optimized to continuously improve the discriminator's ability to distinguish between real and synthetic data produced by the generator. This is achieved through the following optimization problem:
\begin{equation}
\label{eq:gan_loss}
    \min_{\bm{\Theta}_g} \max_{\bm{\Theta}_d} \mathbb E_\Xi \big[\log d(\Xi, \bm{\Theta}_d)\big] + \mathbb E_Z\big[\log \big(1 - d(g(Z, \bm{\Theta}_g),\bm{\Theta}_d)\big)\big].
\end{equation}

The above problem illustrates the nature of adversarial learning. For a given generator, the discriminator parameters are adjusted so that the discriminator outputs higher values if the input comes from the data distribution and lower values if the input comes from the generator. On the other hand, the generator aims to find a parameterization for which the discriminator cannot successfully perform such a distinction between real and synthetic data.
\medskip

This formulation is equivalent to minimizing the Jensen-Shannon divergence \cite{lin1991divergence} between $\mathbb P_{\Xi}$ and $\mathbb P_g$ where $\mathbb P_g$ denotes the probability distribution of the synthetic data $g(Z, \bm{\Theta}_g)$. Consequently, the cross-entropy formulation in Problem~(\ref{eq:gan_loss}), which we use in the numerical part of the paper, is theoretically justified as a way of \textit{ensuring} that synthetic data will be distributionally close to the real data.\footnote{Other types of divergence or distance measures are also used (e.g., \cite{arjovsky_wasserstein_2017} and \cite{nowozin2016f}). Regarding the use of GANs in finance, \cite{cont_tail-gan_2022} proposes a loss function based on the joint elicitability property of a certain class of risk measures to capture heavy tails in asset returns.} 

\medskip
In addition to the loss function (the objective function in Problem~(\ref{eq:gan_loss})), another critical point is the architecture choice for the generator and discriminator. We use Temporal Convolutional Networks (TCNs) as they have proven to be effective for time series generation, initially proposed by \cite{oord2016wavenet} for raw audio and successfully applied to financial data by \cite{wiese_quant_2020}. They are capable of learning complex relationships between distant time points thanks to their dilation mechanism respecting at the same time the order of data. Another advantage, particularly for the generator, is that they can produce variable-sized outputs which can be determined as a function of the noise dimension. 

\subsubsection{Reducing the number of parameters per data point via factor clustering}
\label{sec:factor_clustering}
Since we aim to model factor returns separately and have already seen how to create a training set of time series of size $s$ from a single time series, the initial idea can be to model each factor using an individual GAN. For each factor $i \in \{1, \dots, m\}$, we can construct the sample $\bm{\Xi}_i$ from the time series $\mathbf{F}_{:,i}$ to obtain $m$ generators, each specific to one factor.

\medskip
However, implementing $m$ separate models may not be ideal, especially when $m$ is large and since GANs are known to be \textit{data-hungry} \cite{karras2020training}, that is worsening the issue detailed in Section \ref{sec:sample_size}. To address this, we propose to group the factors into a smaller number of clusters, specifically $n_c$ clusters ($n_c \leq m$). This strategy enables us to model components with similar characteristics using a single model, thereby improving the ratio between the sample size and the number of parameters of the network.

\paragraph{Rescaling the factors.}
First, we scale the factor returns using the associated eigenvalues to obtain series with identical first two moments. We denote this sample of scaled factor returns by $\bar{\mathbf{F}}$ where $\bar{\mathbf{F}}_{:,i} = \frac{1}{\lambda_i}\mathbf{F}_{:,i}$. As a result, $\bar{\mathbf{F}}$ contains time series in its columns that have zero mean and unit variance, which should allow the clustering method to focus on more nuanced properties. This scaling will also be useful for training the GANs for the clusters, as it ensures that the training data fed into the model have the same mean and variance. 

\paragraph{Clustering time series.}
The range of methods available for clustering time series data is extensive (see \cite{aghabozorgi2015time}). Clustering can be handled in a fully data-driven way or by specifying relevant statistics for the properties we want to capture and running a clustering algorithm on these variables. We adopt a simple approach and use agglomerative clustering for the scaled factors $\bar{\mathbf{F}}_{:,i}$, based on five statistics: skewness, kurtosis (expressed in excess throughout this paper), eigenvalue, volatility clustering score, and leverage effect score, which will be detailed in the following sections.\footnote{The eigenvalue can be interpreted as an importance score, included to decrease the probability that \textit{significant} factors (e.g., $F_1(t)$) and \textit{less significant} factors (e.g., $F_m(t)$) fall into the same cluster unless they are very similar in other properties. This helps avoid the disruption of high-impact factors by much lower-impact factors when modeling.}

\medskip
As each factor now belongs to a cluster, we can build the training set on which each of the $n_c$ GANs will be trained. We first construct the sample for each scaled factor $\bar{\mathbf{F}}_{ :,i}$ using the sliding window method mentioned above to obtain the sample $\bar{\bm{\Xi}}_i \in \mathbb R^{(n-s+1) \times s}$. Then we bring these samples together by concatenating them (by rows) to obtain, for each cluster, a training set  $\bm{\Upsilon}_j \in \mathbb R^{c_j(n-s+1) \times s}$, $\forall j \in \{1, \dots, n_c\}$ where $c_j$ is the number of elements in $\mathcal{C}_j$, the set that holds the indices of the factors in the corresponding cluster. Consequently, the factors within the cluster $j$ will be learned by a single GAN, whose estimated generator parameters are denoted by $\hat{\bm{\Theta}}_j$ , allowing it to learn from a larger dataset composed of factor return time series with \textit{similar} distributions.

\begin{figure}[ht]
\centering
\begin{tikzpicture}[scale=0.75, every node/.style={font=\footnotesize}]
\node[draw, align=center, rounded corners] (Return)[text width=3cm] at (0,0) {Asset returns $\mathbf{X}$};
\node[draw, align=center, rounded corners] (stand ret) at (0,-1.5) {Standardized returns $\bar{\mathbf{X}}$};
\node[draw, align=center, rounded corners] (corr) at (0,-3) {Covariance matrix $\hat{\mathbf{\Sigma}}$};
\node[draw, align=center, rounded corners] (e_vectors) at (0,-4.5) {Eigenvectors $\mathbf{P}$};
\node[draw, align=center, rounded corners] (e_values) at (4,-3) {Eigenvalues $\bm{\Delta}$};
\node[draw, align=center, rounded corners] (n_factors) at (4,-4.5) {\# factors $m$};
\node[draw, align=center, rounded corners] (factor_returns)[text width=2cm] at (0,-6) {Factor returns $\mathbf{F} = \bar{\mathbf{X}}\mathbf{P}_{1:m}$};
\node[draw, align=center, rounded corners] (scaled_factor_returns)[text width=2.8cm] at (3,-8) {Scaled factor returns $\bar{\mathbf{F}} = \mathbf{F} (\bm{\Delta}^{-1})_{1:m}$};
\node[draw, align=center, rounded corners] (residuals)[text width=2.5cm] at (-3,-8) {Residual returns $\mathbf{Z} = \bar{\mathbf{X}} - \mathbf{F}\mathbf{P}_{1:m}^\top$};

\node (res1) at (-4.5,-10) {$\mathbf{Z}_{:,1}$};
\node (res2) at (-3.5,-10) {$\mathbf{Z}_{:,2}$};
\node (res3) at (-2.5,-10) {$\hdots$};
\node (res4) at (-1.5,-10) {$\mathbf{Z}_{:,d}$};

\node[circle,draw] (mod1) at (-4.5,-11.5) {$\mathcal{M}$};
\node[circle,draw] (mod2) at (-3.5,-11.5) {$\mathcal{M}$};
\node (mod3) at (-2.5,-11.5) {$\hdots$};
\node[circle,draw] (mod4) at (-1.5,-11.5) {$\mathcal{M}$};

\node (param1) at (-4.5,-13) {$\hat{\bm{\theta}}_1$};
\node (param2) at (-3.5,-13) {$\hat{\bm{\theta}}_2$};
\node (param3) at (-2.5,-13) {$\hdots$};
\node (param4) at (-1.5,-13) {$\hat{\bm{\theta}}_d$};

\node (pc1) at (1.5,-10) {$\bar{\mathbf{F}}_{:,1}$};
\node (pc2) at (2.5,-10) {$\bar{\mathbf{F}}_{:,2}$};
\node (pc3) at (3.5,-10) {$\hdots$};
\node (pc4) at (4.5,-10) {$\bar{\mathbf{F}}_{:,m}$};

\node[draw, align=center, rounded corners] (clustering)[text width=3cm] at (3,-11.5) {Clustering};

\node (cluster1) at (2,-13) {$\bm{\Upsilon}_1$};
\node (cluster3) at (3,-13) {$\hdots$};
\node (cluster4) at (4,-13) {$\bm{\Upsilon}_{n_c}$};

\node[circle,draw] (gan1) at (2,-14.2) {$\mathcal{NN}$};
\node (gan3) at (3,-14.2) {$\hdots$};
\node[circle,draw] (gan4) at (4,-14.2) {$\mathcal{NN}$};

\node (gan_param1) at (2,-15.5) {$\hat{\bm{\Theta}}_1$};
\node (gan_param3) at (3,-15.5) {$\hdots$};
\node (gan_param4) at (4,-15.5) {$\hat{\bm{\Theta}}_{n_c}$};

\draw[->,>=latex] (Return.south) -- (stand ret.north);
\draw[->,>=latex] (stand ret.south) -- (corr.north);
\draw[->,>=latex] (corr.south) -- (e_vectors.north);
\draw[->,>=latex] (corr.east) -- (e_values.west);
\draw[->,>=latex] (e_values.south) -- (n_factors.north);
\draw[->,>=latex] (e_vectors.south) -- (factor_returns.north);
\draw[->,>=latex, bend left=25] (n_factors.south) to (factor_returns.east);

\draw[->,>=latex] (factor_returns.south) -- (scaled_factor_returns.north);
\draw[->,>=latex] (factor_returns.south) -- (residuals.north);

\draw[->,>=latex] (residuals.south) |- (-4.5,-9) -- (res1);
\draw[->,>=latex] (residuals.south) |- (-3.5,-9) -- (res2);
\draw[->,>=latex] (residuals.south) |- (-1.5,-9) -- (res4);

\draw[->,>=latex] (res1) -- (mod1);
\draw[->,>=latex] (res2) -- (mod2);
\draw[->,>=latex] (res4) -- (mod4);

\draw[->,>=latex] (mod1) -- (param1);
\draw[->,>=latex] (mod2) -- (param2);
\draw[->,>=latex] (mod4) -- (param4);

\draw[->,>=latex] (scaled_factor_returns.south) |- (1.5,-9) -- (pc1);
\draw[->,>=latex] (scaled_factor_returns.south) |- (2.5,-9) -- (pc2);
\draw[->,>=latex] (scaled_factor_returns.south) |- (4.5,-9) -- (pc4);

\draw[->,>=latex] (pc1.south) -- ([xshift=-1.5cm] clustering.north);
\draw[->,>=latex] (pc2.south) -- ([xshift=-0.5cm] clustering.north);
\draw[->,>=latex] (pc4.south) -- ([xshift=1.5cm] clustering.north);

\draw[->,>=latex] (clustering.south) |- (2,-12.2) -- (cluster1);
\draw[->,>=latex] (clustering.south) |- (4,-12.2) -- (cluster4);

\draw[->,>=latex] (cluster1) -- (gan1);
\draw[->,>=latex] (cluster4) -- (gan4);

\draw[->,>=latex] (gan1) -- (gan_param1);
\draw[->,>=latex] (gan4) -- (gan_param4);

\end{tikzpicture}

\caption{Training pipeline of the generative model.}
\label{fig:pipeline_learning}
\end{figure}
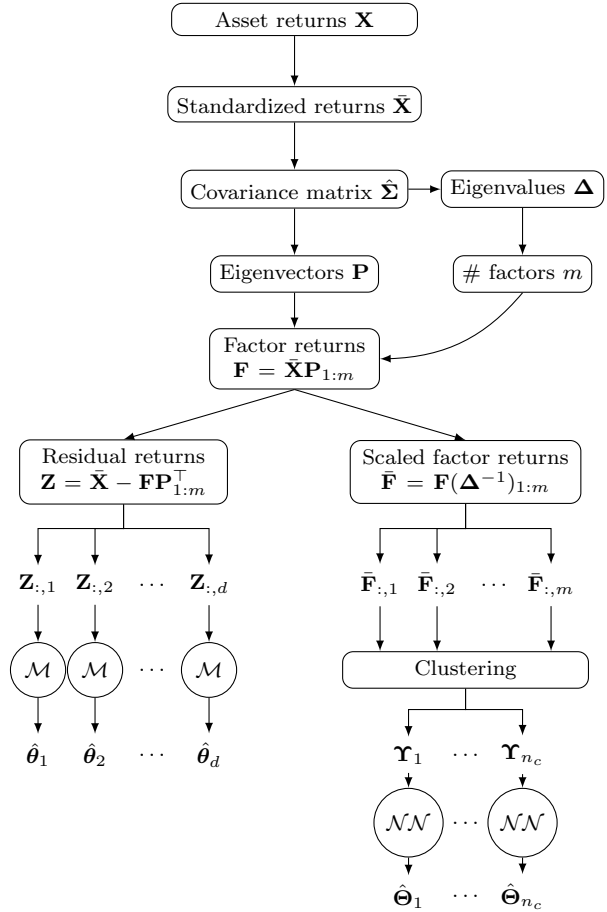

\subsubsection{Variance correction with non-normal white noise}
\label{sec:residuals}
In the preceding subsection, we studied the problem of modeling the first component of the random term in Model~(\ref{eq:main_model}). We now need to focus on the second component, which might account for a significant portion of the variance observed in asset returns. 

\medskip
In Section~\ref{extract_factors}, we related the residual returns to eigenvalues smaller than $\lambda_+$, which correspond to the part of the spectral distribution of the covariance matrix that would be formed by i.i.d. entries. Based on this choice, we model each $Z_i(t)$ as if they are i.i.d. without cross-dependencies. However, being aware that they can possess non-normal distributional structures, we should model the marginal distribution of $Z_i(t)$ in a way that is flexible enough to capture the distributional properties that can arise in univariate distributions of residual returns.

\medskip
A mixture of two Student-t distributions can be a good candidate among the endless classes of parametric models. A random variable $Z$ is said to follow a univariate Student-t mixture distribution with two components if $$Z = 1_{C = 1} Z_1 + 1_{C = 2} Z_2, $$ 
where for all $i \in \{1,2\}$, $Z_i$ follows a Student-t distribution with location $\mu_i$, scale $s_i$, and degrees of freedom $\nu_i>1$ parameters and $C$ is a discrete random variable with values in $\{1, 2\}$ and $\Pb(C=1) = p$ and $\Pb(C=2) = 1-p$ and $C, Z_1, Z_2$ are mutually independent. 

\medskip
The probability density function is given by 
\begin{equation}
    f(x|\bm{\theta}) := p f_s(x|\mu_1, s_1, \nu_1) + (1-p) f_s(x|\mu_2, s_2, \nu_2)
\label{uni_t_density}
\end{equation}
where $f_s$ is the density of Student-t distribution and $\bm{\theta} = (p, \mu_1, s_1, \nu_1, \mu_2, s_2, \nu_2)$.\footnote{The density of Student-t distribution is $f_s(x|\mu, s, \nu)=\frac{\Gamma\left(\frac{\nu+1}{2}\right)}{\Gamma\left(\frac{\nu}{2}\right) \sqrt{\nu \pi}s}\left(1+\frac{1}{\nu}\left[\frac{x-\mu}{s}\right]^2\right)^{-\frac{\nu+1}{2}}$.}

\medskip
The advantage of Student-t mixtures is that they allow for the modeling of skewness and heavy-tails, which are two characteristics that might be observed in residual returns. We should therefore estimate $\hat{\bm{\theta}}_i$ from $\mathbf{Z}_{:,i}$ to have a model for $Z_i(t)$, for all $i \in \{1, \dots, d \}$. A common approach to estimating the parameters of mixture models involves using the expectation-maximization method \cite{dempster1977maximum}.

\medskip
Another reason why we propose Student-t mixtures is that they encompass other \textit{simpler} distributions, such as Gaussian mixtures (the case when $\nu_1 \rightarrow +\infty$ and $\nu_2 \rightarrow +\infty$), Student-t ($p=1$) and Gaussian ($p=1$ and $\nu_1 \rightarrow +\infty$). These simpler distributions can be chosen to minimize the effective number of parameters to be estimated by fixing the necessary parameters in advance.

\medskip

Figure~\ref{fig:pipeline_learning} provides a schematic illustration of the entire pipeline depicted in this section.\footnote{The proposed pipeline is by no means the only architecture, nor necessarily the best one, for addressing the concerns raised in Section~\ref{sec:chap32}. For example, for the issue mentioned in Section~\ref{sec:learning_and_application}, a promising direction would be to modify the loss function so that it places more emphasis on low-variance factors. An entropy-based regularization term, for example, could serve this purpose, in line with the findings of \cite{collas2023entropic}.}

\subsubsection{The market generator}
The above subsections are devoted to describing the transformations applied to the original data and the learning process of the generative model. Here, we focus on how to simulate a synthetic sample of length $\Tilde{n}$ once we have the estimates for all the necessary parameters $\{\hat{\bm{\theta}}_1, \dots, \hat{\bm{\theta}}_d\}$, $\{\hat{\bm{\Theta}}_1, \dots, \hat{\bm{\Theta}}_{n_c}\}$, $\hat{\bm{\beta}}$, $\hat{\bm{\sigma}}$ and $\hat{\bm{\mu}}$. 

\medskip
As a matter of fact, a simulated sample $\Tilde{\mathbf{X}} \in \mathbb R^{\Tilde{n} \times d}$ is generated by

\begin{equation}
\label{eq:market_generator}
    \Tilde{\mathbf{X}}_{i,j} = \sum_{k=1}^m \hat{\bm{\sigma}}_j\hat{\bm{\beta}}_{j,k} \lambda_k \sum_{l=1}^{n_c} 1_{k \in \mathcal{C}_l} g(\mathbf{z}_k, \hat{\bm{\Theta}}_l)_i + \hat{\bm{\sigma}}_j F^{-1}(u_i|\hat{\bm{\theta}}_j) + \hat{\bm{\mu}}_j
\end{equation}

where $F(.|\bm{\theta})$ is the cumulative distribution function associated with the density in (\ref{uni_t_density}), $\mathbf{z}_k$ is a sample (of a specific size such that the output of $g$ is of size $\Tilde{n}$) with elements drawn from a standard normal distribution, $g(.,.)_i$ is the $i^{th}$ element of the output of $g$ and $u_i$ is a point drawn from a uniform distribution on $[0,1]$. 

\medskip
The proposed generative pipeline and modeling framework, abstracting from the technical details of this section, can be summarized as follows:
\begin{itemize}
\item \underline{Training / learning phase:}
\begin{itemize}
\item Start from a dataset of multivariate asset returns (standardized) and estimate its covariance matrix.
\item Select a subset of eigenvectors of this covariance matrix and project the data onto the corresponding orthogonal factor space.
\item Rescale the factor time series using the associated eigenvalues so that they have zero mean and unit variance, and cluster factors with similar statistical properties.
\item Train one GAN per cluster, rather than one GAN per factor, in order to improve the sample-size-to-parameter ratio.
\item Recover residuals by subtracting the factor-based reconstruction from the original returns.
\item Fit a simple parametric distribution independently to the residual series of each asset.
\end{itemize}
\item \underline{Generation phase:}
\begin{itemize}
\item Simulate factor returns by sampling from the GAN associated with each factor’s cluster and rescaling by the corresponding eigenvalue.
\item Project the simulated factor time series back to the asset space using the estimated eigenvectors.
\item Repeat this procedure for all retained factors.
\item Independently sample synthetic residuals from the fitted parametric distributions and add them to the reconstructed factor-driven returns.
\end{itemize}
\end{itemize}

\section{A fit on the S\&P500 universe}
\label{sec:a_fit_on_spx}
In this section, we test the above pipeline on the specific dataset of daily returns of 433 stocks from Jan-2010 and May-2024 selected from the S\&P500 Index.\footnote{We use the index components as of Sep-2023 and obtain 433 stocks after removing stocks that has no data prior to Jan-2010. Note that the goal of this section is to be illustrative, not to account for subtle details of survivorship biases.}
We split the dataset into two parts: a training set from Jan-2010 to Dec-2021 and a test set from Jan-2022 to May-2024. The aim of this section is to demonstrate that the reasoning behind our proposed modeling framework is supported by the data and that the model's ability to generate synthetic data is reasonably satisfying at first glance under conventional evaluation measures.

\subsection{Uncovering stylized facts in asset return components}
In this subsection, we analyze our specific dataset to assess the adequacy of the proposed pipeline prior to any training or simulation performed in Section~\ref{sec:synthetic_data}. The goal is to examine the market data and evaluate the suitability of our modeling philosophy. Moreover, our analysis can be useful for understanding the stylized facts observed across different components of asset returns.

\medskip

\begin{figure}[ht]%
    \centering
    \subfloat{{\includegraphics[width=6.4cm]{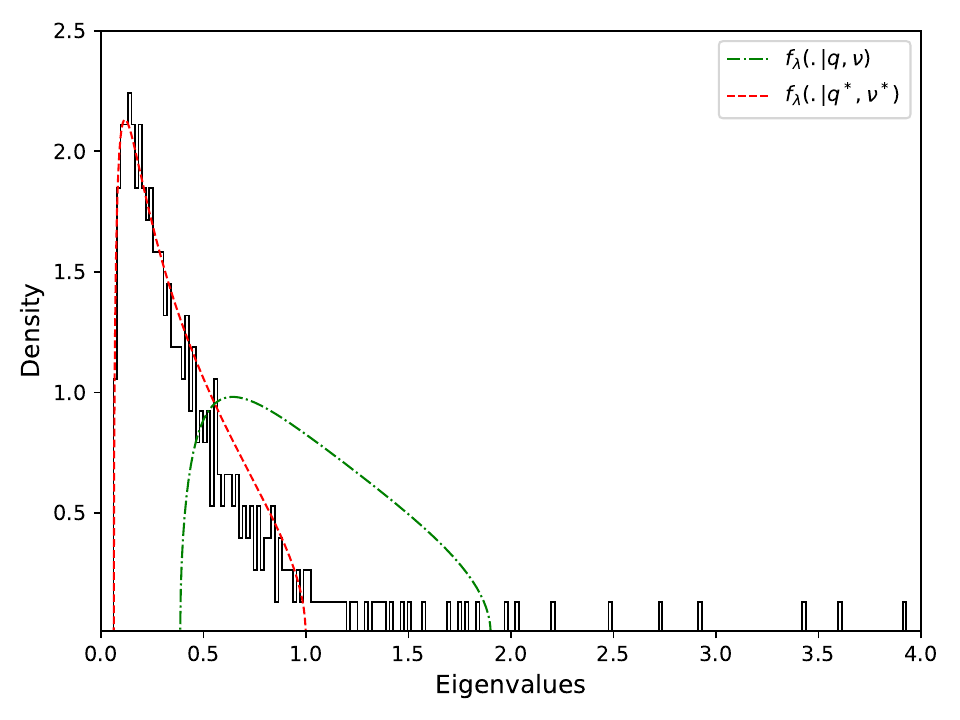} }}%
    \qquad
    \subfloat{{\includegraphics[width=6.4cm]{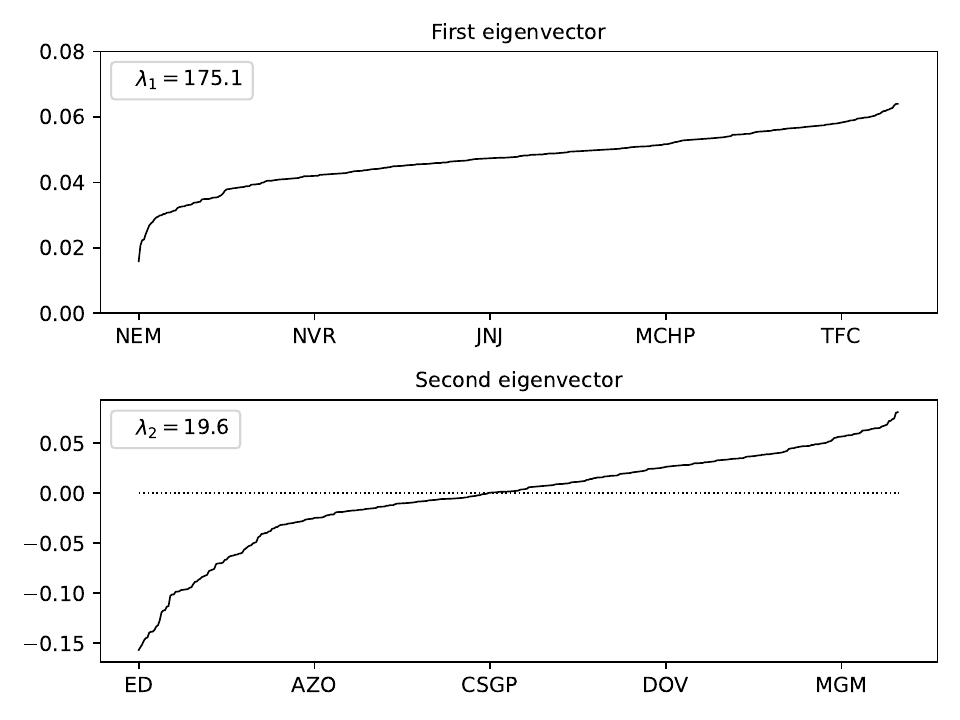} }}%
    \caption{(Left) Eigenvalue distribution for the selected universe (black). Marcenko-Pastur density for $\hat{q}=6.97$ and $\hat{\sigma}^2=1$ (green). A fit obtained with an optimal pair $(q^*,\sigma^*)$ different from those associated with the training set for illustration purposes (red). (Right) Components of the first and second eigenvectors in ascending order.}
    \label{fig:marcenko_pastur}
\end{figure}

Let us start by applying the steps described in Section~\ref{extract_factors} to the training set $\mathbf{X} \in \mathbb{R}^{3020 \times 433}$. We begin by standardizing the data using the sample means $\hat{\bm{\mu}}$ and standard deviations $\hat{\bm{\sigma}}$, followed by computing the sample covariance matrix on which a full principal component decomposition is applied. The resulting eigenvalues should be used to make the distinction between \textit{factors} and \textit{residuals} using the upper bound $\lambda_+=1.90$ which is estimated as a function of the shape of the training set and the variance of its elements, for which our dataset yields $\hat{q}=6.97$ and $\hat{\sigma}^2=1$.\footnote{To choose a \textit{better} $\lambda_+$, one could try to find the parameters providing the best fit as proposed by \cite{de2020machine}. We do not use this method in this paper for the sake of robustness and clarity although we illustrate such a fit in Figure~\ref{fig:marcenko_pastur} which would yield $\lambda_+=0.99$ and thus 44 factors.} There are 16 factors associated with eigenvalues greater then $\lambda_+$ explaining 58.9\% of the variance, as part of the distribution of eigenvalues illustrated in Figure~\ref{fig:marcenko_pastur}. The chart also displays the elements of the first two eigenvectors, showing that the first factor is the market factor, with positive components for all stocks, representing a long-only portfolio of the given universe. This factor is associated with an eigenvalue significantly higher than the second one, which contains both positive and negative components and can be interpreted as a long-short portfolio, though its precise interpretation would require further analysis.

\medskip
With the chosen number of factors, a decomposition (as in Equation~(\ref{eq:sample_return_decomp})) of standardized asset returns to the level of factors and residuals can be performed. In other words, the return of a stock $j \in \{1,\dots, 433\}$ for a given date with the index $i \in \{1,\dots, 3020\}$, $\bar{\mathbf{X}}_{i,j}$, can be expressed as the sum of a factor-based return $\mathbf{Y}_{i,j} = \mathbf{F}_{i,:}^\top\mathbf{P}_{j,1:16}$ and a residual return $\mathbf{Z}_{i,j}$. We adopt different modeling approaches for these components, as detailed in Section~\ref{sec:GANs} and Section~\ref{sec:residuals}, based on the premise that factor-based returns are responsible for the dependence properties of returns, both cross-sectionally and across time, observed in financial markets \cite{cont_empirical_nodate}. 

\medskip
A time-dependency analysis for a given $X(t)$ and lag $\tau \in \mathbb{Z}$ is usually performed using an autocorrelation function defined as $$\rho_{X, (g_1, g_2)}(\tau) = \rho(g_1(X(t)), g_2(X(t+\tau))).$$
where $g_1 : \mathbb R \rightarrow \mathbb R$ and $g_2 : \mathbb R \rightarrow \mathbb R$ denote specific functions chosen according to the analysis in question and $\rho$ is the correlation function between two random variables. 

\medskip

For example, it is well documented that asset returns generally do not exhibit significant linear autocorrelation \cite{fama1970efficient}, unless analyzed at the microstructure level, with $\rho_{X, (x, x)}(\tau) \approx 0$, for $\tau>0$.
But non-linear relationships emerge for different choices of $g_1$ and $g_2$ that we are aiming to preserve on synthetic data. 

\paragraph{Volatility clustering.}
One notable phenomenon is the tendency for large price changes to be followed by other large price changes, also known as volatility clustering, as shown by the fact that $\rho_{X, (x^2, x^2)}$ or $\rho_{X, (|x|, |x|)}$ is significantly positive, at least for non-large $\tau$ \cite{ding1993long}. Another important but less obvious relationship is known as the leverage effect \cite{bouchaud2001leverage}, which implies that negative returns lead to an increase in future volatility, often evidenced by $\rho_{X, (x, |x|)}(\tau)<0$, for $\tau>0$. Absence of this relationship for $\tau<0$ pointing out to the time-reversal asymmetry in asset returns \cite{zumbach_time_2007}.

\medskip

\begin{figure}[ht]%
    \centering
    \subfloat{{\includegraphics[width=6.4cm]{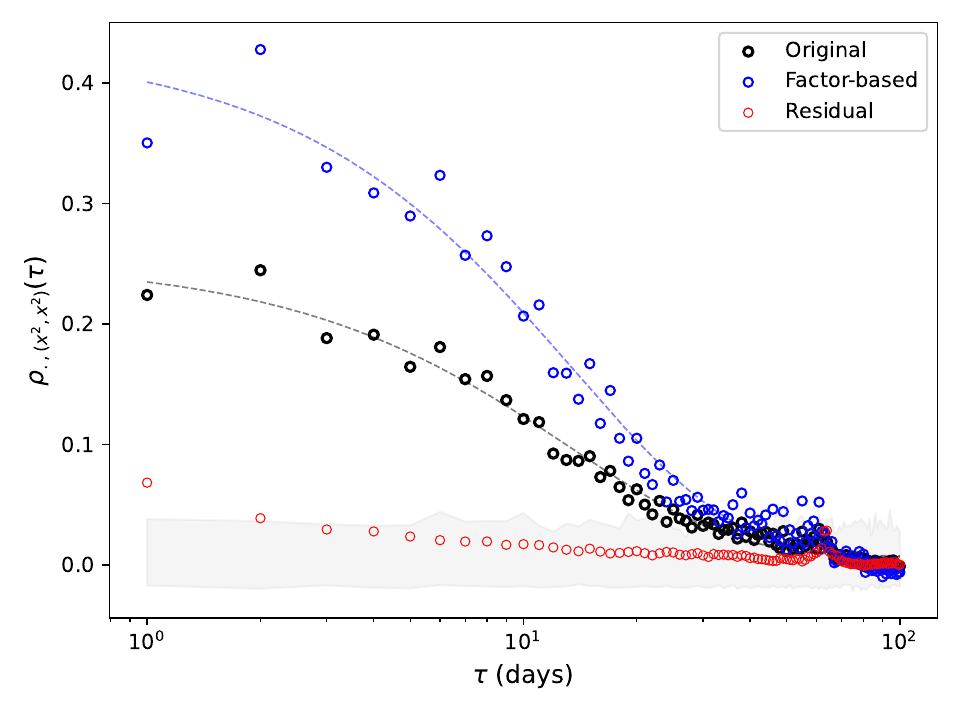} }}%
    \qquad
    \subfloat{{\includegraphics[width=6.4cm]{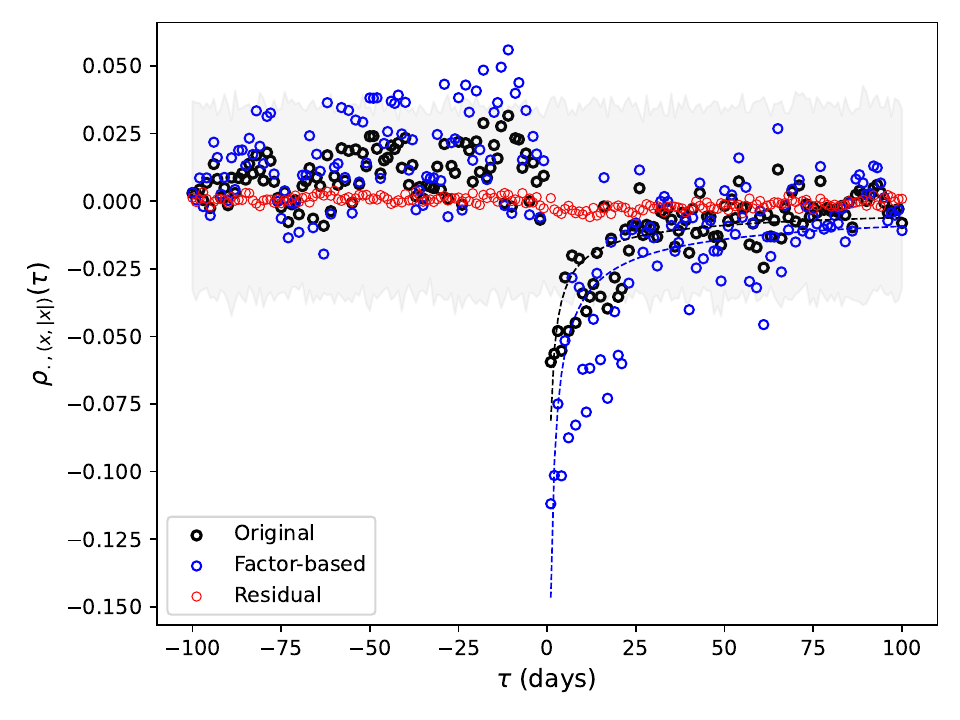} }}%
    \caption{Median non-linear time dependence at the level of different components. Volatility clustering (left) and leverage effect (right) is confirmed in original asset returns and more strongly observed in factor-based returns with exponential $\propto a e^{-\gamma}$ and power law $\propto a x^{-\gamma}$ fits, respectively. Grey areas represent 95\% bootstrap confidence interval for an i.i.d sample of the length of the training set from a standard Student-t with 3 degrees of freedom.}%
    \label{fig:autocorrelations}%
\end{figure}

As illustrated in Figure~\ref{fig:autocorrelations}, we carry out similar statistical analyses at the level of factor-based and residual returns, as well as at the level of original asset returns, to investigate how these two elements, which make up asset returns, contribute to the temporal dependence structure observed in financial markets.\footnote{Precisely, the relevant autocorrelation value is calculated for each of the 433 stocks for a range of $\tau$. Then, for each $\tau$, the median value of these 433 values is plotted.} Firstly, although not shown in the figure, we found that the two components do not exhibit significant linear autocorrelation, nor do the asset returns themselves. However, more interesting insights are revealed from a non-linear analysis. Indeed, factor-based returns tend to show a more severe volatility clustering effect, while this effect is relatively negligible (but not totally absent) on the residual side.\footnote{It should be noted that these findings are obtained for the specific choice of 16 factors. The non-significant correlation observed in the squared residual returns for small lags may disappear after the inclusion of a larger number of factors.}  In the literature, the function $\rho_{X, (x^2, x^2)}$ is often found to be a slow-decay function as a power law with $\gamma \in [0.2,0.4]$, especially when computed on intraday data \cite{cont1997scaling}. In our case, for daily returns, exponential function ($a e^{-\gamma}$) provides significantly better fits  both for the original returns ($a=0.25$ and $\gamma=0.07$) and the factor-based returns ($a=0.42$ and $\gamma=0.07$).  

\paragraph{Leverage effect.}
Similar conclusion can be drawn regarding the leverage effect which is evident not only in the original asset returns ($a=-0.08$ and $\gamma=0.56$) but also and more in the components of factor-based returns ($a=-0.15$ and $\gamma=0.60$) with power law ($a x^{-\gamma}$) fits indicating a significant but fast-decay of this effect.

\paragraph{Low dimension of the multivariate returns.}
The correlation matrix of asset returns is 
often preferred over the covariance matrix for pure co-movement analysis, as it is unaffected by individual asset volatilities. For example, in the case of equities, stock returns within the same market tend to exhibit high correlations, with even higher correlations observed between stocks in the same sector, resulting in clusters within the correlation matrix.

\begin{figure}[ht]
    \centering
    \includegraphics[width=6.4cm]{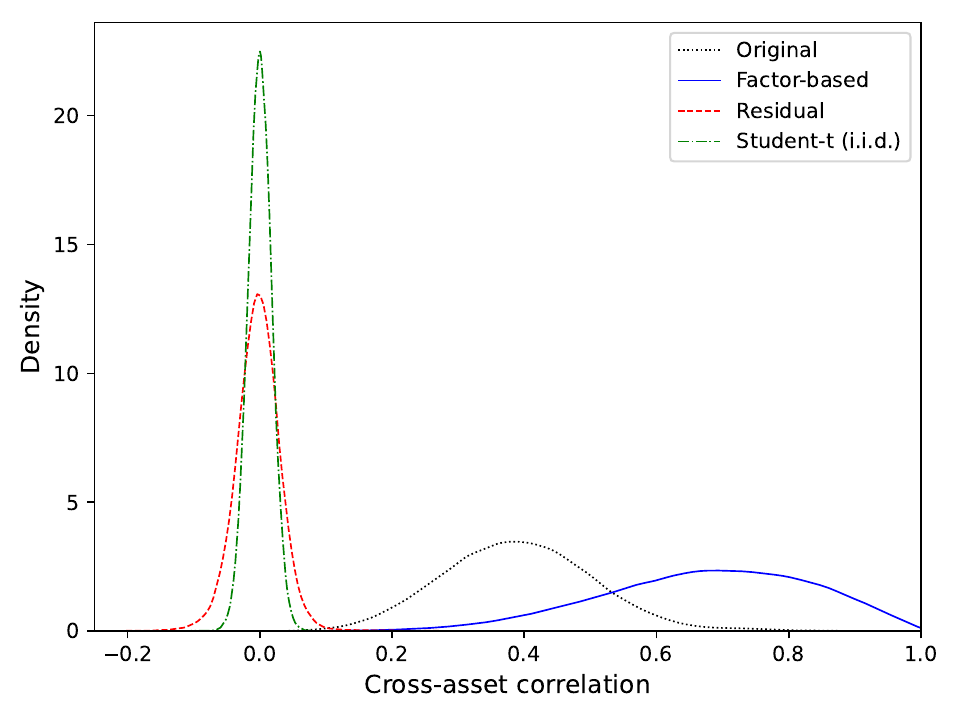}
    \caption{Distribution of cross-asset correlations obtained from the lower triangle of the correlation matrix computed for different components.}
    \label{fig:correlations}
\end{figure}


\medskip 

As indicated in Section \ref{extract_factors}, the distribution of eigenvalues associated with residuals fits with the spectral distribution of a correlation matrix calculated on random entries. The fact that the correlations between asset returns are induced by a few factors 
is demonstrated in Figure~\ref{fig:correlations} and Table~\ref{table:correlations} where we see that the correlations between residual returns are small and centered around zero. Figure~\ref{fig:correl_of_three_comps} in Appendix~\ref{appendix:visual} provides an illustration of the three correlation matrices computed on historical data.

\begin{table}[ht]
\centering
\begin{tabular}{lr}
\toprule
 & Cross-asset correlations \\
\midrule
Original & 0.39 [0.18, 0.63] \\
Factor-based & 0.69 [0.34, 0.94] \\
Residual & 0.00 [-0.07,  0.06] \\
\bottomrule
\end{tabular}
\captionof{table}{Median cross-asset correlations computed for different components with the interval covering 95\% of the cross-asset correlations.}
\label{table:correlations}
\end{table}

\paragraph{Preserving heavy-tailedness (skewness and kurtosis) of returns.}
In addition to inter-temporal and cross-asset dependencies, the marginal distributions of asset returns also exhibit unique properties. Among the most notable are asymmetry in gains and losses and the heavy-tailedness of the asset return distribution. The former is often characterized by the negative skewness of asset returns, indicating a tendency for more extreme downward movements than upward movements, with downward moves being less frequent. While kurtosis can provide insight into heavy-tailedness, a more nuanced analysis on losses involves estimating the tail-index $\gamma$ of the marginal distribution of asset returns using alternative estimators, such as Hill's tail-index estimator: 
\begin{equation}
\label{hills}
  \hat{\xi}(k) = \frac{1}{k} \sum_{i=1}^k \log{(-\mathbf{r}_{(i)})} - \log{(-\mathbf{r}_{(k)})}  
\end{equation}

where $\mathbf{r}_{(k)}$ denotes the $k^{th}$ order statistic of a given sample of asset returns $\mathbf{r}$. Loosely speaking, the parameter $k$, specified based on the sample size, sets the threshold beyond which it is believed the tail begins.\footnote{This is why the term inside the logarithm in Equation~(\ref{hills}) is often positive, preventing any issues from arising.} However, the estimated value of $\hat{\xi}$ can be highly dependent on the choice of $k$. Therefore, it is common practice to compute it for different values of $k$. Intuitively, this estimator evaluates how losses beyond a certain quantile deviate from the selected level on average, offering an estimation of the \textit{heaviness} of the left-tail where the tail-index is estimated by $\hat{\gamma} = 1/\hat{\xi}$.

\begin{figure}[ht]
    \hspace{-0cm}
    \includegraphics[width=1\linewidth]{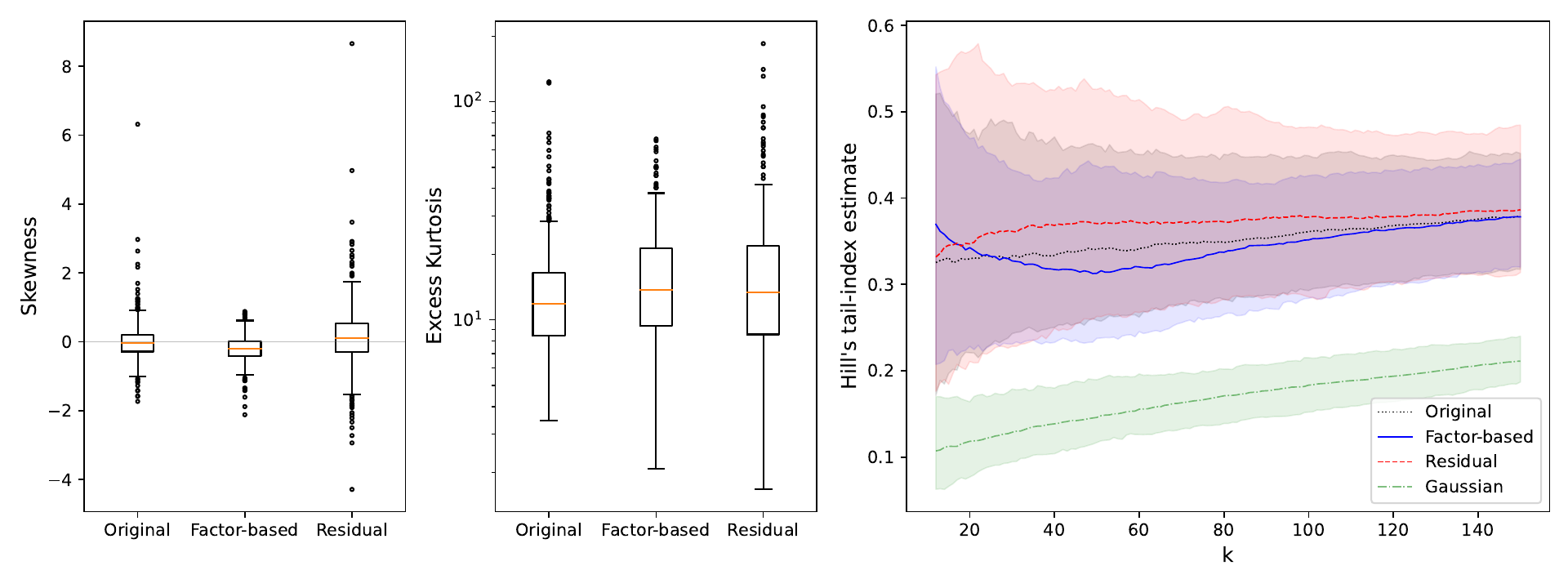}
    \caption{Distribution of the empirical skewness (left) and kurtosis (middle) of the 433 stocks for the different components. Median Hill's index estimates of different components for different values of $k$ with the inverval covering 95\% of the values across the universe (right). Estimates from a normal distribution are added for comparison.}
    \label{fig:tail_figs}
\end{figure}

\medskip
Figure \ref{fig:tail_figs} provides some interesting insights into the marginal distributions of the different components of returns. First, factor-based returns appear to exhibit a consistent negative skewness for most assets, while skewness estimates for residual and original returns tend to vary around zero, although they can reach significantly high positive or negative levels. This suggests that the widely observed negative asymmetry in asset returns may be attributed more to common components that drive returns rather than being solely a feature of individual assets. However, the same argument may not hold for extreme returns and losses, as both components of asset returns exhibit particularly high kurtosis for almost all assets. A similar analysis focusing on the left tail can be carried out by estimating the tail index. Essentially, the well-known phenomenon of heavy tails in asset returns cannot be attributed solely to the presence of a single component, since the factor-based returns and residuals for the majority of stocks simultaneously exhibit this phenomenon in their distributions, as supported by Table~\ref{table:dist_stats}.

\begin{table}[ht]
\centering
\begin{tabular}{llll}
\toprule
 & Skewness & Kurtosis & Tail Index \\
\midrule
Original & -0.03  [-1.01  1.18] & 11.8 [4.4 42.9] & 2.84 [2.19 3.62] \\
Factor-based & -0.21 [-1.10   0.55] & 13.8 [3.8 46.1] & 2.91 [2.30 3.75] \\
Residual &  0.10 [-1.82  2.21] & 13.3 [4.0  64.2] & 2.68 [1.99 3.67] \\
\bottomrule
\end{tabular}
\captionof{table}{Median values for different statistics at the level of different components across the universe with 95\% confidence intervals in brackets.}
\label{table:dist_stats}
\end{table}

\paragraph{Empirical conclusion: 16 factors in 3 clusters with a memory of 63 days, i.i.d. residuals.}
In conclusion, the observations above serve as motivation for our approach to modeling asset returns from the perspective of two distinct components with differing characteristics. We have observed that most of the stylized facts of asset returns are present in the factor-based returns, sometimes more prominently, suggesting that they may warrant modeling using complex tools capable of capturing properties at both the temporal and cross-sectional levels. For the former, we use a memory of length of $s=63$ days, and for the latter, the cross-sectional properties are incorporated exogenously via the projection matrix $\hat{\bm{\beta}} = \mathbf{P}_{1:16}$. 

\medskip

Conversely, residuals can be treated as independent and identically distributed (i.i.d.) given the relative absence of linear and non-linear dependence over time, and the insignificant correlations. However, the model should still better account for their skewed and heavy-tailed nature, making a mixture of Student-t distributions a reasonable choice. For the sake of simplicity, we omit the asymmetry of residual returns and use a Student-t distribution (fixing $p=1$ in Equation~(\ref{uni_t_density})) and estimate the parameters for residuals of each asset using the maximum likelihood approach.

\medskip

Recall from Section~\ref{sec:factor_clustering} that the selected factors are to be modeled by a smaller number of GANs trained on clusters of factors. In our numerical case, 16 factors are grouped into $n_c=3$ clusters based on 5 \textit{features}. Specifically, these features include sample skewness and kurtosis computed on standardized daily factor returns, the eigenvalue associated with each factor, the volatility clustering score, and the leverage effect score computed over 63 days according to the following formula, with appropriate choices of $(g_1, g_2)$ being $(x^2, x^2)$ and $(x, x^2)$, respectively:\footnote{The sign of the final score is adjusted based on the slope of the curve, if necessary, to prevent bias caused by the symmetric nature of the score function around the x-axis.}

\begin{equation}
\label{eq:autocorrel_scores}
    \sum_{\tau=1}^{63} \rho_{F, (g_1, g_2)}(\tau)^2.
\end{equation}

The agglomerative clustering algorithm applied to the above features results in the sets $\mathcal{C}_1 = \{1\}$, $\mathcal{C}_2 = \{3,6,12,14, 15\}$ and $\mathcal{C}_3 = \{2,4,5,7,8,9,10,11,13,16\}$. Therefore, 3 GANs with the same architecture and initial hyperparameters are trained using the training sets constructed based on the given clusterings. Details about the neural network architecture and training design are provided in Appendix~\ref{appendix:TCN}, along with figures related to the evaluation of data generated by each of the three generators in Appendix~\ref{appendix:visual}.

\medskip
The training pipeline depicted in this subsection allows us to obtain the necessary parameters for the market generator as described by Equation~(\ref{eq:market_generator}). In the next section, we will assess the quality of the simulated data to determine whether it can effectively reproduce the learned properties present in asset returns.

\subsection{A first look at simulated data}
\label{sec:synthetic_data}

Although generative models have attracted attention and been used in a wide range of fields, there is no consensus on how to evaluate and validate what is produced by the trained model (see \cite{borji_pros_2018} for a review of evaluation measures). It is natural to expect different evaluation measures in different domains, but the lack of a common set of \textit{intra-domain} measures makes it difficult to compare and horse-race models that have proved instrumental in the theoretical progress of deep learning over the last decade, as we have seen with ImageNet \cite{deng2009imagenet}. The considerable efforts being made to benchmark and evaluate LLMs also underline the importance of the issue \cite{hendrycks2020measuring}.

\medskip
We start by an investigation of the characteristics of the obtained time series at the level of each simulated stock: we conduct an assessment of the marginals of the multivariate (433-dimensional) generated distribution.
We provide not only a visual comparison, but tables with confidence intervals too, that should be the standard for evaluating synthetic data generation.



\subsubsection{A careful look at the marginals of the generated multivariate time series}

If we are in a context of generating multivariate asset returns, a generative model should ideally be capable of capturing the joint dynamics of the universe both cross-sectionally and inter-temporally. It is often more appropriate to carry out such an assessment with multiple evaluation steps, given the difficulty of doing a one-shot evaluation of the model as a whole.\footnote{An asset-by-asset evaluation can also be performed, as demonstrated for a specific stock in Figure~\ref{fig:jpm_eval} in Appendix~\ref{appendix:visual}, however this approach is not scalable and does not provide insight about the correlation structure.} 

\medskip
First, we generate 100 simulated samples of the length of the training set, $n=3020$, from the market generator following the ideas developed in Section~\ref{sec:sample_size} regarding the balance between initial sample size and generated data.\footnote{It corresponds to generating 100 different $\Tilde{\mathbf{X}} \in \mathbb{R}^{3020 \times 433}$.} We begin by understanding whether the marginal distributions of asset returns in the generated scenarios are close to those observed historically, both in-sample (training set) and out-of-sample (test set). We use 1-Wasserstein distance as a measure of distance between simulated and historical distributions. For the empirical distributions, it boils down to a simple function of order statistics of the historical and simulated samples.

\begin{figure}[ht]%
    \centering
    \subfloat{{\includegraphics[width=6.4cm]{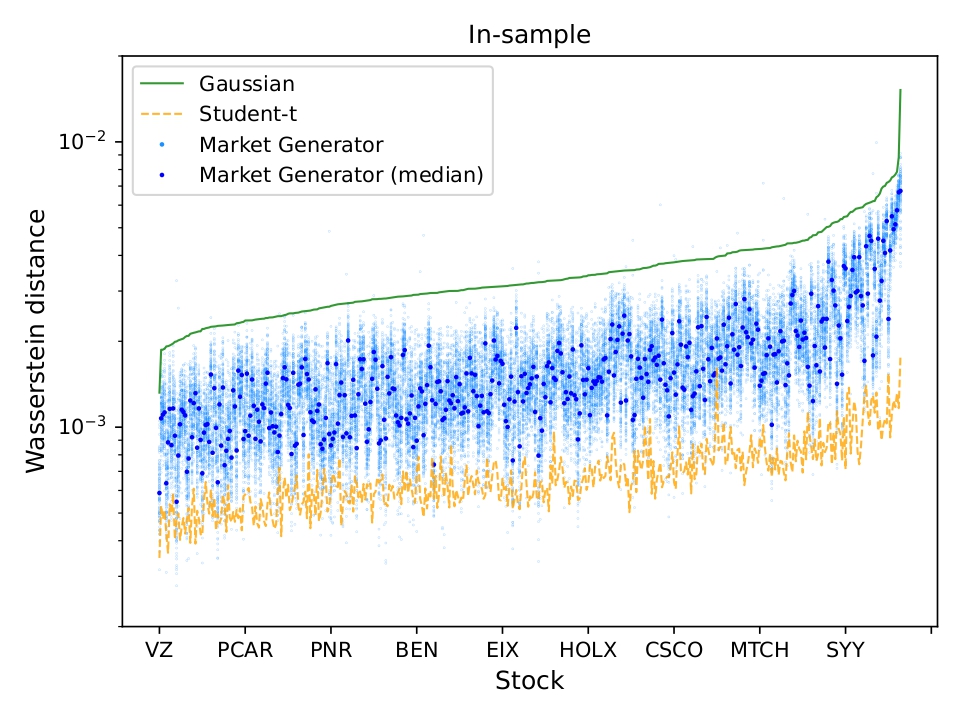} }}%
    \qquad
    \subfloat{{\includegraphics[width=6.4cm]{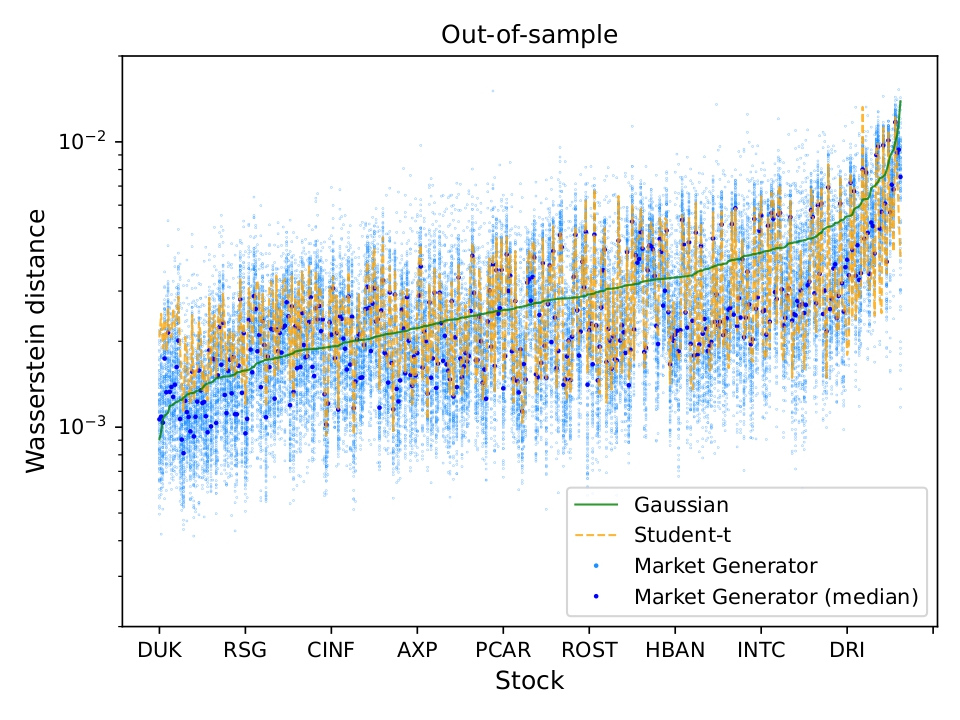} }}%
    \caption{Wasserstein distances between simulated and historical samples based on different models. They are arranged in ascending order on the basis of distance from Gaussian fits for a clearer visualization.}
    \label{fig:wasserstein_dists}
\end{figure}

\medskip

Figure~\ref{fig:wasserstein_dists} illustrates the distances between the distribution of returns for each asset, in all scenarios, and the in-sample / out-of-sample distributions.\footnote{For a fair comparison with out-of-sample scenarios, we generate another 100 simulated samples of the length of the test set, $n=605$.} For ease of comparison, the distances obtained by Gaussian and Student-t fits (obtained marginally for each asset using the training set) are also included. We see that the market generator provides better out-of-sample results on average, while Student-t distributions fit well in-sample but underperform out-of-sample due to likely overfitting issues, as shown in Table~\ref{table:wass_table}. The results look reasonably satisfactory in terms of marginal distributions for a modeling framework in which we do not learn directly from marginal distributions.

\begin{table}[ht]
\centering
\begin{tabular}{lll}
\toprule
 & In-sample & Out-of-sample  \\
\midrule
Gaussian & 3.19 [2.02, 6.84]& 2.69 [1.23, 7.59]\\
Student-t & 0.65 [0.43, 1.24]& 2.40 [1.21, 7.21]\\
Market Generator &  1.45 [0.79, 4.35]&  2.26 [1.07, 7.10]\\
\bottomrule
\end{tabular}
\captionof{table}{Median Wasserstein distance ($\times 10^{4}$) across the universe with the interval covering  95\% of the scores.}
\label{table:wass_table}
\end{table}

\medskip

\begin{figure}[ht]%
    \centering
    \subfloat{{\includegraphics[width=6.4cm]{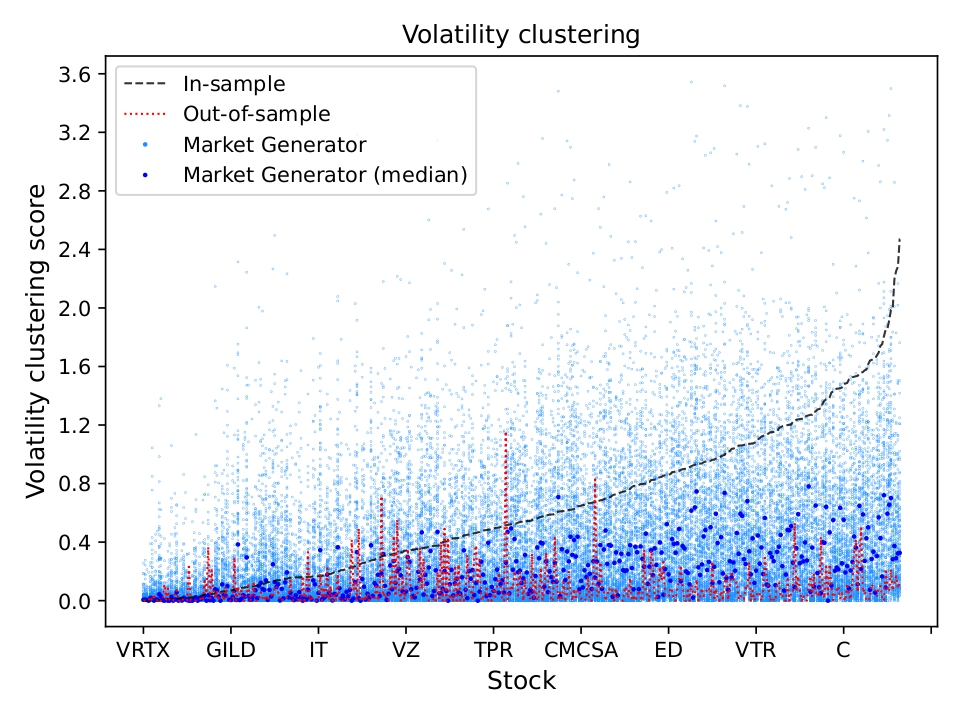} }}%
    \qquad
    \subfloat{{\includegraphics[width=6.4cm]{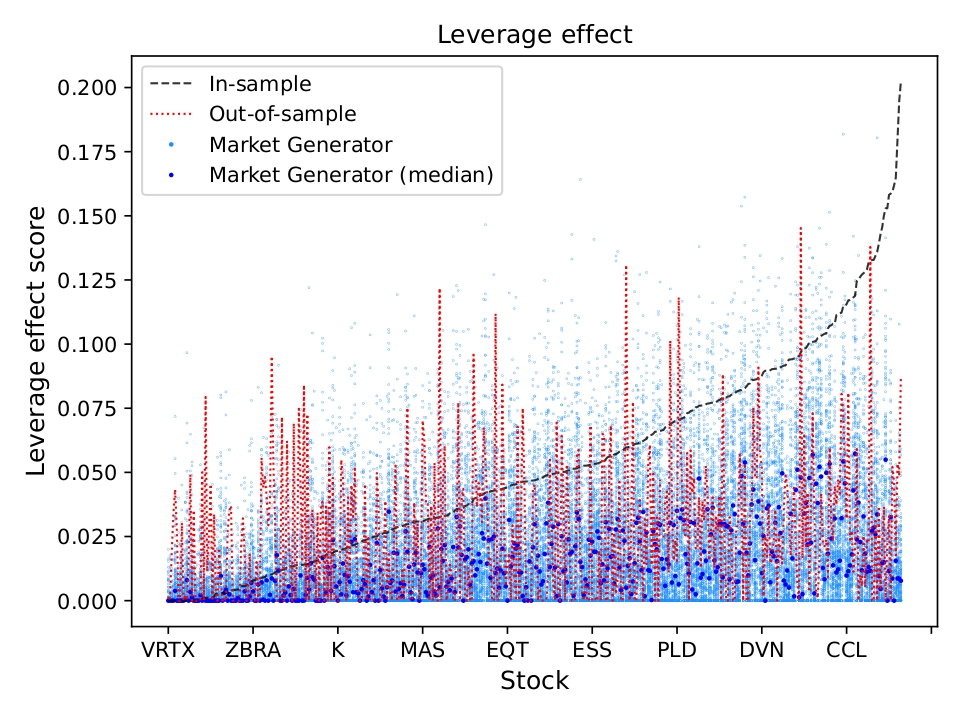} }}%
    \caption{Volatility clustering and leverage effect scores computed on historical and simulated samples for each asset. They are arranged in ascending order on the basis of scores computed in-sample for a clearer visualization.}
    \label{fig:time_dependency_eval}
\end{figure}

To understand the model's ability to reproduce inter-temporal effects at the asset level, we can consult the volatility clustering and leverage effect scores as defined in Equation~(\ref{eq:autocorrel_scores}). Figure~\ref{fig:time_dependency_eval} illustrates the scores computed on historical and simulated samples for each asset. It seems that the upward trend in in-sample scores is captured by the market generator up to a certain level, as shown by the increasing dispersion of the calculated scores and the upward trend in median scores for the simulated samples. Table~\ref{table:temporal_table} also shows that, in the market scenarios produced by the market generator, asset returns exhibit a time-dependency structure similar to that which we observe historically.

\begin{table}[ht]
\centering
\begin{tabular}{lll}
\toprule
 & Volatility Clustering & Leverage Effect  \\
\midrule
In-sample & 53.88 [0.69, 174.34]& 4.69 [0.0, 14.65]\\
Out-of-sample &3.60 [0.0, 42.94]&2.08 [0.0, 8.77]\\
Market Generator & 14.66 [0.0, 63.77]	& 1.16 [0.0, 4.84]\\
\bottomrule
\end{tabular}
\captionof{table}{Median scores ($\times 10^{2}$) across the universe with the interval covering  95\% of the scores.}
\label{table:temporal_table}
\end{table}

Another aspect that deserves particular attention is the tail of asset return distributions, which is one of the most significant properties of asset returns, particularly when it comes to risk. Consequently, a market generator unable to generate extreme losses misses out one of the most crucial phenomena of asset returns.

\begin{figure}[ht]%
    \centering
    \subfloat{{\includegraphics[width=6.4cm]{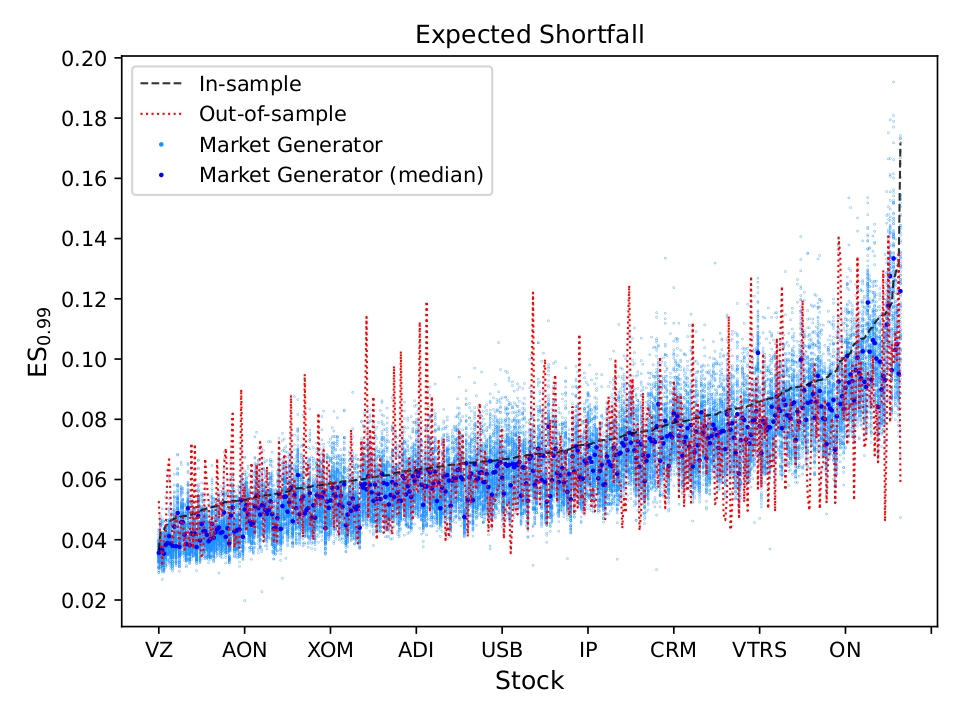} }}%
    \qquad
    \subfloat{{\includegraphics[width=6.4cm]{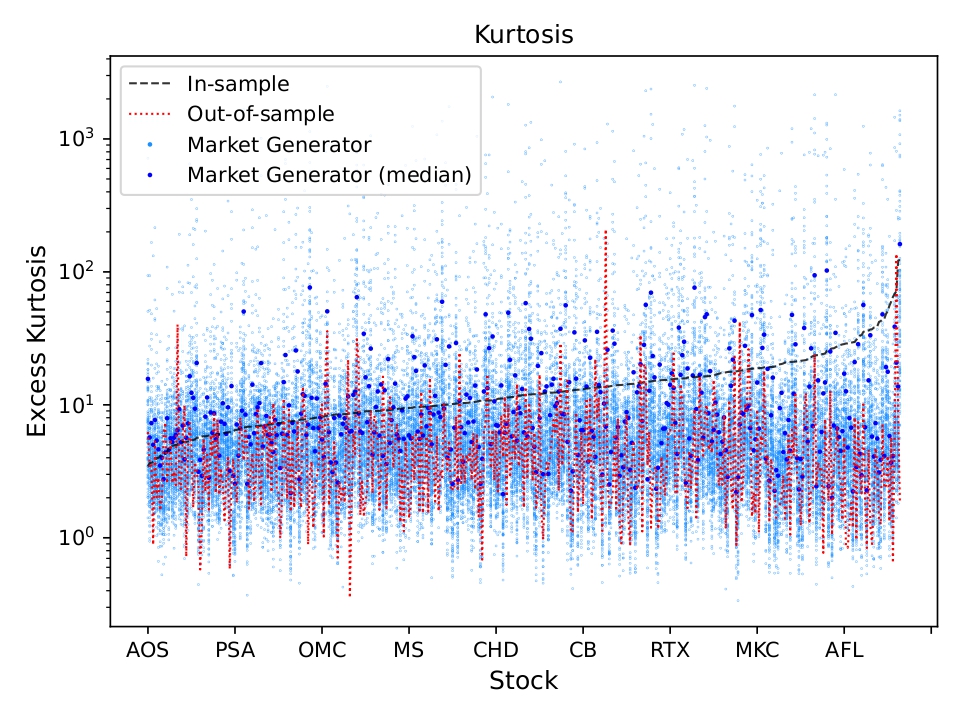} }}%
    \caption{Expected Shortfall at the level 99\% and kurtosis computed on historical and simulated samples for each asset. They are arranged in ascending order on the basis of values computed in-sample for a clearer visualization.}
    \label{fig:tails_eval}
\end{figure}

\medskip
A standard way to conduct an analysis of the left-tail of return distributions is to use Value-at-Risk (VaR) and Expected Shortfall. VaR at the level $\alpha \in (0,1)$ is simply a quantile of loss distribution of a real-valued random variable $X$ (regarded as a loss) and Expected Shortfall, on the other hand, is the average loss given that the loss exceeds $\text{VaR}_\alpha(X)$.\footnote{Formally, $\text{VaR}_\alpha(X) := \inf \{ x \in \mathbb R  \, | \, \mathbb P(X \le x) \ge \alpha \}$ and $\text{ES}_\alpha(X) := \frac 1{1-\alpha} \int_{\alpha}^1 \,\text{VaR}_s(X) \,ds.$} Consequently, the similarity of VaR and Expected Shortfall (at high levels of $\alpha$) computed on historical and simulated samples, along with high kurtosis, should give an indication of the \textit{heavy-tailedness} of the simulated marginal distributions, as summarized in Table~\ref{table:tail_stats} and illustrated in Figure~\ref{fig:tails_eval}.  

\medskip

\begin{table}[ht]
\centering
\begin{tabular}{llll}
\toprule
 & In-sample & Out-of-sample & Market Generator \\
\midrule
$\text{VaR}_{95\%}$ &2.5 [1.7, 4.1] &2.7 [1.8, 4.8]	&2.6 [1.8, 4.5] \\
$\text{VaR}_{99\%}$ &4.6 [3.1, 7.5] &4.6 [3.0, 8.0]	 &4.5 [2.9, 7.5] \\
$\text{ES}_{95\%}$ & 4.0 [2.7, 6.6] &4.0 [2.7, 7.0] &3.9 [2.6, 6.6] \\
$\text{ES}_{99\%}$ & 6.8 [4.7, 11.4] &6.1 [3.9, 11.9] &6.2 [3.8, 10.2] \\
Kurtosis & 11.8 [4.4, 42.9] &3.4 [0.9, 24.2] &4.7 [1.9, 14.2] \\
\bottomrule
\end{tabular}
\captionof{table}{Tail statistics across the universe with the interval covering  95\% of the scores.}
\label{table:tail_stats}
\end{table}

Until now, we have tried to verify whether we could get close to the marginal properties of individual asset returns by omitting the relationships between them, which are one of the most critical parts of multivariate modeling. To understand if at least linear relationships exist within the universe in a similar way to those observed historically, we analyze the distance between the sample correlation matrix calculated on simulated and historical returns:

$$ \sum_{i=2}^{d-1} \sum_{j=1}^{i-1} (\mathbf{C_{sim}}_{i,j} - \mathbf{C_{hist}}_{i,j})^2$$

where $\mathbf{C_{sim}}$ denotes a sample correlation matrix calculated on a simulated sample and $\mathbf{C_{hist}}$ denotes a sample correlation matrix calculated on a historical sample (in-sample or out-of-sample). We also include two other estimators in our analysis to make the distance values meaningful. The first is what we call the one-factor model, which is simply a correlation matrix in which all correlations between assets are equal to the average correlations calculated in-sample. Although overly structured, such an approach is not completely unrealistic. The second correlation matrix estimator we consider is obtained using the covariance matrix estimate resulting from the Ledoit-Wolf estimator:

$$\Tilde{\bm{\Sigma}} = (1-\gamma) \hat{\bm{\Sigma}} + \gamma \frac{\text{tr}(\hat{\bm{\Sigma}})}{d}\mathbf{I}_d$$

where $\hat{\bm{\Sigma}}$ is the sample covariance matrix and $\text{tr}(\hat{\bm{\Sigma}})$ is the trace of the covariance matrix. The optimal $\gamma$ can then be found such that the distance to the true covariance matrix is minimized (see~\cite{ledoit2003honey}).

\begin{table}[ht]
\centering
\begin{tabular}{lll}
\toprule
 & In-sample & Out-of-sample  \\
\midrule
One-Factor & 13.27 & 20.98 \\
Ledoit-Wolf & 0.04 & 13.23 \\
Market Generator & 2.27	[0.79, 11.71] & 16.39 [11.60, 112.10]\\
\bottomrule
\end{tabular}
\captionof{table}{Distance ($\times 10^{3}$) to historical correlation matrix for different estimators. Median value with 95\% confidence interval for 100 realizations is shown for the market generator.}
\label{table:corr_matrix_tables}
\end{table}

\medskip 

Table~\ref{table:corr_matrix_tables} shows the similarity of the correlation matrices to the historical ones obtained using different estimators. The Ledoit-Wolf estimator naturally produces the closest values to the in-sample correlations, since this information is already directly used. On the other hand, the market generator delivers competitive results to the Ledoit-Wolf estimator, particularly from an out-of-sample point of view. These findings underline the market generator's ability to capture the correlation structure between asset returns. In Appendix~\ref{appendix:visual}, we provide an illustration of correlation matrices computed in-sample and on synthetic data (see Figure~\ref{fig:corr_syn_vs_hist}).

\medskip 

\begin{figure}[ht]
    \centering
    \includegraphics[width=6.4cm]{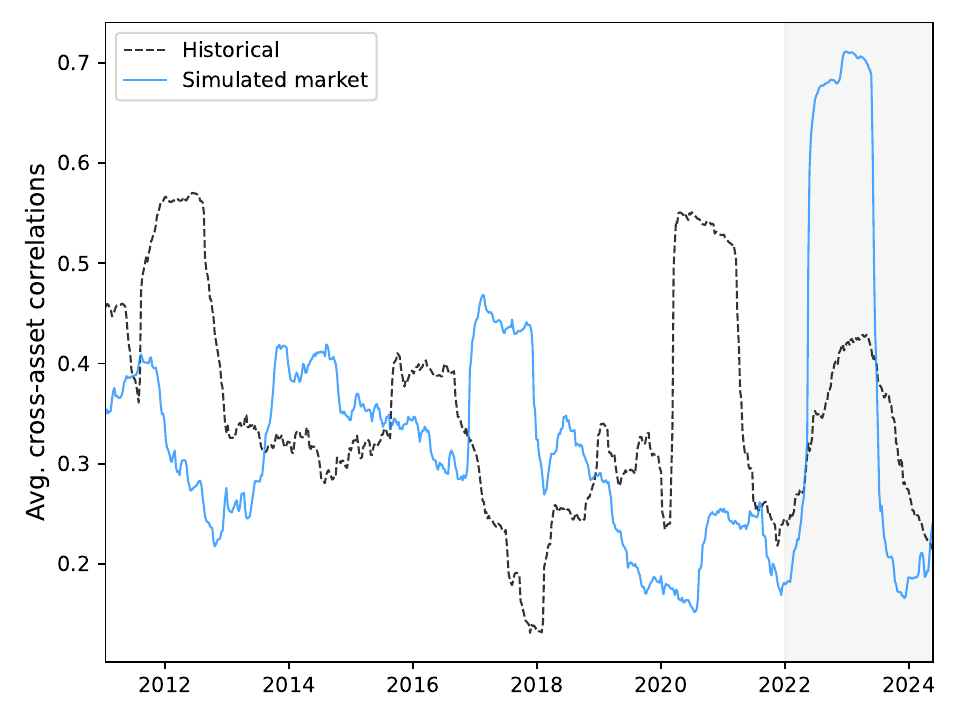}
    \caption{Evolution of average correlations (using a rolling window of one-year) between assets over time, calculated historically and on a generated sample of the same size. Shaded area represent the out-of-sample period.}
    \label{fig:dynamic_correlations}
\end{figure}

In addition to the existence of a co-movement structure between asset returns, another important fact is that this structure is not static, but evolves over time. This means that correlations between assets are dynamic, and that the market has periods when correlations rise and fall, sometimes very suddenly. Increasing correlations between assets indicate a concentration of risk, reducing the scope for diversification within the universe. Such concentration is often monitored by examining the level of average correlations, with a high average indicating concentrated risk. Consequently, a market generator should also be tested for its ability to produce dynamic correlations as illustrated in Figure~\ref{fig:dynamic_correlations} using a simulated sample versus historical levels. This behavior arises from capturing volatility dynamics at the factor level, which induces dynamic correlations at the asset level despite the projection matrix being static.

\subsubsection{Evaluation based on linear combinations}

Given the challenge of evaluating each individual asset and the relationship between them, which we have attempted to achieve above, a natural idea would be to perform such an evaluation for a portfolio of these assets, in order to reduce the problem from high to one-dimensional. To this end, we consider an equally-weighted portfolio of the assets in question and calculate the returns of this portfolio historically and using our 100 simulated samples.\footnote{We rebalance the portfolio every day and assume no transaction cost since it is irrelevant for the purpose of evaluation.} We are now confronted with the problem of comparing two univariate time series (simulated and historical) in a financially meaningful manner.   

\begin{figure}[ht]%
    \centering
    \subfloat{{\includegraphics[width=6.4cm]{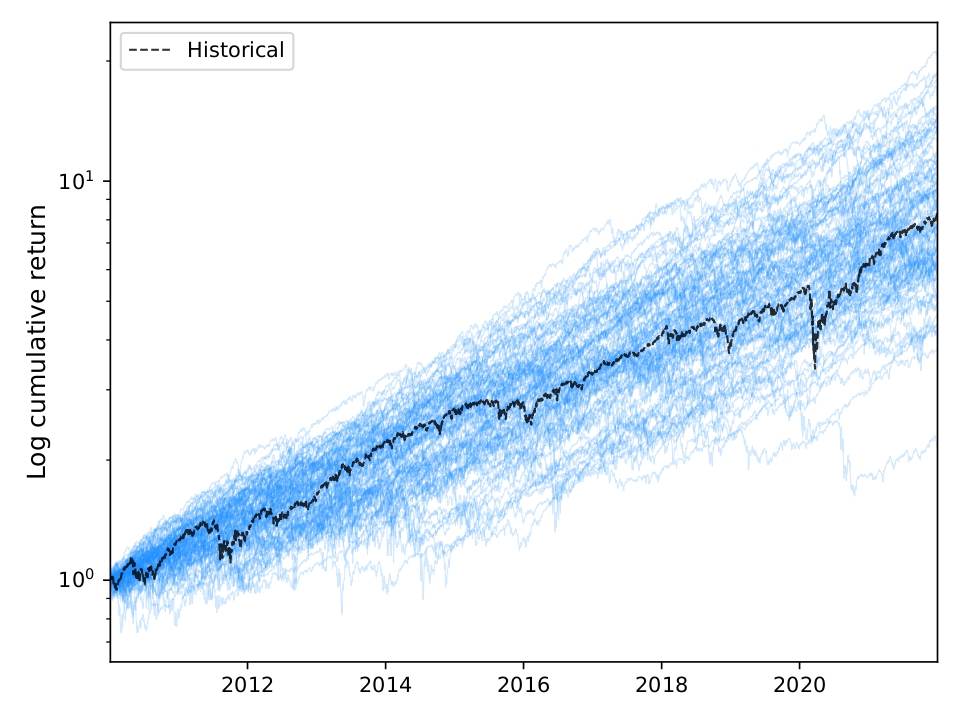} }}%
    \qquad
    \subfloat{{\includegraphics[width=6.4cm]{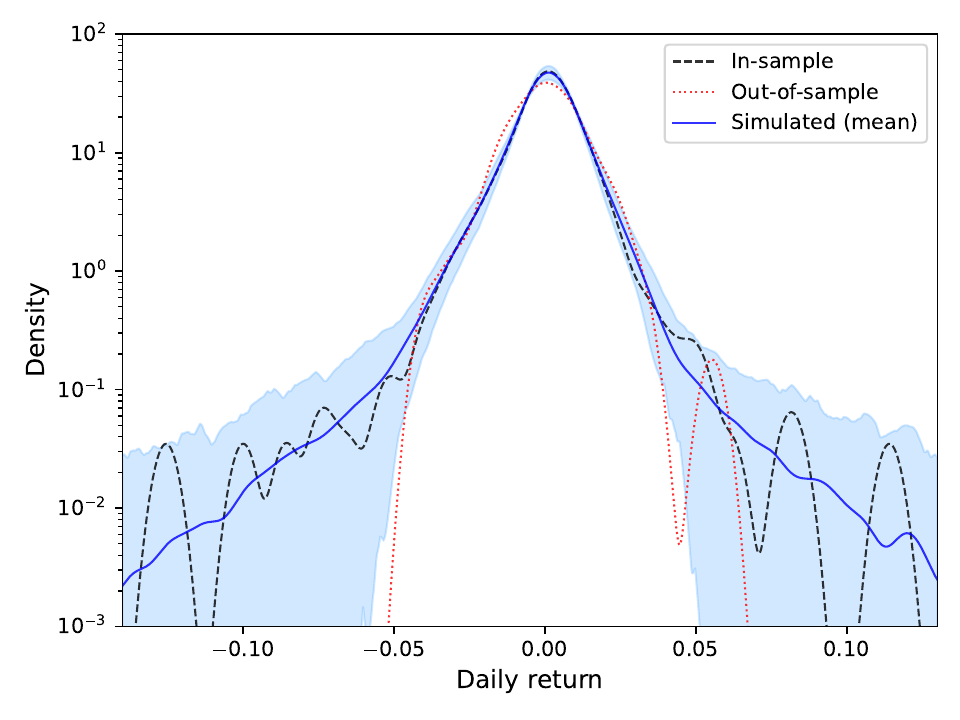} }}%
    \caption{Returns of the equal-weighted portfolio computed on different samples.}
    \label{fig:syn_spx_returns}
\end{figure}

\medskip

Figure~\ref{fig:syn_spx_returns} illustrates the paths of the equal-weighted portfolio based on historical and simulated returns. The simulated trajectories look plausible and diffuse around the historical trajectory. The distribution of returns also appears to be on target, capturing quite well the heavy-tailed nature of portfolio returns. A more systematic approach would be to compute certain statistics of interest to gain a deeper understanding of the results obtained for the simulated scenarios versus the in-sample and out-of-sample scenarios. 

\begin{table}[h]
\centering
\begin{tabular}{llll}
\toprule
 & Market Generator & In-sample & Out-of-sample \\
\midrule
Ann. return (\%) & 19.86 [14.73, 26.19] & 19.90 & 5.98 \\
Ann. volatility (\%) & 18.53 [15.19, 22.52] & 18.50 & 17.97 \\
Sharpe ratio & 1.09 [ 0.70 ,  1.61]& 1.08 & 0.33 \\
Skewness & -0.45 [-0.95,  0.23] & -0.57 & 0.02 \\
Kurtosis & 11.40 [ 2.33, 30.78] & 15.89 & 1.59 \\
Maximum drawdown (\%) & 29.48 [15.45, 44.09] & 38.00 & 19.45 \\
$\text{VaR}_{95\%}$ (daily, \%) & 1.71 [1.39,  2.06] & 1.66 & 1.67 \\
$\text{VaR}_{95\%}$ (weekly, \%) & 3.75 [2.96,  4.50]& 3.50 & 4.10 \\
$\text{VaR}_{95\%}$ (monthly, \%) & 5.87 [3.75,  8.34] & 5.52 & 8.09 \\
$\text{VaR}_{99\%}$ (daily, \%) & 3.32 [2.63,  4.13]& 3.19 & 3.06 \\
$\text{VaR}_{99\%}$ (weekly, \%) & 7.10 [5.28,  8.68] & 6.67 & 6.64 \\
$\text{VaR}_{99\%}$ (monthly, \%) & 12.90 [7.72, 20.15] & 13.09 & 10.39 \\
$\text{ES}_{95\%}$ (daily, \%) & 2.80 [2.23,  3.47] & 2.77 & 2.42 \\
$\text{ES}_{95\%}$ (weekly, \%) & 5.90 [4.43,  7.20] & 5.77 & 5.78 \\
$\text{ES}_{95\%}$ (monthly, \%) & 10.16 [6.13, 14.73] & 10.54 & 9.54 \\
$\text{ES}_{99\%}$ (daily, \%) & 4.75 [3.43,  6.54] & 4.80 & 3.47 \\
$\text{ES}_{99\%}$ (weekly, \%) & 9.50 [6.83, 12.71] & 10.25 & 8.72 \\
$\text{ES}_{99\%}$ (monthly, \%) & 17.32 [9.81, 27.50]& 21.96 & 11.66 \\
Volatility clustering score & 1.26 [0.34,  2.42] & 1.69 & 0.43 \\
Leverage effect score & 0.11 [0.06,  0.19] & 0.15 & 0.11 \\
\bottomrule
\end{tabular}
\captionof{table}{Sample statistics of the equal-weighted portfolio computed on different samples with 95\% confidence interval in brackets.}
\label{table:spx_results}
\end{table}

\medskip

\begin{figure}[H]%
    \centering
    \subfloat{{\includegraphics[width=6.4cm]{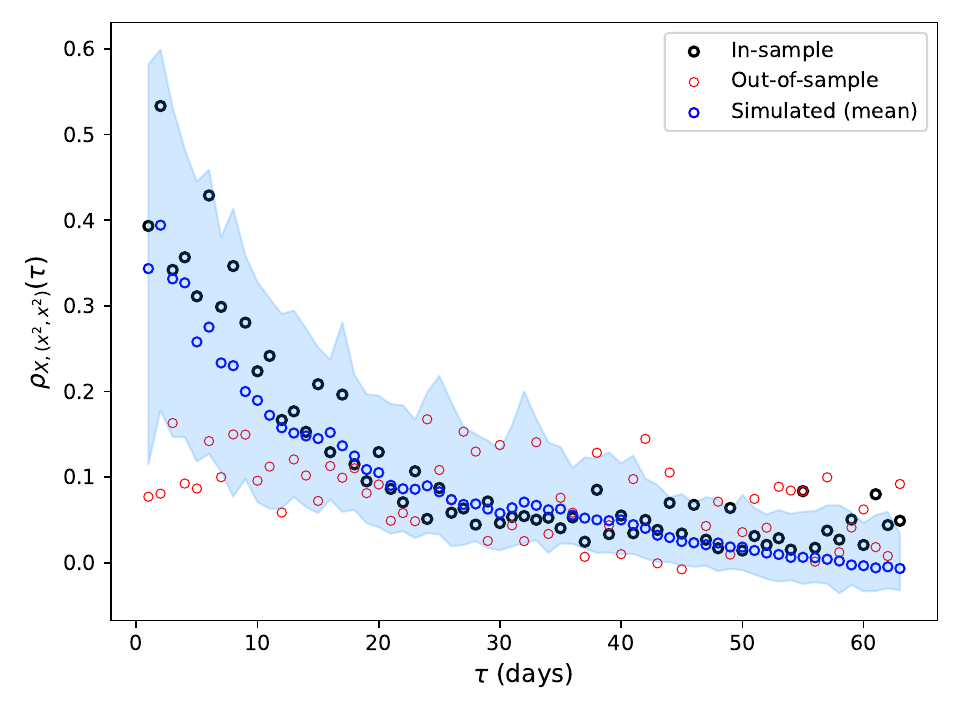} }}%
    \qquad
    \subfloat{{\includegraphics[width=6.4cm]{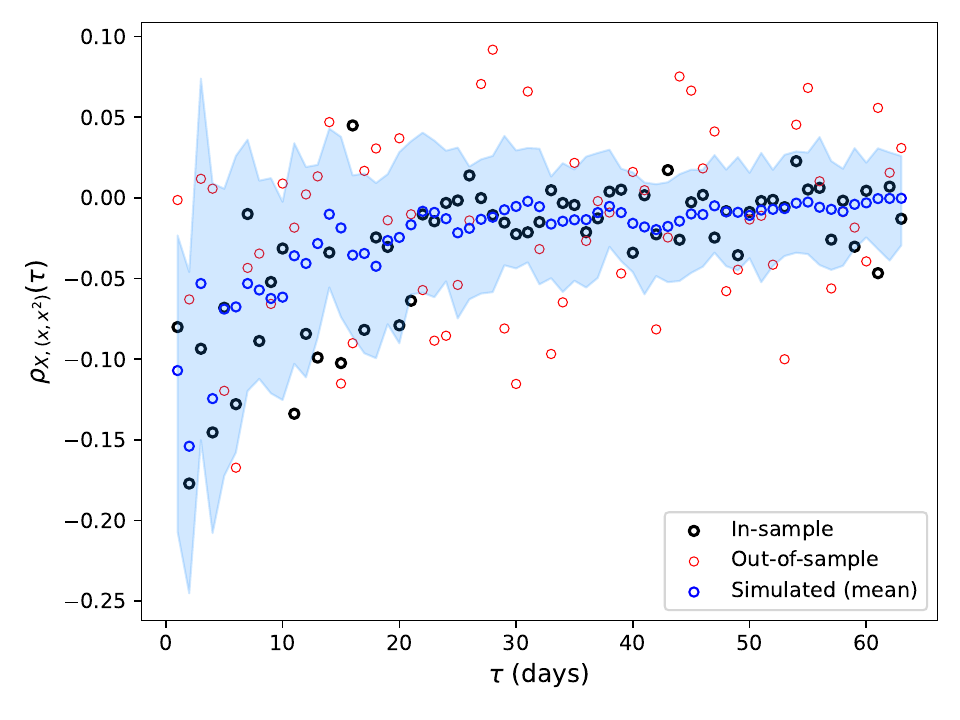} }}%
    \caption{Inter-temporal properties of the equal-weighted portfolio computed on different samples.}
    \label{fig:syn_spx_time_dependence}
\end{figure}

Table~\ref{table:spx_results} lists some statistics that can summarize the behavior of portfolio returns and allow for an evaluation of the quality of the simulated samples. The market generator gives rather punctual (from an in-sample point of view) Sharpe ratio estimates, but this is natural since the market generator uses sample means and sample variances of individual assets estimated in-sample.\footnote{The risk-free rate is assumed to be 0 throughout the paper.} More interesting results concern higher moments and tails. The negative skewness profile of the portfolio is well captured, although this phenomenon is absent out-of-sample, probably due to the fact that the test set is very limited, which also explains why in-sample and out-of-sample results differ substantially. Similar conclusions apply to kurtosis, although the mean value is lower than the in-sample value, but a wide confidence interval indicates the ability to generate various scenarios. Maximum drawdown being a path-dependent measure is at realistic levels on average where the confidence interval indicates the existence of scenarios where the strategy's maximum drawdown can be (at least) as small as 15.45\% and as large as 44.09\% covering both in-sample and out-of-sample values. The VaR and Expected Shortfall calculated for different time horizons prove that the model not only works on a daily scale, but is also suitable for longer time horizons. Furthermore, the model's performance does not decrease much for VaR and ES at high levels, demonstrating its ability to generate extreme losses. The volatility clustering and leverage effect scores indicate that these phenomena exist at the level of portfolio returns historically, which are also observed in the simulated scenarios. As these scores are not very revealing, we illustrate in Figure~\ref{fig:syn_spx_time_dependence} how these properties are reproduced in the simulated samples. 

\medskip

In this section, we have carried out a general analysis to understand the behavior of our market generator and show that it can produce realistic scenarios according to some generic measures of evaluation. However, does it mean that it can now be used for the desired application safely? The next section is devoted to this question.

\section{In-depth evaluation of generative models}
\label{sec:evaluating_generative_models}
The previous section checked whether the generative model produces known stylized facts.
Nevertheless, the goal of a generator is not only to reproduce such facts, it targets a specific usage, a typical goal is to produce more data to tune hyperparameters of a downstream algorithm.
Examples of downstream tasks are reinforcement learning algorithms or any portfolio construction algorithm.
Thanks to Section~\ref{sec:sample_size}, we know that it is useless to generate too many synthetic data points, but keeping it in a reasonable range, we can generate data points hoping it would help the downstream algorithm to generalize.
In this spirit, underline it is known for long that adding noise to data provides some robustness to downstream optimizers (see \cite{bishop1995training}), 
hence even if the virtue of a generative model is no more than to add noise a sound way to the information existing in the initial sample point, it would not be useless.


\medskip
However, as mentioned, mimicking financial returns of a list of tradable instruments is exposed to a paradox. On the one hand the more the learning process of a generative model focuses on the distribution of the marginals (i.e. the returns of each instrument) and on the joint distribution of the returns (i.e. their covariance), the more it is attracted to the components of large variance of price moves. On the other hand, as soon as the generated trajectories are used to study portfolios, the exposure to these large-variance-carrying components is generally neutralised (or \textit{projected out} by the process of portfolio construction). This is particularly true for long-short market neutral portfolios. As a consequence the focus of the generative model on the components with large variance is counterproductive. Our architecture aims to overcome this issue, and the only way to determine if it succeeds is by testing it within the ultimate use case: portfolio construction,
%
and particularly in the case of backtesting long-short strategies.

\subsection{The market generator and the Sharpe ratio of mean-reversion strategies: a comparison with block bootstrap}
\label{sec:lev_mg_for_mr}

We propose to evaluate the quality of synthetic data on long-short portfolios invested in the cross-section of returns during the last $h$ days of the considered universe of stocks. The direction of the portfolio is made as if it is betting on the \emph{mean-reversion} of the returns. In addition to the long-short case, we also consider the long-only version of the strategy for benchmarking purposes.\footnote{We deliberately limit ourselves to strategies without optimization, such as mean-reversion. Other portfolio construction methods would require optimization that is very resource-intensive, particularly for the large universe considered here and in the context of dynamic strategies involving frequent rebalancing during backtesting.}


\medskip

\begin{figure}[ht!]
    \hspace{-0cm}
    \includegraphics[width=1\linewidth]{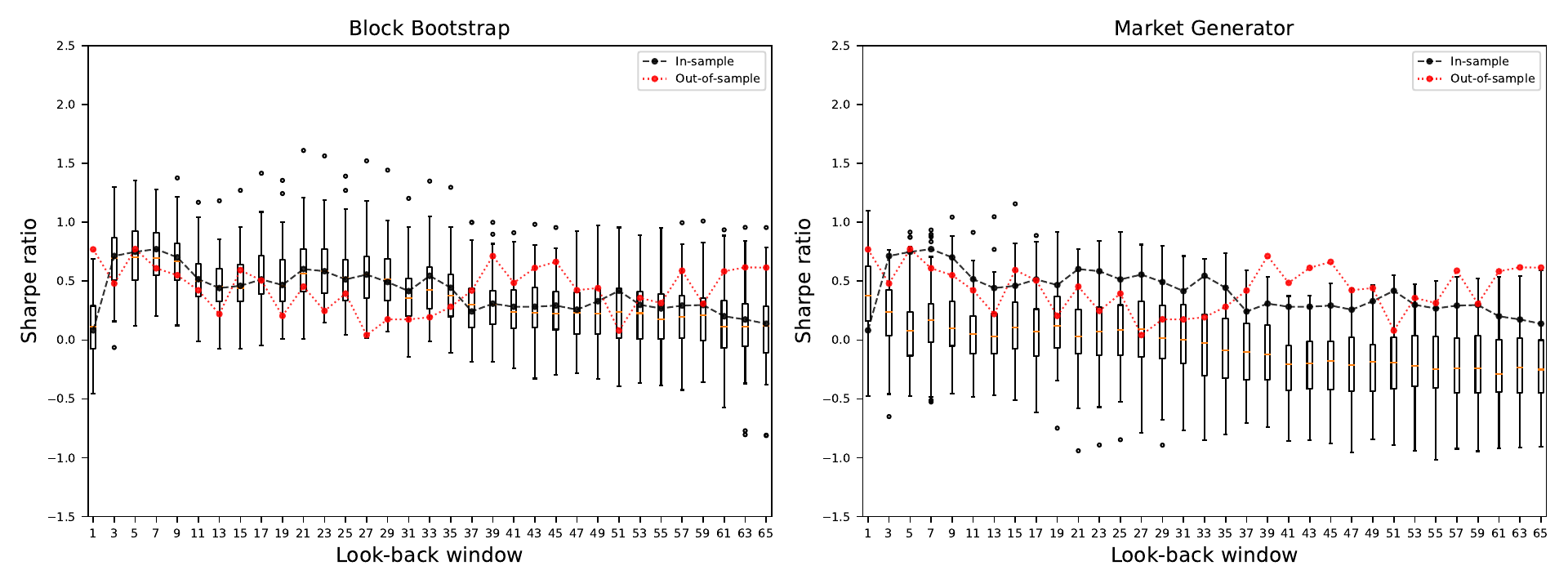}
    \caption{Boxplots of Sharpe ratios of the long-short mean-reversion strategy for different look-back window parameters $h$ backtested on samples from block bootstrap approach (left) and the market generator (right). Note the in-sample curve exhibits the expected behaviours identified in the literature: from (positive) short-term mean-reversion to (negative) momentum.}
    \label{fig:mean_reversion_sharpes}
\end{figure}

The exact computations are detailed below, but first the reader should keep in mind stylized facts about the cross-section of returns: at short time scales (around one week, cf. for instance \cite{yeo2017risk}), the returns tend to mean-revert, generating a positive Sharpe ratio, then it slowly decreases down to the point it becomes the opposite of the cross-sectional momentum (one year plus one month, cf. \cite{asness2013value}), then the sign of the profits and losses slowly inverts again to capture slow economic cycles (around a time scale of 3 to 5 years).
This effect is partially visible in Figure \ref{fig:mean_reversion_sharpes}.
Beyond the test of one specific generative model, \emph{this effect can be considered a valuable test for synthetic returns of universes of stocks}.

\begin{figure}[ht]
    \hspace{-0cm}
    \includegraphics[width=1\linewidth]{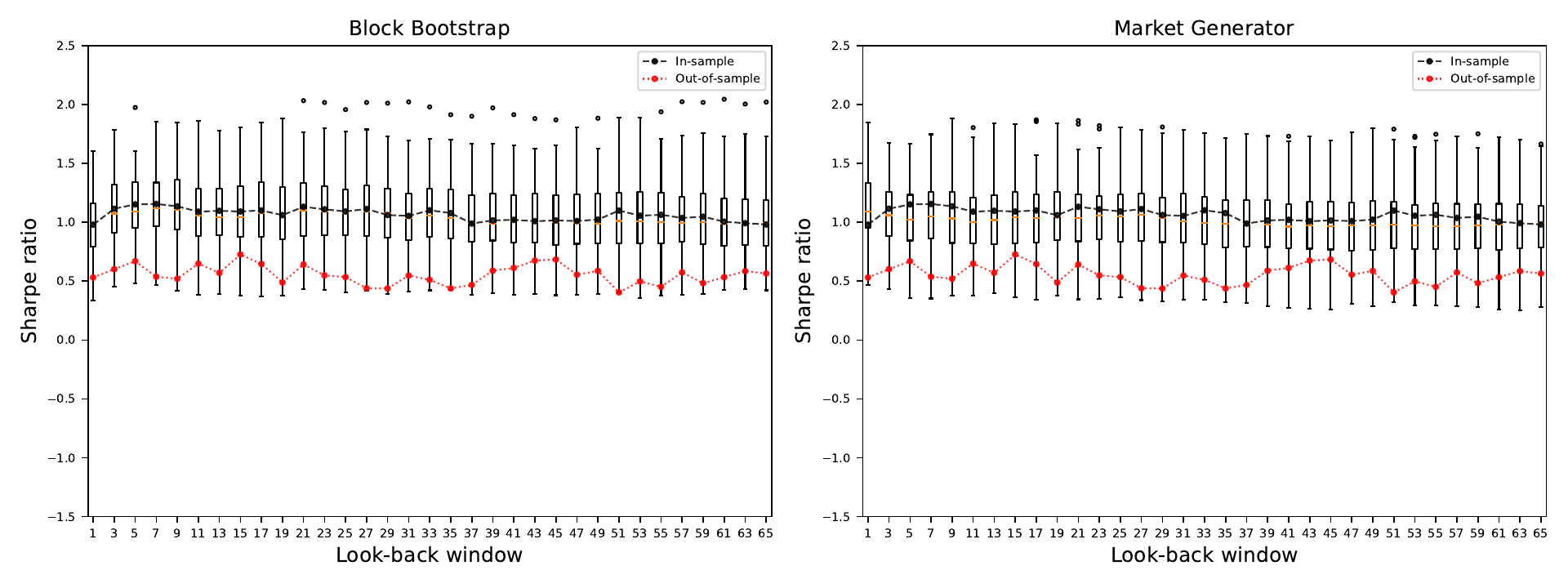}
    \caption{Boxplots of Sharpe ratios of the long-only mean-reversion strategy for different look-back window parameters $h$ backtested on samples from block bootstrap approach (left) and the market generator (right).}
    \label{fig:long_only_mean_reversion_sharpes}
\end{figure}

\paragraph{Details of the methodology.}
More specifically, let us consider a case where the market generator will be used to capture the real shape of \emph{the Sharpe ratio profile across time scales}---a curve obtained by plotting the Sharpe ratio of a given strategy as a function of varying window size parameter $h$.

\medskip

To implement the long-short mean-reversion strategy, we compute $h$-day returns for each asset in the universe, we rank them in descending order, short the first quintile and long the last quintile, assigning equal weight to each asset in both legs such that the portfolio is dollar-neutral. We also use an implementation lag of $\max(\frac{h}{10},1)$ days.\footnote{This lag guarantees that the investment strategy is causal; moreover, the usual definitions of mean-reversion and momentum involve such a lag.} 
 Long-only version of the strategy corresponds to an equal-weighted portfolio of the last quintile, all other things being equal. We assume a daily rebalancing and no transaction costs.\footnote{Transactions costs are not required to check the generated portfolio exhibit the stylized facts of mean-reversion and momentum; on the opposite, it makes the validation more straightforward.} As mentioned, we want to observe the evolution of the Sharpe ratio for different $h$ under simulated scenarios.

\medskip

We use the block bootstrap method to obtain confidence intervals for the statistics of strategies, preserving the intertemporal structure of stock returns \cite{carlstein1986use}. 
As a consequence, block bootstrapped confidence intervals on historical returns can be considered as a baseline model to compare any generative model to.

\medskip

We run the long-only and the long-short mean reversion strategies on 100 block-bootstrapped historical samples (using a window size of 63 days) and 100 simulated samples from the market generator of the size of the training set. For each sample, we test for $h \in \{1,3,5, \dots, 65\}$. It means that for each~$h$, we obtain an empirical distribution of Sharpe ratios which is illustrated in Figure~\ref{fig:long_only_mean_reversion_sharpes} for the long-only strategy under both methods. 

\begin{table}[ht!]
\centering
\begin{adjustbox}{width=\textwidth}
\begin{tabular}{c|cccc|cccc}
\toprule
 & \multicolumn{4}{c|}{Long-Only Mean Reversion} & \multicolumn{4}{c}{Long-Short Mean Reversion} \\
 \midrule
$h$ & Block Boot. & Market Gen. & IS & OoS & Block Boot. & Market Gen. & IS & OoS \\
\midrule
1 & 0.99 [0.47, 1.35] & 1.09 [0.54, 1.75] & 0.98 & 0.53 & 0.11 [-0.32, 0.57] & 0.38 [-0.36, 1.05] & 0.08 & 0.77 \\
5 & 1.09 [0.68, 1.60] & 1.02 [0.56, 1.59] & 1.15 & 0.67 & 0.70 [0.21, 1.23] & 0.08 [-0.46, 0.83] & 0.75 & 0.77 \\
9 & 1.10 [0.67, 1.62] & 1.03 [0.56, 1.60] & 1.14 & 0.52 & 0.67 [0.30, 1.18] & 0.10 [-0.38, 0.85] & 0.70 & 0.55 \\
13 & 1.05 [0.61, 1.62] & 1.02 [0.58, 1.61] & 1.10 & 0.57 & 0.45 [0.13, 0.82] & 0.03 [-0.40, 0.56] & 0.44 & 0.22 \\
17 & 1.07 [0.62, 1.60] & 1.04 [0.56, 1.55] & 1.10 & 0.65 & 0.51 [0.17, 1.01] & 0.07 [-0.36, 0.72] & 0.51 & 0.51 \\
21 & 1.10 [0.52, 1.65] & 1.03 [0.56, 1.60] & 1.13 & 0.64 & 0.56 [0.18, 1.11] & 0.03 [-0.52, 0.75] & 0.60 & 0.45 \\
25 & 1.08 [0.54, 1.61] & 1.05 [0.56, 1.67] & 1.09 & 0.53 & 0.53 [0.13, 1.09] & 0.09 [-0.45, 0.82] & 0.51 & 0.39 \\
29 & 1.07 [0.53, 1.60] & 1.04 [0.57, 1.61] & 1.06 & 0.44 & 0.51 [0.10, 1.00] & 0.02 [-0.52, 0.70] & 0.49 & 0.18 \\
33 & 1.06 [0.55, 1.59] & 1.00 [0.54, 1.58] & 1.10 & 0.51 & 0.42 [0.03, 0.90] & -0.03 [-0.64, 0.50] & 0.55 & 0.19 \\
37 & 0.99 [0.50, 1.58] & 0.99 [0.52, 1.59] & 0.99 & 0.47 & 0.30 [-0.11, 0.78] & -0.10 [-0.67, 0.49] & 0.24 & 0.42 \\
41 & 1.01 [0.51, 1.57] & 0.96 [0.52, 1.53] & 1.02 & 0.61 & 0.24 [-0.16, 0.77] & -0.21 [-0.76, 0.34] & 0.28 & 0.49 \\
45 & 1.00 [0.54, 1.58] & 0.97 [0.52, 1.55] & 1.02 & 0.68 & 0.22 [-0.22, 0.77] & -0.18 [-0.79, 0.32] & 0.29 & 0.66 \\
49 & 0.99 [0.51, 1.59] & 0.97 [0.52, 1.55] & 1.02 & 0.59 & 0.22 [-0.22, 0.75] & -0.19 [-0.81, 0.36] & 0.33 & 0.44 \\
53 & 1.01 [0.53, 1.58] & 0.97 [0.53, 1.58] & 1.06 & 0.50 & 0.23 [-0.27, 0.81] & -0.22 [-0.77, 0.32] & 0.30 & 0.36 \\
57 & 0.99 [0.51, 1.56] & 0.97 [0.50, 1.53] & 1.04 & 0.57 & 0.20 [-0.26, 0.73] & -0.24 [-0.76, 0.39] & 0.29 & 0.59 \\
61 & 0.99 [0.50, 1.52] & 0.98 [0.49, 1.55] & 1.01 & 0.53 & 0.11 [-0.36, 0.72] & -0.29 [-0.84, 0.37] & 0.20 & 0.58 \\
\bottomrule
\end{tabular}
\end{adjustbox}
\captionof{table}{Median Sharpe ratios (with 95\% confidence intervals in brackets) of the long-only and long-short mean-reversion strategies for different look-back window parameters $h$ backtested on samples from block bootstrap approach and the market generator along with in-sample (IS) and out-of-sample values (OoS).}
\label{table:sharpes}
\end{table}

\paragraph{Commenting the results.}
As expected for the long-only portfolios since their PnL is driven by the directions carrying to most variance, the median values of the block bootstrap approach align closely with the in-sample Sharpe ratios, and the out-of-sample values falling within the 95\% confidence intervals. Similarly, the market generator shows a coherent pattern, with the exception of small differences at small $h$.
The distance between empirical out-of-sample Sharpe ratios and the in-sample ones give an idea of the level of non-stationarity of financial time series: the out-of-sample ones are at extremes of the wide confidence intervals (generated by the bootstrap or by the synthetic data).

\medskip
The main features of Figure~\ref{fig:mean_reversion_sharpes} and Table~\ref{table:sharpes} is that the confidence intervals produced by the bootstrap and the synthetic data are different.
First it is expected from the theoretical results of Section \ref{sec:learning_and_application} that the risk-reward profile of Long-Short portfolio is more subtle to reproduce: it is not enough to capture the few factors with a large variance.

\medskip

By construction, block bootstrap on historical data is clearly closer to the history than what synthetic data can provide. The out-of-sample Sharpe ratios are nevertheless not in the center of its confidence bounds for time windows (i.e. time-scale) from 25 to 43 days and after 60 days.
As opposite, the time-scales of 25 to 33 of the out-of-sample Sharpe ratio are inline with the confidence intervals produced by the synthetic data.

\medskip
These evidences confront us with a question that appears each time one produces synthetic data to augment a (relatively) small sample size: how to judge the result in a context where historical out-of-sample are different from in-sample ones?

\medskip

The answer lies in the natural uncertainty due to the generative model:
it is difficult to know the truth without having access to the true generative model that generated the data (here the economic context and the way it shapes the returns of the considered stocks).
In the remainder of the paper, we propose a natural way to make such an assessment.
\emph{Once we trained a generative model, we can use it as a ground truth: we can use it to judge the capability of our family of generative models to retrieve some of these features.}

\medskip

This is, of course, not as informative as having access to the true generative model, but it can shed light on the general capability of the considered class of generative models.



\subsection{Testing for identifiability}
\label{sec:test_for_iden}




Let us recall the problem encountered at the end of Section \ref{sec:lev_mg_for_mr} regarding the boxplots for the market generator not being centered around in-sample estimates and exhibiting a different trend. Although this observation might be significant, we cannot be certain whether the Sharpe ratios computed on simulated scenarios are also far from the true values associated with the underlying data-generating process, which remains unknown in reality.

\begin{figure}[H]%
    \centering
    \subfloat{{\includegraphics[width=6.4cm]{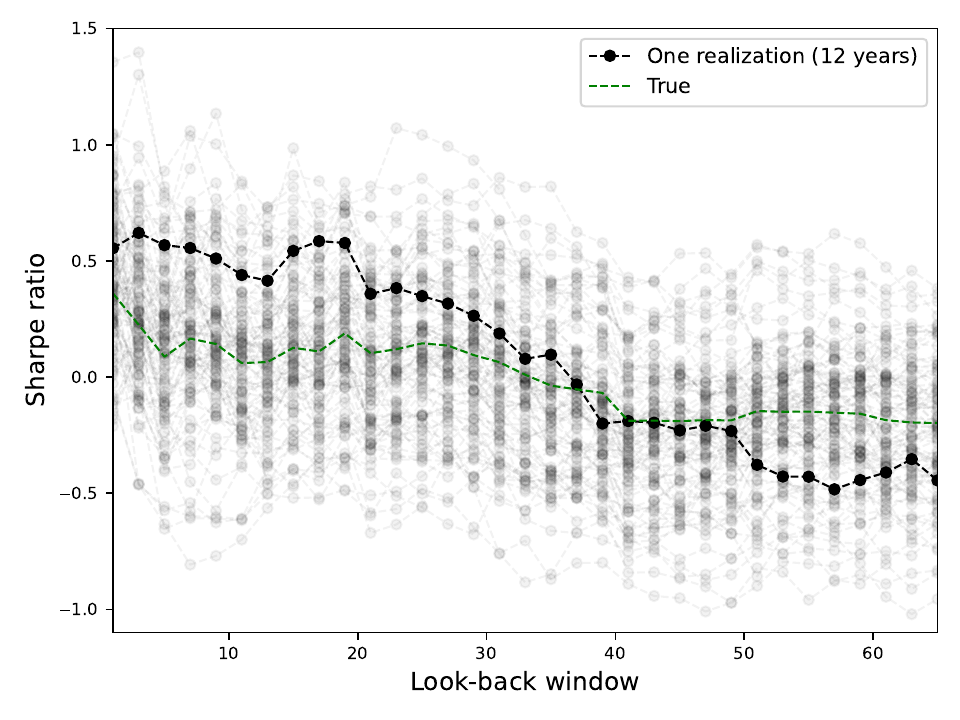} }}%
    \qquad
    \subfloat{{\includegraphics[width=6.4cm]{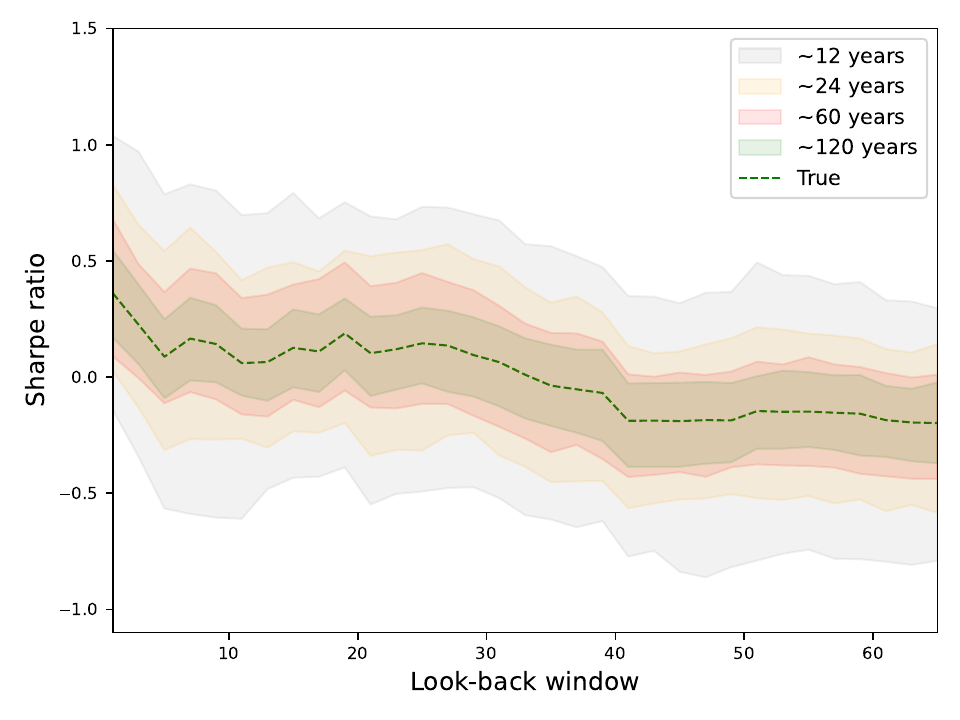} }}%
    \caption{(Left) The Sharpe ratio profile across time scales, computed on 100 simulated samples of the same size as the training set. One of the curves is randomly highlighted in bold. The true Sharpe ratios are represented by the green line. (Right) The Sharpe ratio profile across time scales, computed on 100 simulated samples of varying sizes to visualize the standard error as a function of the sample size using 95\% confidence intervals.}
    \label{fig:sharpe_errors}
\end{figure}

\paragraph{Crafting a ground truth to evaluate the modelling capability of a GAN.}
Once our generative model is trained on the in-sample historical data, we can consider it as a ground truth since we know exactly the expectation of its Sharpe ratio over time scales.
The right-hand side of Figure~\ref{fig:sharpe_errors} shows that 120 years of data are necessary to get a tight confidence interval on this ground truth. We can see that with 12 years of data (that is what we used for in-sample dataset) the sign of the Sharpe ratio itself is very difficult to assess.
These 120 years have to be put in perspective of the non-stationarity of the time series of stocks returns, i.e. on the changes of the economic context that can occur in 120 years.
Table~\ref{table:sharpes_standard_errors} shows how the accuracy of in-sample statistics in approximating the true values improves with increasing sample size. These results may highlight the challenge of optimizing long-short strategies with relatively small data sets.

\medskip



\begin{table}[h!]
\centering
\begin{tabular}{ccccc}
\toprule
$h$ & 12 years & 24 years & 60 years & 120 years \\
\midrule
1 & 0.44 (0.31) & 0.41 (0.22) & 0.39 (0.15) & 0.36 (0.10) \\
5 & 0.15 (0.33) & 0.11 (0.24) & 0.11 (0.14) & 0.09 (0.10) \\
9 & 0.17 (0.35) & 0.16 (0.22) & 0.16 (0.14) & 0.14 (0.10) \\
13 & 0.09 (0.31) & 0.07 (0.20) & 0.09 (0.12) & 0.06 (0.09) \\
17 & 0.12 (0.30) & 0.11 (0.21) & 0.13 (0.13) & 0.11 (0.09) \\
21 & 0.08 (0.31) & 0.10 (0.22) & 0.11 (0.14) & 0.10 (0.09) \\
25 & 0.12 (0.32) & 0.15 (0.21) & 0.16 (0.15) & 0.15 (0.09) \\
29 & 0.08 (0.32) & 0.09 (0.20) & 0.11 (0.13) & 0.09 (0.09) \\
33 & -0.03 (0.32) & -0.00 (0.21) & 0.02 (0.13) & 0.01 (0.09) \\
37 & -0.10 (0.31) & -0.07 (0.22) & -0.05 (0.14) & -0.05 (0.10) \\
41 & -0.25 (0.29) & -0.21 (0.19) & -0.18 (0.12) & -0.19 (0.10) \\
45 & -0.24 (0.29) & -0.21 (0.19) & -0.19 (0.12) & -0.19 (0.10) \\
49 & -0.23 (0.29) & -0.20 (0.18) & -0.18 (0.12) & -0.19 (0.09) \\
53 & -0.18 (0.29) & -0.16 (0.19) & -0.15 (0.12) & -0.15 (0.09) \\
57 & -0.18 (0.30) & -0.16 (0.19) & -0.15 (0.13) & -0.15 (0.09) \\
61 & -0.22 (0.29) & -0.19 (0.20) & -0.19 (0.12) & -0.19 (0.09) \\
\bottomrule
\end{tabular}
\captionof{table}{Mean Sharpe ratios (with standard errors in parentheses) for the long-short mean-reversion strategy on simulated samples of varying sizes (in years) by the market generator.}
\label{table:sharpes_standard_errors}
\end{table}

The left-hand chart in Figure~\ref{fig:sharpe_errors} generates 100 simulated samples, each with the same size as the training set (3,020 points, approximately 12 years of daily returns), spanning the natural uncertainty of such a short dataset.


\paragraph{Testing the identifiability of a model: the \emph{regurgitative approach}.}
Identifiability is the capability to recover the parameters of a model of the same class using data generated by this secondary model.
\medskip

What we propose as a generic method to assess the identifiability of a class of GANs:
\begin{enumerate}
\item Train an instance of this class of GAN on a given history of market data (in our illustrative case: 12 years); call it the \emph{reference market generator}.
\item Take this trained reference market generator and simulate a sample of the same size as the training set used to train it (in our case generate 12 years of data over 433 stocks). Note that you know everything about the underlying generative law for these data: you have a ground truth.
\item Train a second generative model---let us call it the \emph{regurgitative market generator}---on this synthetic sample.
\item Now, one can assess how these two models (the reference one and the regurgitated one) are close to each other.
\end{enumerate}

\begin{figure}[h!]
    \hspace{-0cm}
    \includegraphics[width=1\linewidth]{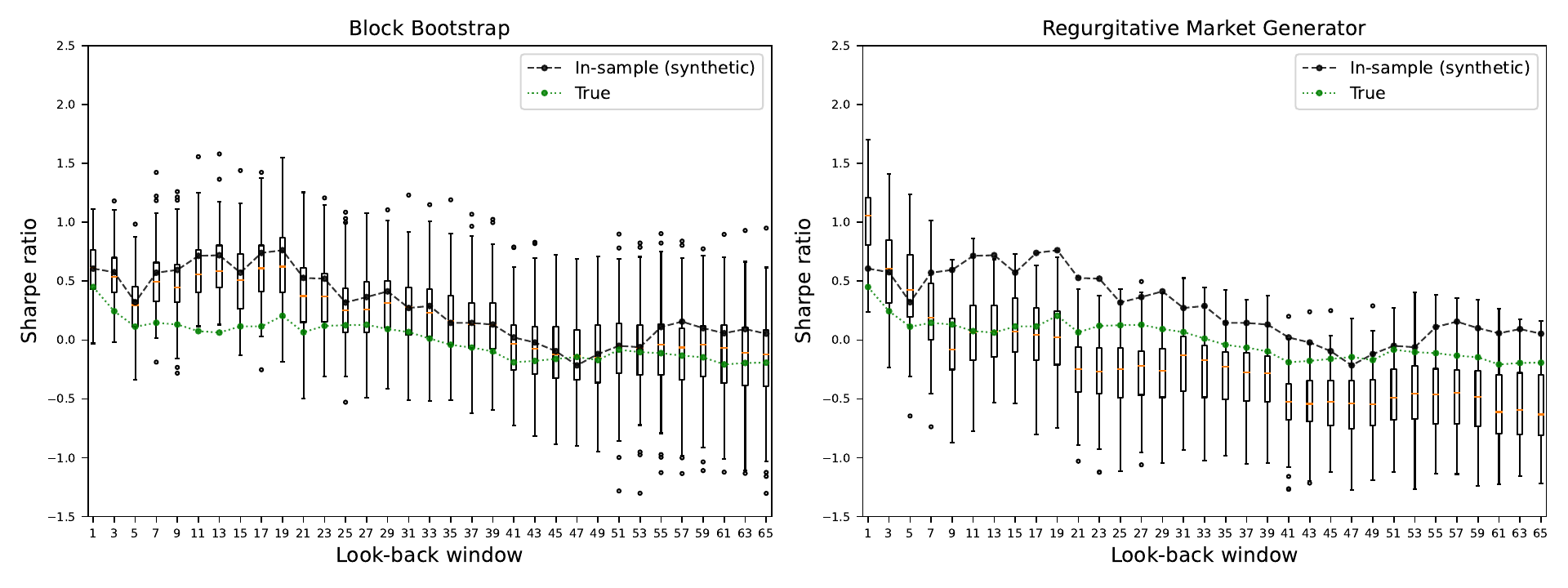}
    \caption{Boxplots of Sharpe ratios of the long-short mean-reversion strategy for different look-back window parameters $h$ backtested on samples from the block bootstrap approach (left) and the regurgitative market generator (right).}
    \label{fig:mean_reversion_sharpes_regurgitative}
\end{figure}

\begin{table}[h!]
\centering
\begin{tabular}{ccccc}
\toprule
$h$ & Block Boot. & Reg. Market Gen. & IS & True \\
\midrule
1 & 0.61 [0.07, 1.00] & 1.06 [0.29, 1.60] & 0.61 & 0.45 \\
5 & 0.30 [-0.12, 0.83] & 0.42 [-0.29, 1.13] & 0.32 & 0.11 \\
9 & 0.45 [-0.08, 1.15] & -0.08 [-0.64, 0.61] & 0.60 & 0.13 \\
13 & 0.58 [0.16, 1.17] & 0.07 [-0.46, 0.63] & 0.72 & 0.06 \\
17 & 0.61 [0.06, 1.15] & 0.05 [-0.58, 0.57] & 0.74 & 0.12 \\
21 & 0.37 [-0.11, 1.07] & -0.25 [-0.83, 0.25] & 0.53 & 0.07 \\
25 & 0.25 [-0.28, 1.00] & -0.25 [-0.93, 0.27] & 0.32 & 0.13 \\
29 & 0.32 [-0.25, 0.93] & -0.26 [-0.94, 0.32] & 0.41 & 0.09 \\
33 & 0.23 [-0.31, 0.88] & -0.17 [-0.86, 0.35] & 0.29 & 0.01 \\
37 & 0.13 [-0.40, 0.86] & -0.27 [-0.87, 0.24] & 0.15 & -0.06 \\
41 & -0.03 [-0.66, 0.60] & -0.52 [-1.12, -0.05] & 0.02 & -0.19 \\
45 & -0.12 [-0.79, 0.57] & -0.53 [-1.05, -0.00] & -0.09 & -0.16 \\
49 & -0.15 [-0.78, 0.61] & -0.54 [-1.10, 0.03] & -0.12 & -0.17 \\
53 & -0.07 [-0.84, 0.69] & -0.45 [-1.05, 0.17] & -0.06 & -0.10 \\
57 & -0.06 [-0.94, 0.68] & -0.45 [-1.04, 0.25] & 0.16 & -0.13 \\
61 & -0.07 [-0.95, 0.63] & -0.61 [-1.13, 0.12] & 0.06 & -0.21 \\
\bottomrule
\end{tabular}
\captionof{table}{Median Sharpe ratios (with 95\% confidence intervals in brackets) of the long-short mean-reversion strategies for different look-back window parameters $h$, backtested on samples from both the block bootstrap approach and the regurgitative market generator. The regurgitative market generator was trained on a synthetic sample (IS) generated by the market generator where the true Sharpe ratios are known (True).}
\label{table:regurgitative_mg}
\end{table}

We can compare their risk-reward curve for mean-reverting portfolios.
  In our case, Figure \ref{fig:mean_reversion_sharpes_regurgitative} and the corresponding Table~\ref{table:regurgitative_mg} exhibit that:
  \begin{itemize}
  \item The left panel of Figure \ref{fig:mean_reversion_sharpes_regurgitative} exhibits the block bootstrapped confidence interval on the generated 12 years of data (step 2): the known ground truth (the dotted green line) is not really in the center of the obtained boxplots (especially for short time scales).
  \item On the right panel, one can see the confidence intervals produced by the regurgitative model (noting that it was learned on the time series corresponding to the dark dashed line): it is better centered on the green dotted line (the ground truth) for small time scales.
  \end{itemize}


\medskip
The above process is not designed to validate good models but to detect bad ones. The aim is to put forward a necessary condition, not a sufficient one, and to subject the generative model to a test it must pass to be considered a viable candidate for further use. This test allows to understand whether the regurgitative market generator can capture some of the true characteristics of the risk-reward profile of a benchmark strategy (for instance the Sharpe ratio of the Equity Mean-Reversion factor across time scales).

\medskip
For our illustrative example, this test shows that our class of model should probably not be used for time scales that are longer than one month (21 business days).
It shows also that it can rediscover underlying patterns at shorter time scales, since the dotted green curve is in the center of the confidence intervals of the right panel, Figure \ref{fig:mean_reversion_sharpes_regurgitative}. 

\section{Conclusion}
In this paper, we first show rigorously the importance of the initial sample size when training generative models. It is crucial to recognize that generating more synthetic data does not necessarily improve the accuracy of estimate, on the contrary the more synthetic data are generated, the closer to the initial learning bias the results will be, leading to biased conclusions. We then prove why applying a generative model to financial data for portfolio construction, especially long-short portfolios, is problematic. This issue is fundamentally tied to the principles of both portfolio construction and the standard approach for generative modeling, that is based on matching distributions, and hence attracted by subsets with large variance.

\medskip

Next, we propose a generative pipeline that is better suited for portfolio construction following a literature review of existing models. We conclude the paper by different ways to assess the quality of generative models for financial time series, based on the identifiability through the evaluation of a regurgitative model, learned on one sample generated by the original generative model. Throughout our exploration of generative models for financial applications, we propose to use the risk-reward profile of the mean-reversion strategy at different time scales, since it combines a lot of properties of stock returns.


\newpage
\section*{Acknowledgment}
Warm thanks go to Laurence Carassus for her role as discussant and for her comprehensive discussion following a presentation of this work. The paper was presented at the University of Oxford, AMLD2025 in Lausanne, the 18th Financial Risks International Forum, and the London-Paris Bachelier Workshop in Paris. Interactions with the audience at these seminars helped us improve the paper.

\section*{Data Availability Statement}
The data that support the findings of this study are available from the corresponding author upon reasonable request.

\section*{Disclosure of Interest}
The authors declare that they have no relevant or material financial interests that relate to the research described in this paper.

\section*{Disclosure of AI Use}
The authors used ChatGPT (OpenAI, GPT-5.2) for minor language editing and text improvement. All scientific content and conclusions are the responsibility of the authors.

\section*{Funding Statement}
The authors declare that no funding was received for the conduct of this study.

\newpage

\addcontentsline{toc}{section}{References}
\bibliographystyle{apalike}
\bibliography{biblio_main}

\newpage

\addcontentsline{toc}{section}{List of notations}
\renewcommand\nomname{List of notations}

\nomenclature[01]{$X$}{a random variable (scalar, vector or matrix)}
\nomenclature[02]{$x$}{a scalar (constant)}
\nomenclature[03]{$\mathbf{x}$}{a vector (constant)}
\nomenclature[04]{$\mathbf{X}$}{a matrix (constant)}

\nomenclature[05]{$\text{diag}(\mathbf{x})$}{a diagonal matrix with diagonal entries given by $\mathbf{x}$}
\nomenclature[05]{$\text{diag}^{-1}(\mathbf{X})$}{a vector formed using the diagonal of the square matrix $\mathbf{X}$}
\nomenclature[06]{$\mathbf{I}_n$}{Identity matrix with dimensions $n \times n$}
\nomenclature[07]{$\mathbf{x}_i$}{$i^{th}$ element of the vector $\mathbf{x} = (x_1, \dots, x_n)$}
\nomenclature[08]{$\mathbf{x}_{i:j}$}{slice of $\mathbf{x}$ from $i^{th}$ to $j^{th}$ element (inclusive)}
\nomenclature[08]{$\mathbf{X}_{i:j}$}{a matrix formed by $i^{th}$ to $j^{th}$ columns (inclusive) of $\mathbf{X}$}
\nomenclature[09]{$\mathbf{X}_{i,j}$}{element  associated with $i^{th}$ row and $j^{th}$ column of the matrix $\mathbf{X}$}
\nomenclature[10]{$\mathbf{X}_{i,:}$}{$i^{th}$ row of the matrix $\mathbf{X}$}
\nomenclature[11]{$\mathbf{X}_{:,i}$}{$i^{th}$ column of the matrix $\mathbf{X}$}
\nomenclature[12]{$\mathbf{1}_n$}{a vector of size $n$ whose every element is 1}
\nomenclature[12]{$\mathbf{0}_n$}{a vector of size $n$ whose every element is 0}
\nomenclature[13]{$\mathbf{A} \odot \mathbf{B}$}{element-wise product of $\mathbf{A}$ and $\mathbf{B}$ ($\mathbf{A}_{i,j}\mathbf{B}_{i,j}$)}
\nomenclature[14]{$\mathbf{A} \oslash \mathbf{B}$}{element-wise division of $\mathbf{A}$ and $\mathbf{B}$ ($\mathbf{A}_{i,j}/\mathbf{B}_{i,j}$)}
\nomenclature[15]{$\{X(t)\}_{t \in T}$}{a stochastic process (scalar or vector-valued)}
\nomenclature[16]{$X_{i}(t)$}{$i^{th}$ element of $X(t) = (X_1(t), \dots, X_d(t))$, the random variable with the index $t$}
\nomenclature[17]{$\hat{\theta}$}{estimate of a parameter $\theta$}
\nomenclature[18]{$\bar{\mathbf{X}}$}{a matrix obtained by standardizing $\mathbf{X}$ column by column $\Bigg(\bar{\mathbf{X}}_{i,j} = \frac{\mathbf{X}_{i,j} - \frac{1}{n}\sum_{k} \mathbf{X}_{k,j}}{\sqrt{\frac{1}{n}\sum_{k} \mathbf{X}_{k,j}^2 - (\frac{1}{n}\sum_{k} \mathbf{X}_{k,j})^2}}\Bigg)$}
\nomenclature[19]{$\mathbb P_X$}{probability distribution of the random variable $X$}
\nomenclature[20]{$X \sim \mathbb P_X$}{the random variable $X$ following the probability distribution $\mathbb P_X$}
\nomenclature[21]{$\mathbb E_X$}{expectation with respect to $\mathbb P_X$ }
\printnomenclature

\newpage



\appendix
\section{Error bounds for $U$-statistics computed on synthetic data}
\label{appendix:error_bounds}
Since $U_{\Tilde{n}}$ is a $U$-statistics with mean $\Tilde{\theta}_n$, we can safely write, by following Property~(\ref{eq:berry_esseen}),
\begin{align*}
          \left|\mathbb{P}\left( \frac{\sqrt{\Tilde{n}}(U_{\Tilde{n}} - \Tilde{\theta}_n)}{r\hat{\sigma}_1} \leq x \right) - \Phi(x) \right| \leq \frac{\hat{c}}{(1+|x|)^\beta\sqrt{\Tilde{n}-r+1}}.
\end{align*}    
Let $b=\frac{xr\hat{\sigma}_1}{\sqrt{\Tilde{n}}}+a_n>0$ where $a_n=\Tilde{\theta}_n - \theta$. The above expression becomes 
\begin{align*}
 \left|\mathbb{P}\left( U_{\Tilde{n}} - \theta \leq b \right) - \Phi\left(\frac{(b-a_n)\sqrt{\Tilde{n}}}{r\hat{\sigma}_1}\right) \right| \leq \frac{\hat{c}}{\left(1+\left|\frac{(b-a_n)\sqrt{\Tilde{n}}}{r\hat{\sigma}_1}\right|\right)^\beta\sqrt{\Tilde{n}-r+1}}=\Tilde{c}(a_n,b,\Tilde{n})
\end{align*}
where $\Tilde{c}(x,y,z) = \frac{\hat{c}}{\left(1+\left|\frac{(y-x)\sqrt{z}}{r\hat{\sigma}_1}\right|\right)^\beta\sqrt{z-r+1}}$.

\medskip

It implies that

\begin{align}
\label{ineq:first}
-\Tilde{c}(a_n,b,\Tilde{n}) + \Phi\left(\frac{(b-a_n)\sqrt{\Tilde{n}}}{r\hat{\sigma}_1}\right) \leq \mathbb{P}\left( U_{\Tilde{n}} - \theta \leq b \right) \leq \Tilde{c}(a_n,b,\Tilde{n}) + \Phi\left(\frac{(b-a_n)\sqrt{\Tilde{n}}}{r\hat{\sigma}_1}\right).
\end{align}

However, we are interested in the absolute distance $b$ from $\theta$. To address this, we first present the following analogous inequality.

\begin{align}
\label{ineq:second}
-\Tilde{c}(a_n,-b,\Tilde{n}) - \Phi\left(\frac{(-b-a_n)\sqrt{\Tilde{n}}}{r\hat{\sigma}_1}\right) \leq -\mathbb{P}\left( U_{\Tilde{n}} - \theta \leq -b \right) \leq \Tilde{c}(a_n,-b,\Tilde{n}) - \Phi\left(\frac{(-b-a_n)\sqrt{\Tilde{n}}}{r\hat{\sigma}_1}\right).
\end{align}

Note that, $$\mathbb{P}\left( U_{\Tilde{n}} - \theta \leq b \right) - \mathbb{P}\left( U_{\Tilde{n}} - \theta \leq -b \right) = \mathbb{P}\left(|U_{\Tilde{n}} - \theta| \leq b \right)$$ where $b>0$ and also
$$\Phi(-x) = 1 - \Phi(x).$$

Therefore, the following bound can be obtained by combining Inequality~(\ref{ineq:first}) and (\ref{ineq:second}),

\begin{equation*}
\left| \mathbb{P}\left(|U_{\Tilde{n}} - \theta| \leq b \right) - \left[ \Phi\left(\frac{(a_n+b)\sqrt{\Tilde{n}}}{r\hat{\sigma}_1}\right) - \Phi\left(\frac{(a_n-b)\sqrt{\Tilde{n}}}{r\hat{\sigma}_1}\right)\right] \right| \leq \Tilde{c}(a_n,b,\Tilde{n}) + \Tilde{c}(a_n,-b,\Tilde{n}).
\end{equation*}

\newpage

\section{Error propagation due to eigenvector perturbation}
\label{appendix:perturb}
Using Equation~(\ref{eq:MPO:simple}), for any arbitrary vector $\mathbf{z}$, 

$$\bm{\Tilde{\Sigma}}_{(k)}^{-1} \mathbf{z} = \frac{1}{\lambda_1}\,\langle \mathbf{z},\mathbf{\Tilde{P}}_{:,1} \rangle\,\mathbf{\Tilde{P}}_{:,1}+ \sum_{\substack{1<i\leq m \\ i\neq k}} \frac{1}{\lambda_i}\,\langle \mathbf{z},\mathbf{P}_{:,i} \rangle\,\mathbf{P}_{:,i} +\frac{1}{\lambda_k}\,\langle \mathbf{z},\mathbf{\Tilde{P}}_{:,k} \rangle\,\mathbf{\Tilde{P}}_{:,k}+ \frac{1}{\lambda_c}\sum_{i\leq d-m}\langle \mathbf{z},\mathbf{Q}_{:,i} \rangle\,\mathbf{Q}_{:,i}$$

where $\bm{\Tilde{\Sigma}}_{(k)}$ is given by (\ref{new_cov}).

\medskip

Recall that for $k>1$, 
 $$
 \mathbf{\Tilde{P}}_{:,1} = -\sin \epsilon \cdot \mathbf{P}_{:,k} + \cos \epsilon \cdot \mathbf{P}_{:,1}
     \quad
\text{and}  \quad \mathbf{\Tilde{P}}_{:,k} = \cos \epsilon \cdot \mathbf{P}_{:,k} + \sin \epsilon \cdot \mathbf{P}_{:,1}.$$  

We are actually interested in, 
$$\bm{\Tilde{\Sigma}}_{(k)}^{-1} \mathbf{z} - \bm{\Sigma}^{-1} \mathbf{z} = \frac{1}{\lambda_1}(\underbrace{\langle \mathbf{z},\mathbf{\Tilde{P}}_{:,1} \rangle\,\mathbf{\Tilde{P}}_{:,1}-\langle \mathbf{z},\mathbf{P}_{:,1} \rangle\,\mathbf{P}_{:,1}}_{a_k}) + \frac{1}{\lambda_k}(\underbrace{\langle \mathbf{z},\mathbf{\Tilde{P}}_{:,k} \rangle\,\mathbf{\Tilde{P}}_{:,k}-\langle \mathbf{z},\mathbf{P}_{:,k} \rangle\,\mathbf{P}_{:,k}}_{b_k})$$
where
$$a_k = \left(\sin^2\epsilon\,\langle \mathbf{z},\mathbf{P}_{:,k} \rangle - \sin\epsilon\cos\epsilon \,\langle \mathbf{z},\mathbf{\mathbf{P}_{:,1}} \rangle\right)\, \mathbf{P}_{:,k} + \left(-\sin\epsilon\cos\epsilon\,\langle \mathbf{z},\mathbf{P}_{:,k} \rangle -\sin^2\epsilon \,\langle \mathbf{z},\mathbf{\mathbf{P}_{:,1}} \rangle\right)\, \mathbf{\mathbf{P}_{:,1}}$$
and 
$$b_k = \left(-\sin^2\epsilon\,\langle \mathbf{z},\mathbf{P}_{:,k} \rangle+ \sin\epsilon\cos\epsilon \,\langle \mathbf{z},\mathbf{P}_{:,1} \rangle\right)\, \mathbf{P}_{:,k} + \left(\sin\epsilon\cos\epsilon\,\langle \mathbf{z},\mathbf{P}_{:,k} \rangle+ \sin^2\epsilon \,\langle \mathbf{z},\mathbf{P}_{:,1} \rangle\right)\,\mathbf{P}_{:,1}$$

using the identity $\cos^2\epsilon+\sin^2\epsilon=1$. Since $b_k=-a_k$, 

$$\bm{\Tilde{\Sigma}}_{(k)}^{-1} \mathbf{z} - \bm{\Sigma}^{-1} \mathbf{z} = \left(\frac{1}{\lambda_1}-\frac{1}{\lambda_k}\right)a_k.$$

Therefore, 
\begin{align*}
\delta_{(k)}(\epsilon, \mathbf{z}) &= \bigl\| \bm{\Tilde{\Sigma}}_{(k)}^{-1} \mathbf{z} - \bm{\Sigma}^{-1} \mathbf{z} \bigl\|^2=\left(\frac{1}{\lambda_1}-\frac{1}{\lambda_k}\right)^2 \bigl\|a_k\bigl\|^2 \\
&=\left(\frac{1}{\lambda_1}-\frac{1}{\lambda_k}\right)^2 \left[\left(\sin^2\epsilon\,\langle \mathbf{z},\mathbf{P}_{:,k} \rangle - \sin\epsilon\cos\epsilon \,\langle \mathbf{z},\mathbf{\mathbf{P}_{:,1}} \rangle\right)^2+\left(\sin\epsilon\cos\epsilon\,\langle \mathbf{z},\mathbf{P}_{:,k} \rangle+ \sin^2\epsilon \,\langle \mathbf{z},\mathbf{P}_{:,1} \rangle\right)^2\right] \\
&=\left(\frac{1}{\lambda_1}-\frac{1}{\lambda_k}\right)^2\left[\sin^4\epsilon\left(\langle \mathbf{z},\mathbf{P}_{:,k}\rangle^2 + \langle \mathbf{z},\mathbf{P}_{:,1}\rangle^2\right)+\cos^2\epsilon\sin^2\epsilon\left(\langle \mathbf{z},\mathbf{P}_{:,k}\rangle^2 + \langle \mathbf{z},\mathbf{P}_{:,1}\rangle^2\right)\right] \\
&= \left(\frac{1}{\lambda_1}-\frac{1}{\lambda_k}\right)^2\sin^2\epsilon\left(\langle \mathbf{z},\mathbf{P}_{:,k}\rangle^2 + \langle \mathbf{z},\mathbf{P}_{:,1}\rangle^2\right).
\end{align*}

\newpage

\section{Architecture and training parameters}
\label{appendix:TCN}

In Table~\ref{table:nn_params}, we outline the architecture and training parameters of the GAN used to model factor returns. While a full description of the architecture is beyond the scope of this paper, it is based on the work \cite{wiese_quant_2020}, which offers a comprehensive explanation of temporal convolutional networks and the elements listed in the table that may be unfamiliar to readers. We recommend consulting that reference for further clarity.

\begin{table}[ht]
\centering
\begin{tabular}{llr}
\toprule
\textbf{Category} & \textbf{Parameter} & \textbf{Value} \\
\midrule

\multirow{3}{*}{Training set dimension} 
& Cluster 1 & $(2958 \times 1)  \times 63 $ \\
& Cluster 2 & $(2958 \times 10) \times 63$ \\
& Cluster 3 & $(2958 \times 5) \times 63 $ \\
\midrule

\multirow{10}{*}{Generator} 
& Architecture & TCN with skip connections \\
& Hidden layers & 6 temporal blocks${}^*$\\
& Hidden layer dimension & 100\\
& Dilations & (1,1,2,4,8,16) \\
& Kernel size & (1,2,2,2,2,2) \\
& Output layer & $1\times1$ Convolution \\
& Noise distribution & Gaussian \\ 
& Input (noise) dimension${}^\ddagger$ & $(s + 63 - 1) \times 3$ \\
& Output dimension & $ s \times 1$ \\
& Batch normalization & True \\
\midrule

\multirow{9}{*}{Discriminator} 
& Architecture & TCN with skip connections \\
& Hidden layers & 6 temporal blocks${}^*$\\
& Hidden layer dimension & 100\\
& Dilations & (1,1,2,4,8,16) \\
& Kernel size & (1,2,2,2,2,2) \\
& Output layer & $1\times1$ Convolution \\
& Input dimension & $63 \times 1$ \\
& Output dimension & $ 1$ \\
& Batch normalization & False \\
\midrule

\multirow{6}{*}{Training parameters} 
& Loss function & Binary cross-entropy \\
& Mini-batch size & 128 \\
& Generator learning rate & $5 \times 10^{-6}$ \\
& Discriminator learning rate & $5 \times 10^{-5}$ \\
& Optimizer & Adam ($\epsilon=10^{-8}$, $\beta_1=0$, $\beta_2=0.9$) \\
& Stopping criteria & 50,000 mini-batch iterations \\
\bottomrule
\end{tabular}
\caption{Overview of the chosen GAN architecture and training parameters. *A temporal block is composed of two dilated causal convolutions, each followed by a PReLU activation function. ${}^\ddagger$During training, $s$ is set to 63. After the training, $s$ can be adjusted so that the generator can produce time series of the desired length. For instance, in the numerical part, $s$ is chosen to be 3020 to match the length of the in-sample set.}
\label{table:nn_params}
\end{table}

\newpage

\section{Supplementary figures for visual analysis}
\label{appendix:visual}

\medskip

\medskip

\begin{figure}[ht]
    \hspace{-0cm}
    \includegraphics[width=1\linewidth]{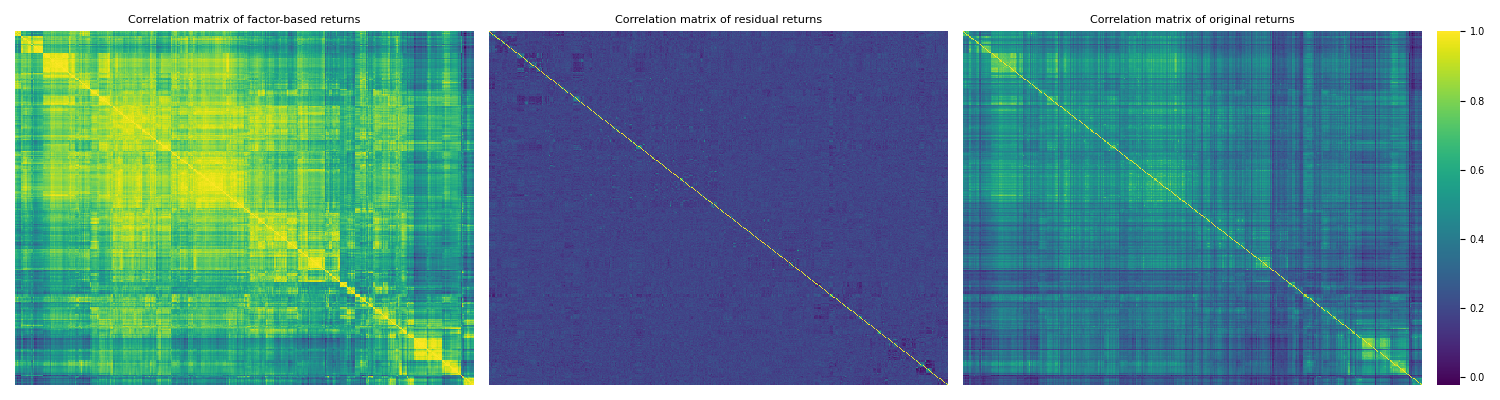}
    \caption{Sample correlation matrices for factor-based, residual and original returns.}
    \label{fig:correl_of_three_comps}
\end{figure}

\medskip

\medskip

\medskip

\medskip

\begin{figure}[ht]
    \hspace{-0cm}
    \includegraphics[width=1\linewidth]{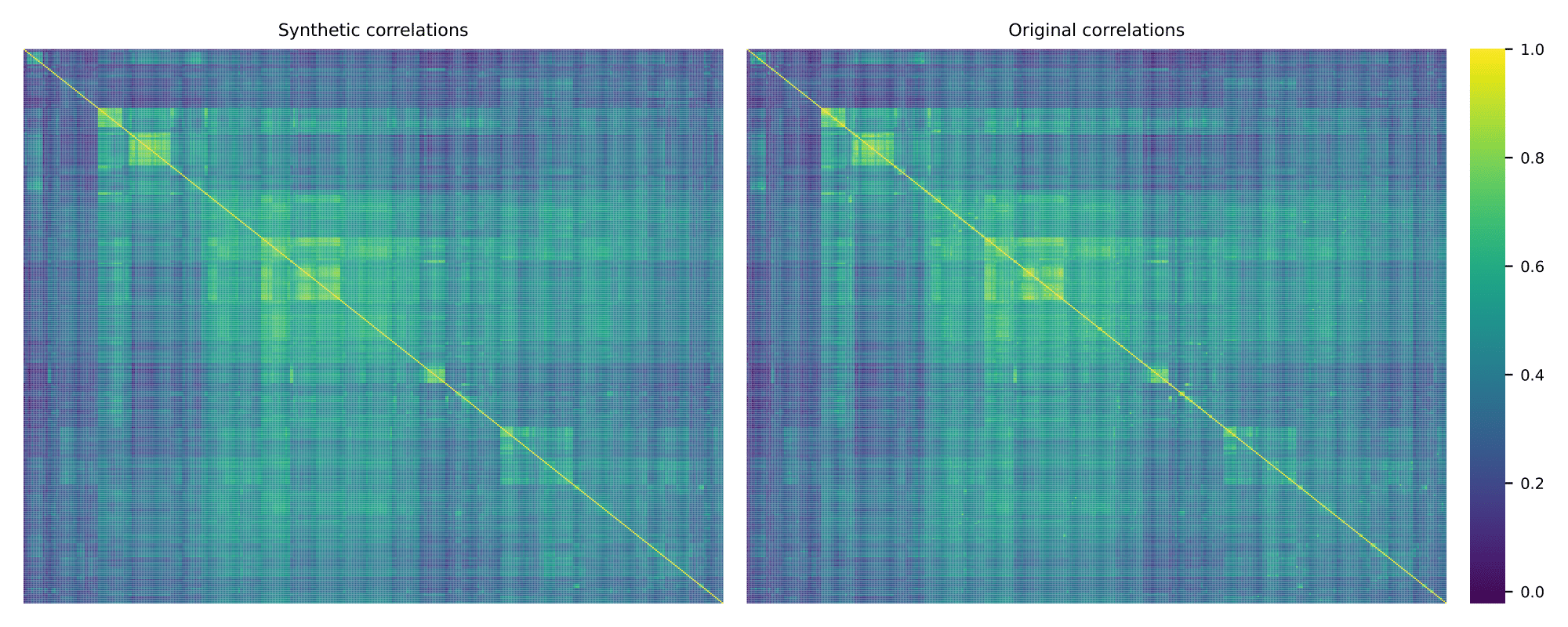}
    \caption{Correlation matrices computed on historical and one simulated sample.}
    \label{fig:corr_syn_vs_hist}
\end{figure}

\begin{landscape}
\begin{figure}[ht]
    \hspace{-0cm}
    \includegraphics[width=1\linewidth]{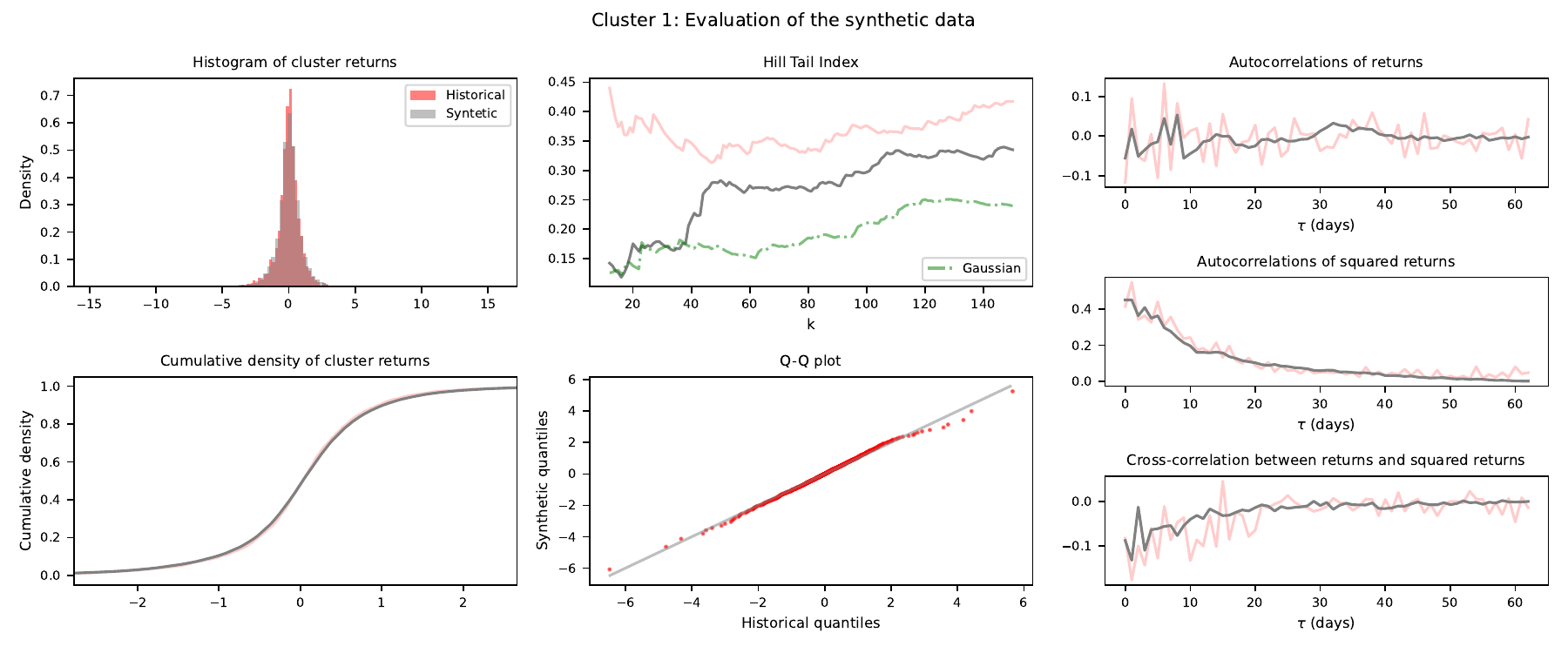}
    \caption{Evaluation of the first GAN.}
    \label{fig:first_gan}
\end{figure}
\end{landscape}

\begin{landscape}
\begin{figure}[ht]
    \hspace{-0cm}
    \includegraphics[width=1\linewidth]{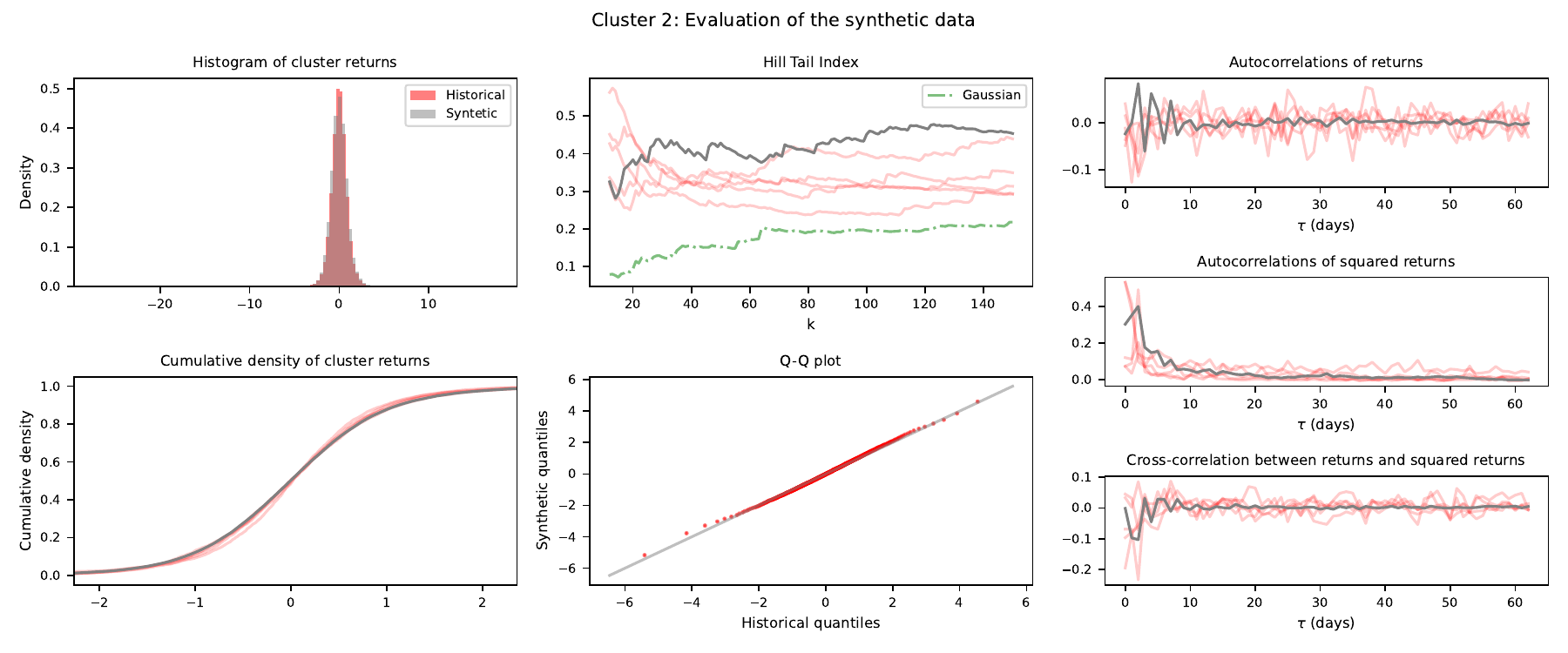}
    \caption{Evaluation of the second GAN.}
    \label{fig:second_gan}
\end{figure}
\end{landscape}

\begin{landscape}
\begin{figure}[ht]
    \hspace{-0cm}
    \includegraphics[width=1\linewidth]{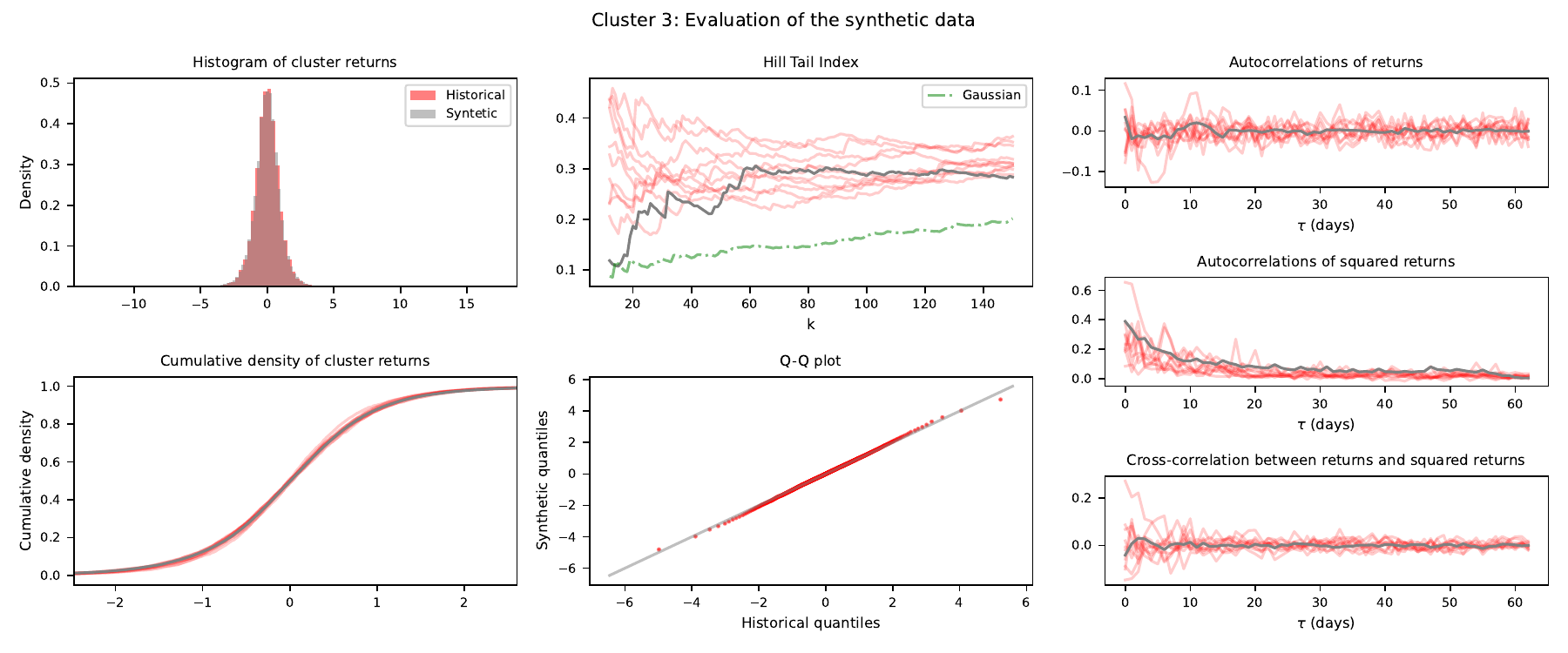}
    \caption{Evaluation of the third GAN.}
    \label{fig:third_gan}
\end{figure}
\end{landscape}

\begin{landscape}
\begin{figure}[ht]
    \hspace{-0cm}
    \includegraphics[width=1\linewidth]{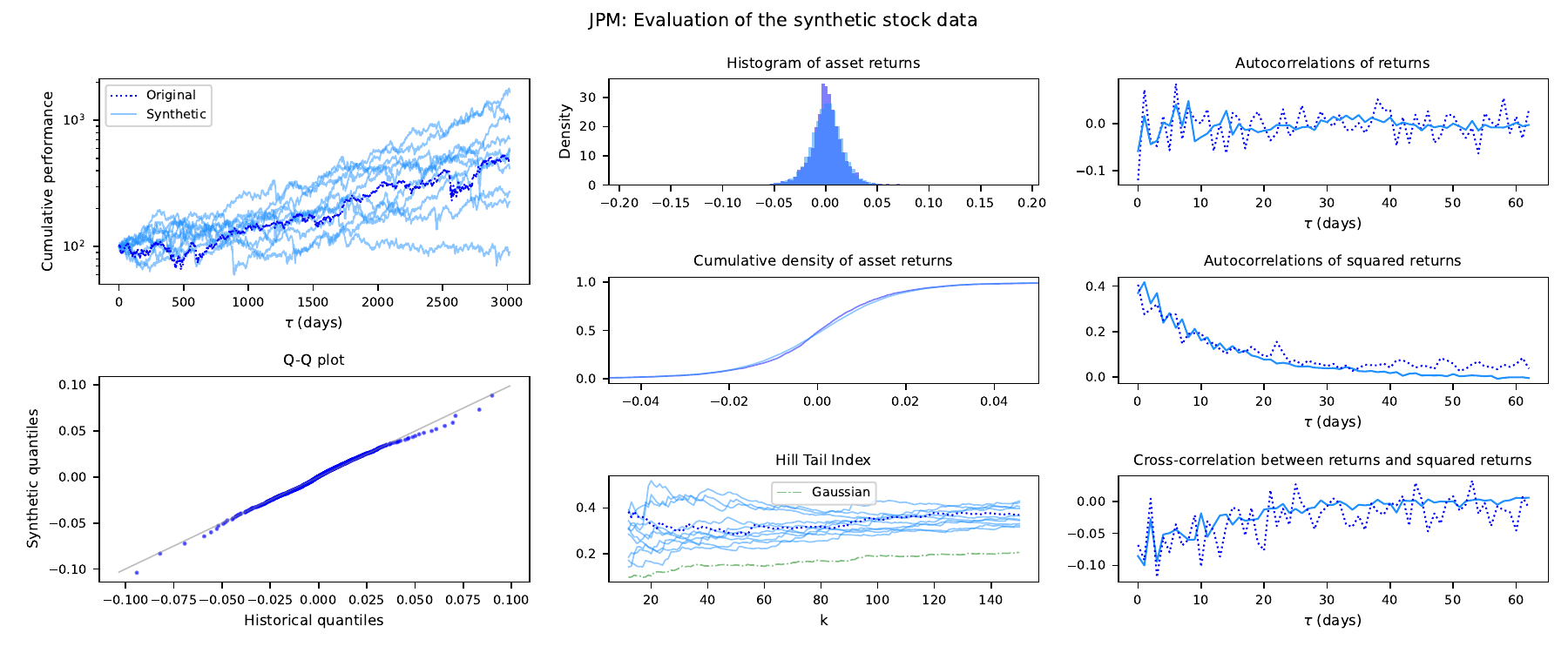}
    \caption{Asset-based monitoring screen for a specific stock.}
    \label{fig:jpm_eval}
\end{figure}
\end{landscape}

\end{document}